\documentclass{article}

\usepackage[colorlinks=true,linkcolor=blue,citecolor=blue,urlcolor=blue,pdfborder={0 0 0}]{hyperref}
\usepackage[utf8]{inputenc}
\usepackage[english]{babel}
\usepackage{amsfonts,amsmath,amssymb,amsthm,commath,wasysym,mathrsfs}
\usepackage{xcolor}
\usepackage{multirow}
\usepackage{multicol}
\usepackage{geometry}
\usepackage{graphicx}
\usepackage{pgf}
\usepackage{ragged2e}
\usepackage{makecell}
\usepackage{enumitem}
\usepackage{calc}
\usepackage{subcaption}

\usepackage{cleveref}

\crefname{theorem}{Theorem}{Theorems}
\crefname{thm}{Theorem}{Theorems}
\crefname{lemma}{Lemma}{Lemmas}
\crefname{claim}{Claim}{Claims}
\crefname{lem}{Lemma}{Lemmas}
\crefname{remark}{Remark}{Remarks}
\crefname{prop}{Proposition}{Propositions}
\crefname{proposition}{Proposition}{Propositions}
\crefname{defn}{Definition}{Definitions}
\crefname{definition}{Definition}{Definitions}
\crefname{corollary}{Corollary}{Corollaries}
\crefname{conjecture}{Conjecture}{Conjectures}
\crefname{question}{Question}{Questions}
\crefname{chapter}{Chapter}{Chapters}
\crefname{section}{\S}{Sections}
\crefname{subsection}{\S}{Sections}
\crefname{subsubsection}{\S}{Sections}
\crefname{part}{Part}{Parts}
\crefname{figure}{Figure}{Figures}
\crefname{table}{Table}{Tables}

\usepackage{fancyhdr}
\renewcommand{\P}{\mathbb P}
\newcommand{\E}{\mathbb E}

\newcommand{\R}{\mathbb R}
\newcommand{\Z}{\mathbb Z}

\newcommand{\cH}{\mathcal H}

\newcommand{\bbG}{\mathbb G}

\newcommand{\bbX}{\mathbb X}

\renewcommand{\abs}[1]{|#1|}

\usepackage[export]{adjustbox}

\graphicspath{ {./images/} }

\newenvironment{Figure}
{\par\medskip\noindent\minipage{\linewidth}}
{\endminipage\par\medskip}

\newtheorem{theorem}{Theorem}

\newtheorem{conjecture}[theorem]{Conjecture}

\theoremstyle{remark}

\makeatletter
\newenvironment{tablehere}
{\def\@captype{table}}
{}

\makeatother

\usepackage{booktabs,caption}
\usepackage[flushleft]{threeparttable}

\title{\Huge{\textsc{What are the limits of universality?}}}
\usepackage{titlesec}

\titleformat{\section}[block]{\Large\bfseries\filcenter }{\S\thesection:}{0.3em}{}
\titleformat{\subsection}[block]{\large\bfseries\filcenter }{\S\thesubsection:}{0.3em}{}
\titleformat{\subsubsection}[block]{\large\bfseries\filcenter }{\S\thesubsubsection:}{0.3em}{}

\author{Noah Halberstam and Tom Hutchcroft\\
\footnotesize{Department of Pure Mathematics and Mathematical Statistics, University of Cambridge}\\
\footnotesize{Email: \href{mailto:nh448@cam.ac.uk}{nh448@cam.ac.uk} and \href{mailto:t.hutchcroft@maths.cam.ac.uk}{t.hutchcroft@maths.cam.ac.uk}}}

\titleformat{\paragraph}
{\normalfont\normalsize\bfseries}{\theparagraph}{1em}{}
\titlespacing*{\paragraph}
{0pt}{3.25ex plus 1ex minus .2ex}{1.5ex plus .2ex}

\begin{document}

\crefformat{section}{#2{}\S#1{}#3}
\crefformat{subsection}{#2{}\S#1{}#3}
\crefformat{subsubsection}{#2{}\S#1{}#3}

\maketitle

\begin{abstract}
It is a central prediction of renormalisation group theory that the critical behaviours of many statistical mechanics models on Euclidean lattices depend only on the dimension and not on the specific choice of lattice. 
We investigate the extent to which this universality continues to hold beyond the Euclidean setting, taking as case studies Bernoulli bond percolation and lattice trees.
We present strong numerical evidence that the critical exponents governing these models on transitive graphs of polynomial volume growth depend only on the volume-growth dimension of the graph and not on any other large-scale features of the geometry. For example, our results strongly suggest that percolation, which has upper-critical dimension six, has the same critical exponents on 
$\Z^4$
 and the Heisenberg group despite the distinct large-scale geometries of these two lattices preventing the relevant percolation models from sharing a common scaling limit.

On the other hand, we also show that no such universality should be expected to hold on fractals, even if one allows the exponents to depend on a large number of standard fractal dimensions. Indeed, we give natural examples of two fractals which share Hausdorff, spectral, topological, and topological Hausdorff dimensions but exhibit distinct numerical values of the percolation Fisher exponent $\tau$. This gives strong evidence against a conjecture of Balankin et al.\ [Phys. Lett. A 2018].

\end{abstract}

\vspace{1em}

	\begin{multicols}{2}

\phantom{.}
\vspace{-2.5em}

		\section{Introduction}

	For many models of statistical physics, the critical behaviour of the system is believed to be dependent solely on the large-scale geometry of the substrate, independently of the microscopic details of its geometry. The behaviour at criticality in encoded in a set of \emph{critical exponents} which describe how properties of the model are dependent on the length scale at which the system is observed. These critical exponents are often summarised as a function of the dimension of the substrate under consideration, and, fascinatingly, apparently unrelated models are often found to share the same critical exponents. This phenomenon is known as \emph{universality}, and systems with identical exponents are grouped together into  \emph{universality classes}. For background on the universality phenomenon and its renormalization group interpretations, see e.g.\ \cite{cardy_1996,doi:10.1142/S0217751X13300500,RevModPhys.45.589}.

	 Underlying the phenomenon of universality is the fact that Euclidean lattices have a single well-defined dimension which determines all their large-scale geometric features via their common scaling limit $\mathbb{R}^d$. In contrast, it is possible in more general settings to have many potentially inequivalent notions of dimension, and even to have multiple substrates for which all these notions of dimension agree but which nevertheless have highly distinct large-scale geometries. This raises several interesting questions: can we characterise the set of geometric features of the substrate on which the critical exponents depend? It is possible that they depend only on the dimensions?
	 To what extent do the answers to these questions depend on the model under consideration? In other words, how universal is universality?

	In this paper, we study these questions in two classes of geometric setting: transitive (possibly non-Euclidean) lattices with polynomial volume growth and self-similar fractals.
	Our results in these two cases push in opposite directions. For transitive lattices we present clear numerical evidence that the critical exponents depend only on the dimension, suggesting that a very strong form of universality should hold in this setting. In stark contrast, we construct two self-similar fractals for which a large number of standard dimensions coincide but which do not appear to have the same critical exponents for Bernoulli bond percolation, showing that no such universality should be expected to hold in this case.

		\begin{figure*}
		\centering

		\hspace{0.5cm}\includegraphics[height=7cm]{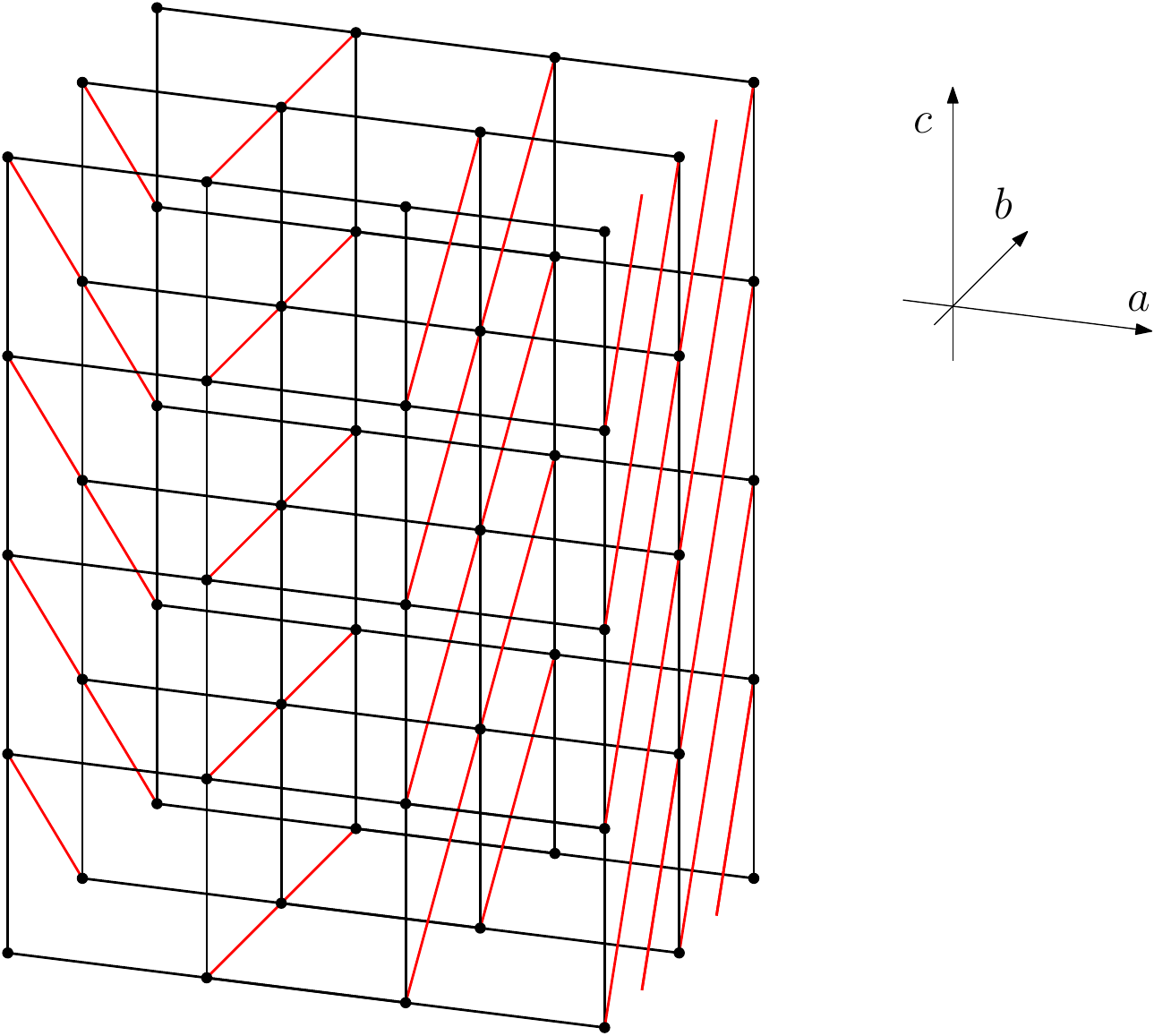} \hspace{1.2cm} \includegraphics[height=7cm]{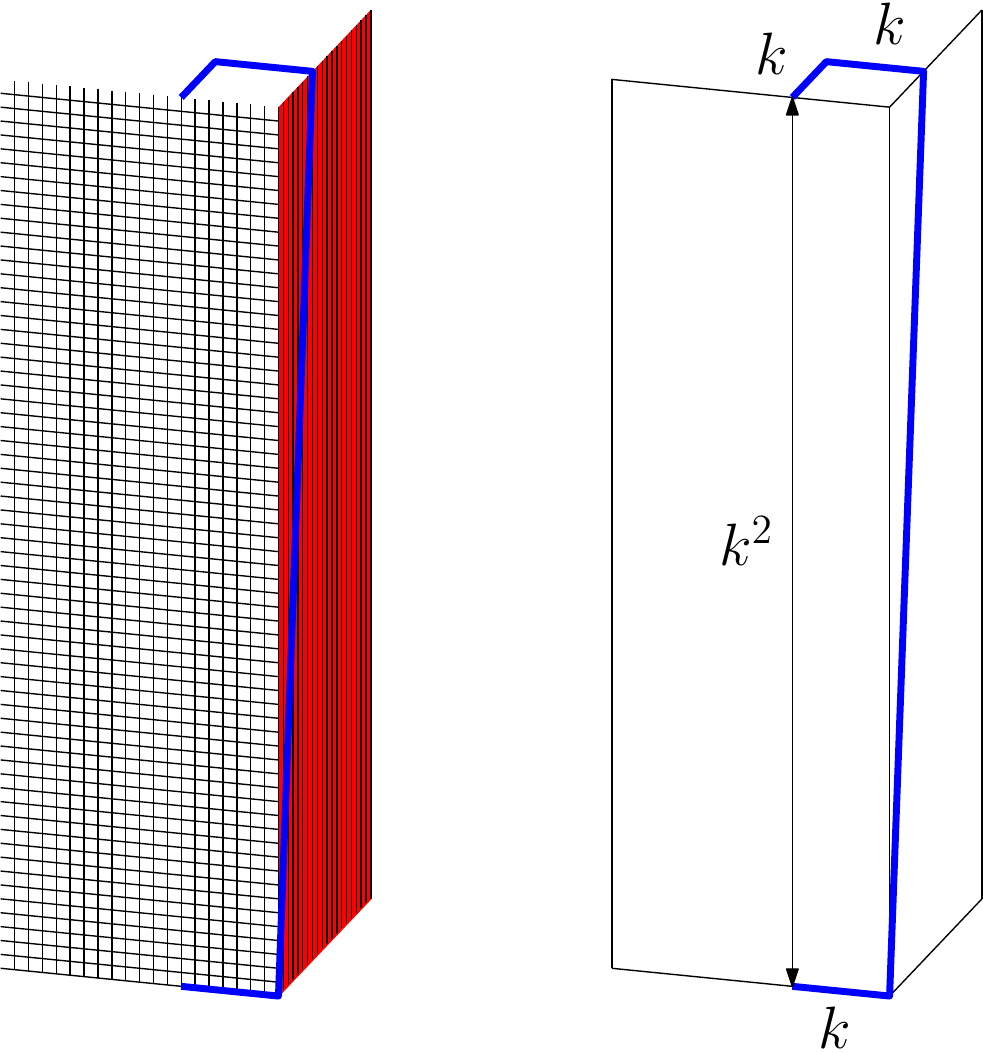}
		\captionof{figure}{The non-Euclidean geometry of the Heisenberg group. Left: A section of a Cayley graph of the Heisenberg group with generators $a$, $b$, and $c=[a,b]$. This graph may be obtained from the cubic lattice $\mathbb{Z}^3$ by applying a vertical shear of coefficient $n$
		to each of the hyperplanes $\{(a,b,c): a=n\}$. (Note that the $a \leftrightarrow b$ asymmetry of this picture arises from our choice to take the right Cayley graph rather than the left Cayley graph.) Right: One may reach $(0,0,k^2)$ from $(0,0,0)$ in $4k$ steps by first going $k$ steps in the $a$ direction, then  $k$ steps in the direction $(0,1,k)$, then  $k$ steps in the negative $a$ direction, then finally coming back $k$ steps in the negative $b$ direction. This leads to the Heisenberg group having volume-growth dimension $4$ rather than $3$. In fact the graph metric on the Heisenberg group is comparable the the quasi-norm $\|(a,b,c)\|=|a|+|b|+|c|^{1/2}$. 
		To illustrate just how alien the geometry of this space is, let us mention a theorem of Monti and Rickley \cite{MR2135806} which states that any three non-colinear points in the continuum Heisenberg group have the entire space as their convex hull.
		}
		\label{fig:Heisenberg}
	\end{figure*}

	\subsection{Transitive graphs of polynomial growth}
  We now introduce the class of transitive graphs that we will study. Recall that a graph is said to be \emph{transitive} if any vertex can be mapped to any other vertex by a symmetry of the graph. A transitive graph has \emph{polynomial volume growth} if there exists a constant $C$ such that
	$\abs{B(v,r)}\leq Cr ^C$
	for every $r\geq 1$, where $B(v,r)$ is the graph distance ball  of radius $r$ around the vertex~$v$.
	While the hypercubic lattices $\mathbb{Z}^d$ are trivially seen to be transitive graphs of polynomial volume growth, there are also many examples of highly non-Euclidean transitive graphs of polynomial volume growth.
	Indeed, the possible large scale geometries of these graphs are classified by famous theorems of Gromov \cite{gromov81poly} and Trofimov \cite{MR811571} which imply that every transitive graph of polynomial volume growth is quasi-isometric to a Cayley graph of a torsion-free nilpotent group.  A theorem of Bass \cite{bass72poly-growth} and Guivarc'h \cite{guivarch73poly-growth} then implies that every transitive graph of polynomial growth has a well-defined \emph{integer} dimension $d$ such that
	\[C^{-1}r^d \leq\abs{B(v,r)}\leq Cr ^d\]
	for some constant $C$ and every $r\geq1$.	
This same dimension $d$ also arises as the spectral and isoperimetric dimensions of the graph by a theorem of Coulhon and Saloff-Coste \cite{MR1232845}.

	 In the low-dimensional cases $1\leq d\leq 3$, it is a consequence
	   of the Bass-Guivarc'h formula and the classification of low-dimensional nilpotent Lie algebras \cite{MR2303198} that there is only one possible large-scale geometry, namely that of $\Z^d \approx \R^d$. For $d=4$, there are exactly two possible large-scale geometries exemplified by the abelian group $\mathbb{Z}^4$ and the \emph{Heisenberg group} (Figure \ref{fig:Heisenberg}), i.e.\ the $3\times3$ matrix group
	\[
	\mathcal{H} =\Biggl\{ \begin{pmatrix}
		1 & a & c\\
		0 & 1 & b\\
		0 & 0 & 1
	\end{pmatrix}:a,b,c\in\mathbb{Z}\Biggr\}.
	\]
The fact that Heisenberg group has distinct large-scale geometry than $\mathbb{Z}^4$ is evidenced by the fact that its scaling limit is not $\mathbb{R}^4$ but is instead the \emph{continuum Heisenberg group} equipped with its \emph{Carnot–Carathéodory metric} -- a self-similar \emph{sub-Riemannian} manifold that is homeomorphic to $\mathbb{R}^3$ but has Hausdorff dimension $4$ \cite{MR3742567}. 
For $d=5$ there are again exactly two quasi-isometry classes, namely those of $\mathbb{Z}^5$ and $\mathcal{H}\times\mathbb{Z}$. 
In higher dimensions the number of possibilities is much larger, and indeed the classification of possible geometries is not completely understood \cite[Section 19.7]{MR3793294}. As with the Heisenberg and continuum Heisenberg groups above, each finitely generated torsion-free nilpotent group has an associated nilpotent Lie group, known as its \emph{Mal'cev completion}, which contains the group as a lattice and which carries a self-similar sub-Riemannian metric arising as the scaling limit of its Cayley graphs by a theorem of Pansu \cite{MR741395}. Further background on these topics can be found in the surveys \cite{MR1421823,MR3793294,MR3742567,MR1902363}.

In this paper we simulate critical Bernoulli bond percolation on $\mathcal{H}$ and $\mathcal{H}\times\mathbb{Z}$, and lattice trees on $\mathcal{H}$, $\mathcal{H} \times \mathbb{Z}$, and two non-Euclidean seven-dimensional geometries known as $G_{4,3}$ and $G_{5,8}$. Here, a \emph{uniform lattice tree} is simply a finite subtree of the lattice chosen uniformly at random among those subtrees that contain the origin and have some  fixed number of vertices $n$; detailed definitions of both models and of the graphs we work with are given in \cref{section:models} and \cref{section:TLattices}. As summarised in Table \ref{table:summary}, the numerical values of the critical exponents we obtain are in good agreement with previous results for Euclidean lattices,
providing strong evidence in favour of the following conjecture:
\begin{conjecture}\label{conj:1}
	The critical exponents describing Bernoulli percolation and lattices trees on transitive graphs of polynomial growth are each determined by the volume-growth dimension of the graph.
\end{conjecture}

We also expect similar conjectures to hold for many other models; see \cref{section:OQD} for further discussion. Be careful to note that the exponent estimates reported in \cref{table:summary} are only one facet of the evidence we provide in favour of \cref{conj:1}, with a more nuanced perspective on the data presented in \cref{section:TLattices}.


\begin{table*}[t!]
\vspace{-0.2cm}
\centering
\begin{subtable}{\textwidth}
	\centering
	\caption{\textsc{Critical exponent estimates for percolation.}}
	\label{table:EvNE}
	\begin{tabular}{|l|l||l|l||l|l||l|}
		\hline
		 &  &$\mathcal{H}$ & $\mathbb{Z}^{4}$ & $\mathcal{H}\times\mathbb{Z}$ & $\mathbb{Z}^{5}$ & $d\geq 6$\\ \hline\hline
		$\tau$& \makecell{cluster-size\\distribution}&2.315
		&\makecell{
			2.313
			\quad\cite{PhysRevE.64.026115} 
			\\2.314
			\quad\cite{PhysRevResearch.2.013067}
			\\2.311
			\quad\cite{PhysRevE.98.022120}
			\\2.314
			\quad\cite{MR4268773}
			\\2.312
			\quad\cite{PhysRevD.92.025012}
			}
		&2.420&
		\makecell{
		2.412
		\quad\cite{PhysRevE.64.026115}
		\\2.422
		\quad\cite{PhysRevE.98.022120}
		\\2.418
		\quad\cite{MR4268773}
		\\2.417
		\quad\cite{PhysRevD.92.025012}
		}
		& 2.5 
		\begin{tabular}{@{}c@{}} \\ \end{tabular}                  \\ \hline
		$\sigma$&\makecell{size of\\large clusters}&0.476&
		\makecell{
		0.480 
		\quad\cite{MR4268773}
		\\0.474 
		\quad\cite{PhysRevD.92.025012}
		}&0.499&
		\makecell{
		0.494
		\quad\cite{MR4268773}
		\\0.493
		\quad\cite{PhysRevD.92.025012} }                & 0.5 \\ \hline
	\end{tabular}
\end{subtable}

\vspace{0.4cm}
\begin{subtable}{\textwidth}
\centering
	\caption{\textsc{Critical exponent estimates for lattice trees.}}
	\begin{tabular}{|l|l|| l | l || l | l || l | l | l || l|} 
		\hline
		&&$\mathcal{H}$ & $\mathbb{Z}^4$ & $\mathcal{H}\times\mathbb{Z}$ & $\mathbb{Z}^5$ & $G_{4,3}$ & $G_{5,8}$ & $\Z^7$ & $d \geq 8$ \\
		\hline
		\hline
		$\rho$&\makecell{intrinsic\\radius}&0.595& \makecell{0.609 \quad\quad(new)\\0.607 \quad\quad \cite{PhysRevE.58.3971}} 
		&0.570& \makecell{0.576 \quad\quad(new)\\0.578 \quad\quad \cite{PhysRevE.58.3971}}&0.526&0.524 & 0.530 \quad\quad \cite{PhysRevE.58.3971} & 0.5
		\\ \hline 
		$\nu$&\makecell{extrinsic\\radius}&0.420&\makecell{0.417 \quad\quad(new)\\0.415 \quad\quad \cite{PhysRevE.58.3971}}&0.358&\makecell{0.358 \quad\quad(new)\\0.359 \quad\quad \cite{PhysRevE.58.3971}}&0.286&0.283 & 0.291 \quad\quad \cite{PhysRevE.58.3971} & 0.25\\
		\hline
		
	\end{tabular}
	\end{subtable}

	\vspace{0.3cm}
	\caption{A summary of our results for transitive graphs of polynomial volume growth. All estimates are presented to three decimal places for ease of comparison.  For percolation, the exponents $\tau$ and $\sigma$ heuristically describe the distribution of the size of the cluster of the origin at and near criticality via the ansatz $\P_p(|K|= s) \approx s^{1-\tau} g(|p-p_c|^{1/\sigma}\cdot s)$ for some rapidly decaying function $g$. These exponents are equivalent to those known as $\delta$ and $\Delta$ by the relations $\tau=2+1/\delta$ and $\sigma=1/\Delta$.
	 For lattice trees, the exponents $\rho$ and $\nu$ are defined so that a typical $n$-vertex lattice tree will have intrinsic and extrinsic radii of order $n^{\rho}$ and $n^{\nu}$ respectively. Note that Gracey's estimates \cite{PhysRevD.92.025012} are obtained using (non-rigorous) renormalization group methods rather than numerically and that the percolation estimates credited to Zhang et al.\ \cite{MR4268773} were computed from their estimates of the exponents $\nu$ and $d_f$ using the scaling relations $\tau-1=d/d_f$ and $\sigma=1/\nu d_f$. In each case our results are consistent with those obtained for the Euclidean lattices of corresponding dimension, with the small differences in numerical values reasonably attributed to finite-size effects and noise.
	}
	\vspace{-0.5cm}
	\label{table:summary}
\end{table*}

\cref{conj:1} is uncontroversial in high dimensional settings: The critical exponents describing percolation and lattice trees are strongly believed to take their \emph{mean-field} values  above the upper-critical dimensions of $d_c=6$ and $d_c=8$ respectively~\cite{MR1063208,MR1043524} and the heuristic arguments in support of this do not rely on the Euclidean geometry of $\mathbb{Z}^d$ in any way\footnote{Hara and Slade's rigorous derivation of mean-field behaviour for these models in high dimensions via the lace expansion  \cite{MR1063208,MR1043524} does however rely on specific features of Euclidean geometry, and it is an open problem to extend their analysis to the non-Euclidean case.}.
For models with $d_c=4$ such as the Ising model, $\varphi^4$ field theory, self-avoiding walk, and the uniform spanning tree, the dearth of possible low-dimensional geometries causes the analogous conjecture to reduce the standard universality principle for Euclidean lattices. Indeed, we chose to study lattice trees in part because their high upper-critical dimension allowed for the analysis of a larger number of interesting examples. (One could however make a similar universality conjecture concerning the \emph{logarithmic corrections to scaling} at the upper-critical dimension, so that our conjectures would remain interesting for, say, the 4d Ising model. We are inclined to believe such a conjecture but have not tested it numerically.)

We note that for percolation our simulations on $\mathcal{H}$ and $\mathcal{H}\times\mathbb{Z}$ exhaust all available 
non-Euclidean geometries below the upper-critical dimension $d_c=6$, so that our results lend particularly strong support to the conjecture in this case.


If the conjecture is true, it may be difficult to explain using existing methodology. Indeed, the equality of critical exponents on different Euclidean lattices of the same dimension is often explained as a consequence of the stronger statement that the two models have the same \emph{scaling limit}. In our setting, however, it is certainly \emph{not} the case that e.g.\ percolation on $\mathbb{Z}^4$ and $\mathcal{H}$ have a common scaling limit, since one scaling limit would be defined on $\mathbb{R}^4$ while the other would be defined on the continuum Heisenberg group. Again, we stress that the continuum Heisenberg group is self-similar and non-Riemannian, so that it is \emph{not} approximated by Euclidean space on any scale. In light of these difficulties, we are optimistic that further investigation into \cref{conj:1} may also significantly deepen our understanding of the original Euclidean models.

\begin{figure*}
\centering
\begin{minipage}[b]{.48\linewidth}
\centering
\includegraphics[width=0.83\linewidth]{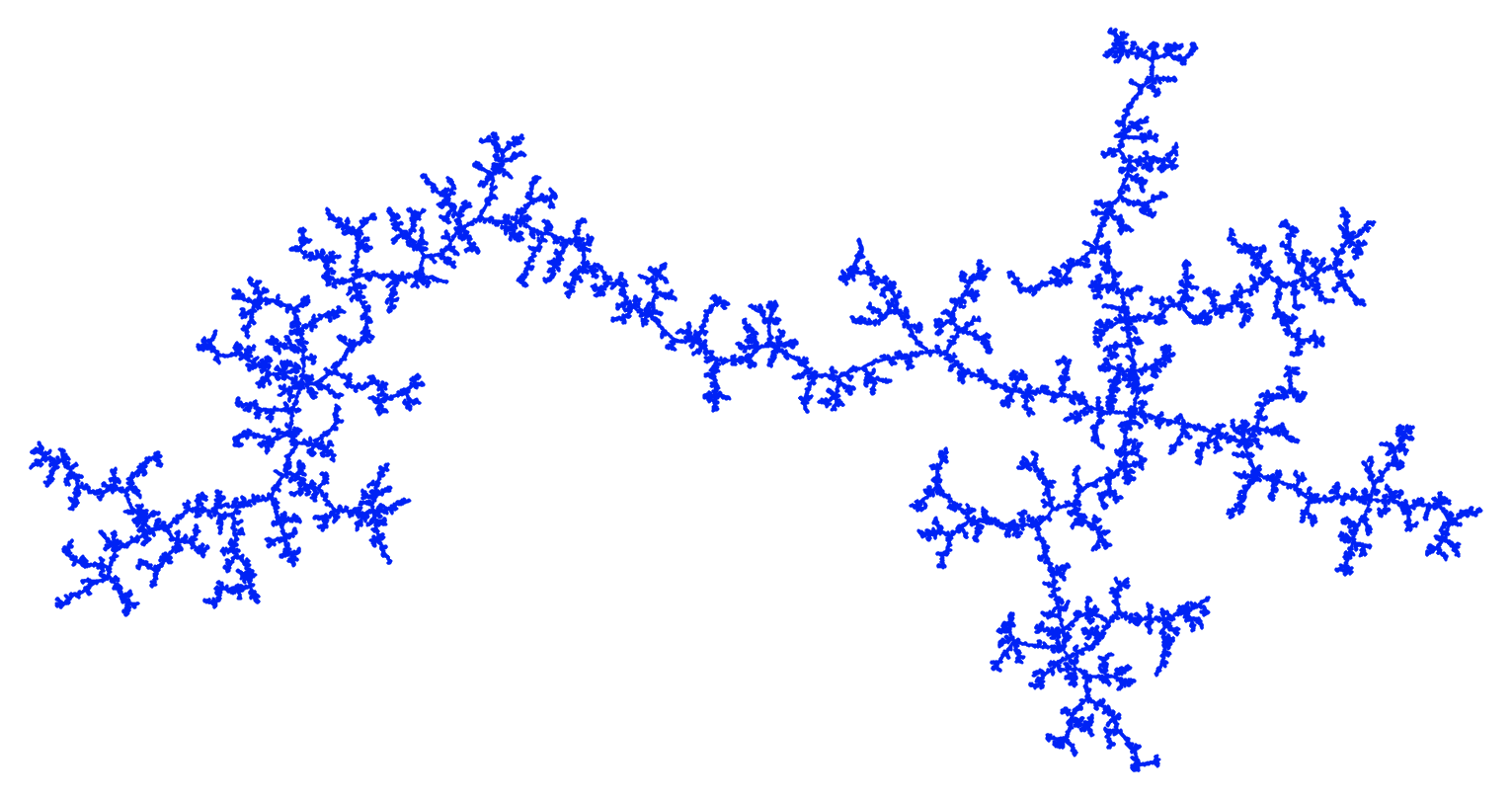}
\vspace{0.3cm}
\subcaption{\textsc{The four-dimensional hypercubic lattice $\mathbb{Z}^4$.}}
\end{minipage}%
\hfill
\begin{minipage}[b]{.48\linewidth}
\centering
\includegraphics[width=0.8\linewidth,angle=180,origin=c]{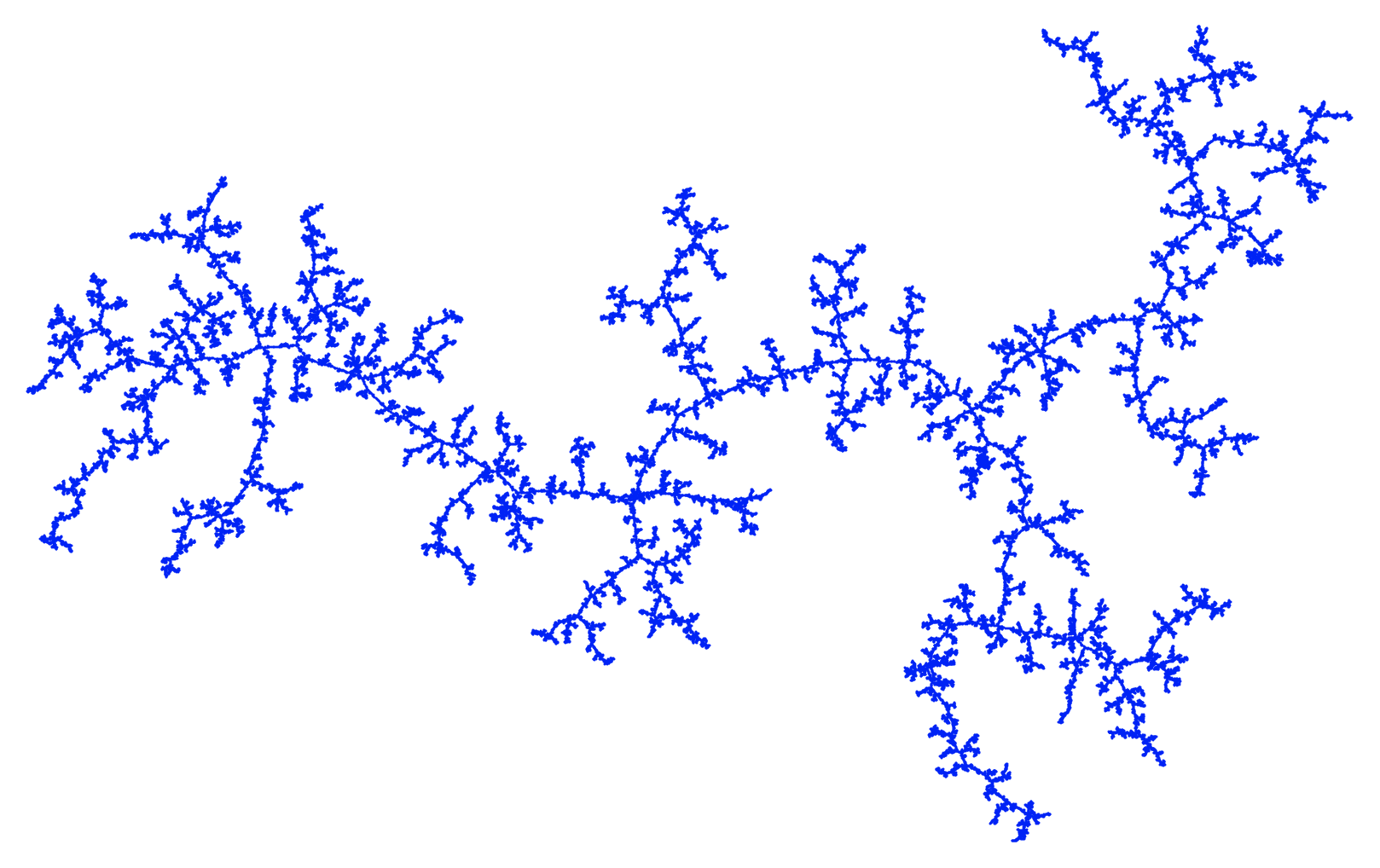}
\vspace{0.2cm}
\subcaption{\textsc{The four-dimensional Heisenberg group $\mathcal{H}$.}}
\end{minipage}\\
\begin{minipage}[b]{.48\linewidth}
\centering
\vspace{0.7cm}
\includegraphics[width=0.83\linewidth]{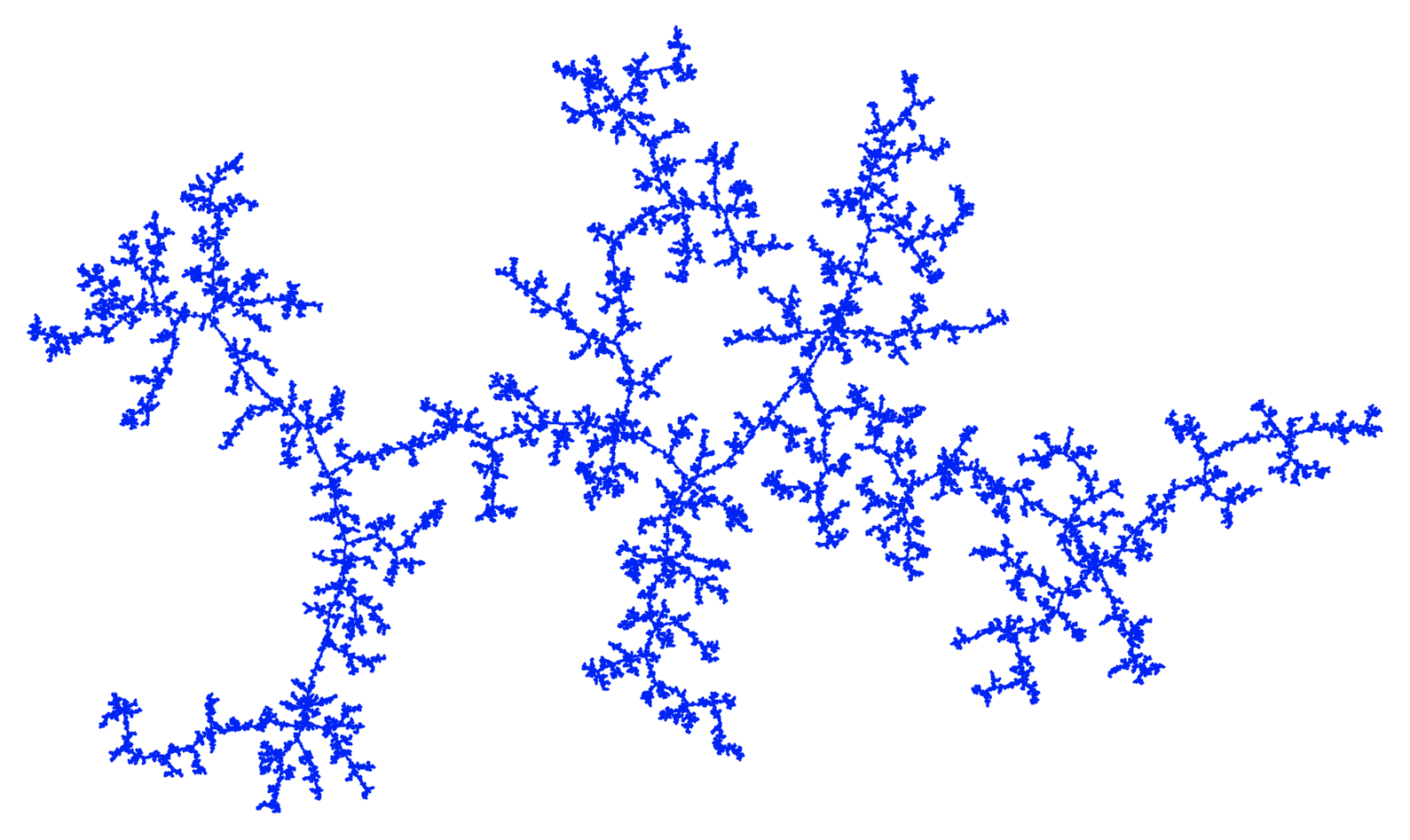}
\vspace{0.5cm}
\subcaption{\textsc{The five-dimensional hypercubic lattice $\mathbb{Z}^5$.}}
\end{minipage}%
\hfill
\begin{minipage}[b]{.48\linewidth}
\centering
\vspace{0.7cm}
\includegraphics[width=0.75\linewidth]{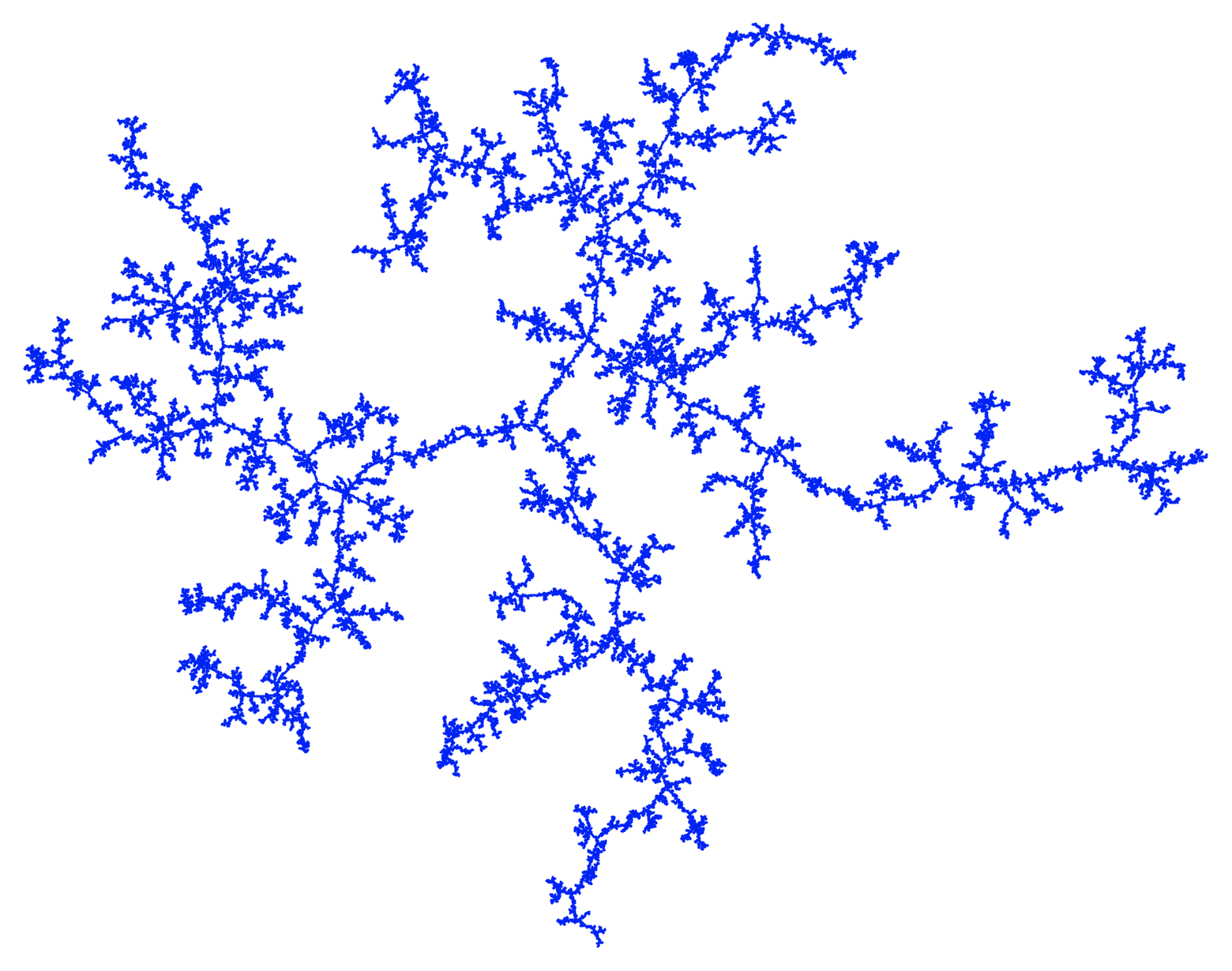}
\vspace{0.2cm}
\subcaption{\textsc{The five-dimensional product space $\mathcal{H}\times \mathbb{Z}$.}}
\end{minipage}\\
\begin{minipage}[b]{.48\linewidth}
\centering
\vspace{0.3cm}
\includegraphics[width=0.8\linewidth]{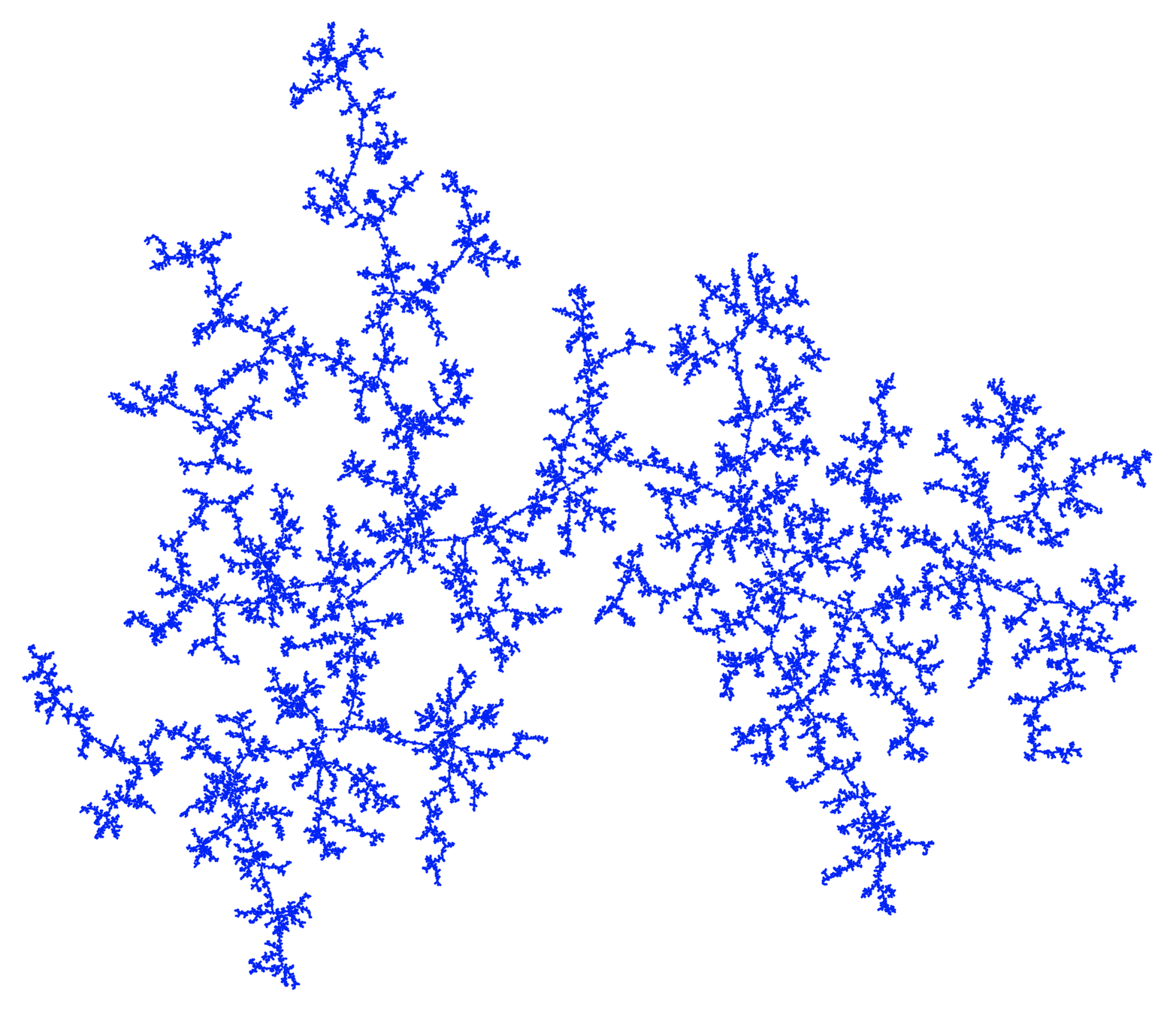}
\vspace{0.2cm}
\subcaption{\textsc{The seven-dimensional geometry $G_{4,3}$.}}
\end{minipage}
\hfill
\begin{minipage}[b]{.48\linewidth}
\centering
\vspace{0.5cm}
\includegraphics[width=0.85\linewidth]{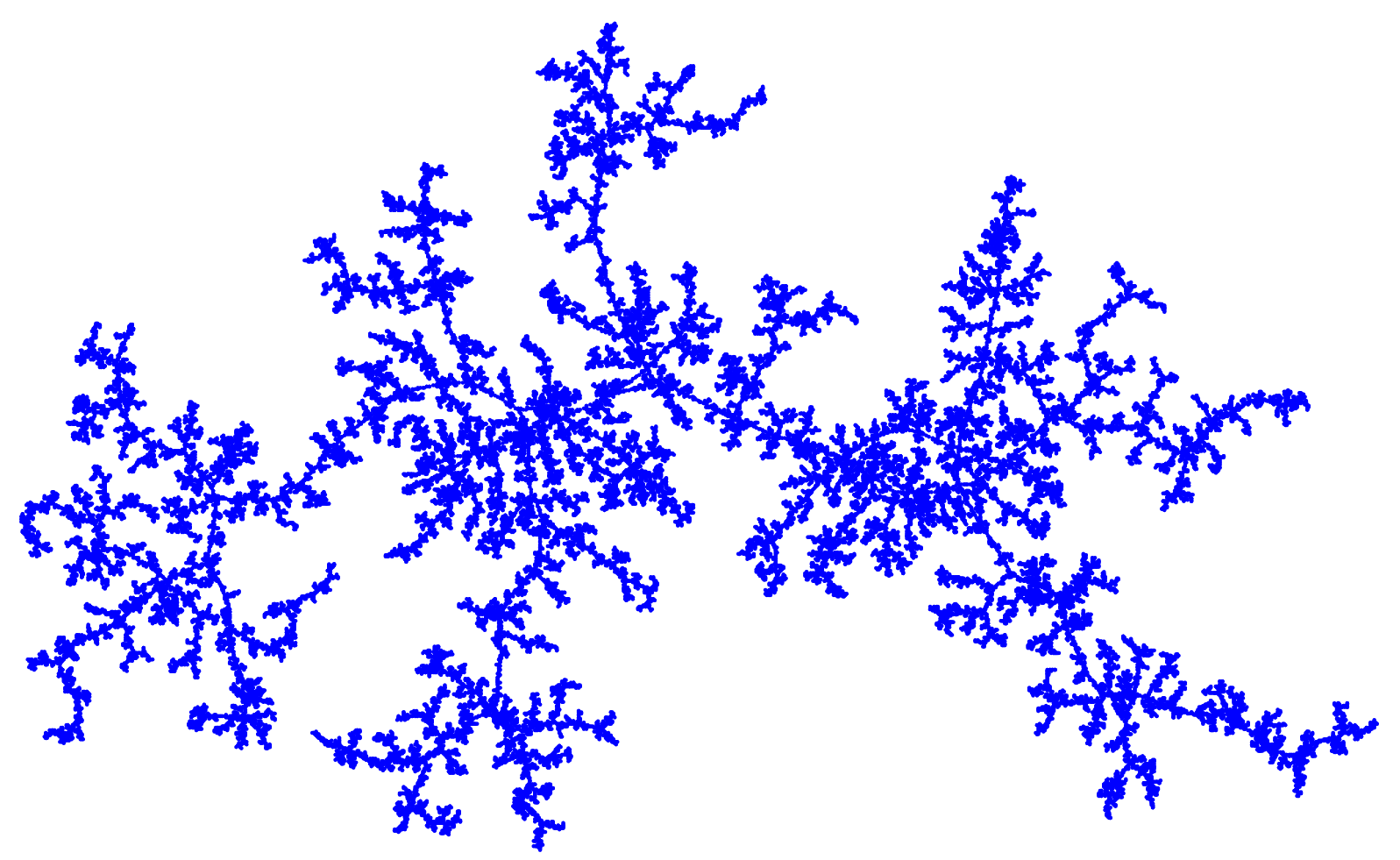}
\vspace{1cm}
\subcaption{\textsc{The seven-dimensional geometry $G_{5,8}$.}}
\end{minipage}
\caption{A visual cross-comparison of large, approximately uniform lattice trees in six different geometries, each with $60,000$ vertices. These trees were sampled via the MCMC method described in \cref{sec:latticetreesbackground} and  drawn in the plane using Mathematica's \textsf{SpringElectricalEmbedding} algorithm with parameter $\textsf{RepulsiveForcePower}=-3$. Note that this is \emph{not} an isometric embedding, and tends to distort distances rather severely. The difference in exponents between the low-dimensional and high-dimensional cases manifests itself in the seven-dimensional lattice trees looking much ``bushier" than their four-dimensional lattice tree counterparts. 
Could it be that the scaling limits of these random trees depend only on the dimension and not on the choice of transitive graph? 
The reader may like to compare these figures to the simulations of Aldous's \emph{continuum random tree} \cite{MR1085326} that are available on e.g.\ Igor Kortchemski's webpage \url{https://igor-kortchemski.perso.math.cnrs.fr/images.html}, noting that the continuum random tree is expected to arise as the scaling limit of large uniform lattice trees in eight dimensions and above.
}\label{fig:lattice_tree_embeddings}
\end{figure*}
\begin{table*}[t!]
\centering
	\textsc{Percolation exponents on self-similar fractals.}
	\vspace{1em}
	\label{table:EvNE}
	\begin{tabular}{|l||l|l|l|l||l|l|}
	\hline
	&\multicolumn{4}{c||}{Dimension}&\multicolumn{2}{c|}{Exponent}\\
		\hline
		 &  Hausdorff & Spectral & Topological & Top.\ Hausdorff & $\tau$ & $\sigma$ \\ \hline\hline
		$H_1$& 3 & 7/3 & 2 & 3 & 2.195 $\pm$ 0.005 & ? \\ 
\hline
$H_2$ & 3 & 7/3 & 2 & 3 & 2.151 $\pm$ 0.001 & 0.385 $\pm$ 0.005\\
\hline
$H_3$ & 8 & 5 & 4 & 8 & 2.66\phantom{1} $\pm$ 0.01 & 0.41\phantom{1} $\pm$ 0.01 \\
\hline
$H_3$ & 8 & 5 & 4 & 8 & ? & 0.41\phantom{1} $\pm$ 0.01 \\
\hline
	\end{tabular}
	\vspace{0.5cm}
	\caption{Summary of results of percolation on self-similar fractals. Note the unambiguous separation in numerical values of the Fisher exponent $\tau$ between the two equidimensional fractals $H_1$ and $H_2$ and the coincidence in the numerical values of $\sigma$ for the two fractals $H_3$ and $H_4$. We found the finite-size effects to be \emph{much} larger on these graphs than on the transitive graphs we considered and -- despite us considering clusters of up to a billion vertices --  for some graphs the relevant log-log plots were too far from linear to reliably extract any exponent value at all. Again, the more detailed data presented in \cref{section:FTrees} gives a much more complete picture of the situation than the raw exponent estimates presented here. In particular, we find the data presented in \cref{fig:H1H2taucomp} to demonstrate very convincingly that $H_1$ and $H_2$ have distinct values of $\tau$.
	}
	\label{table:fractal_summary}
\end{table*}

\begin{figure*}
		\centering
	\includegraphics[width=.45\textwidth]{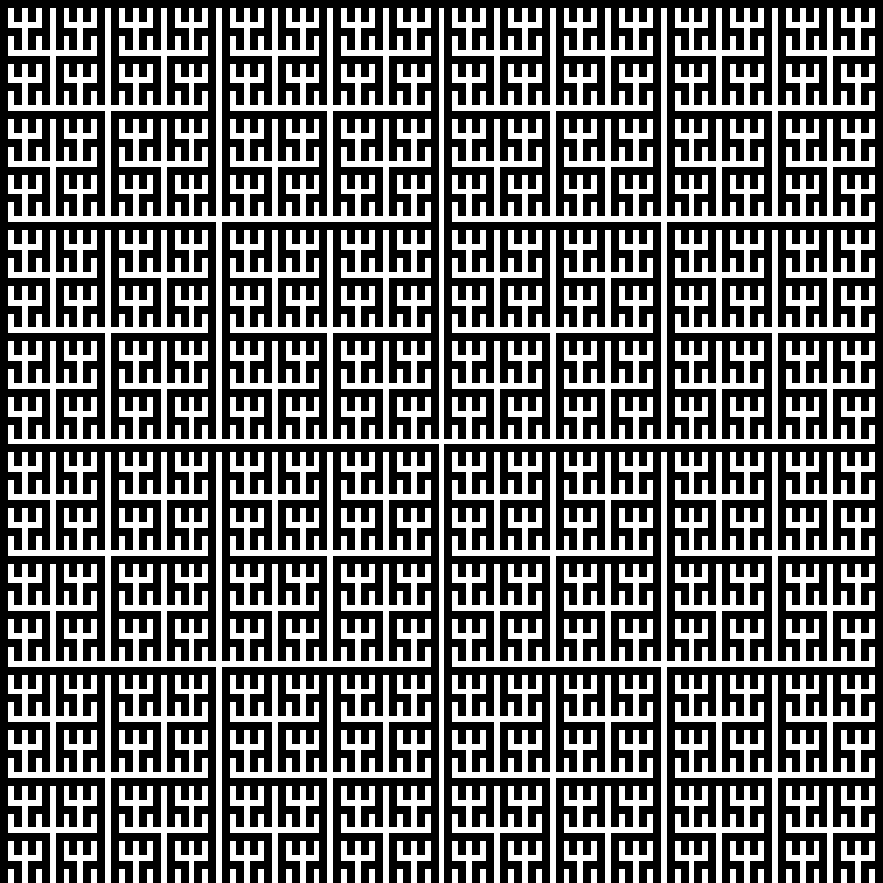}
	\hspace{.05\textwidth}
		\includegraphics[width=.45\textwidth]{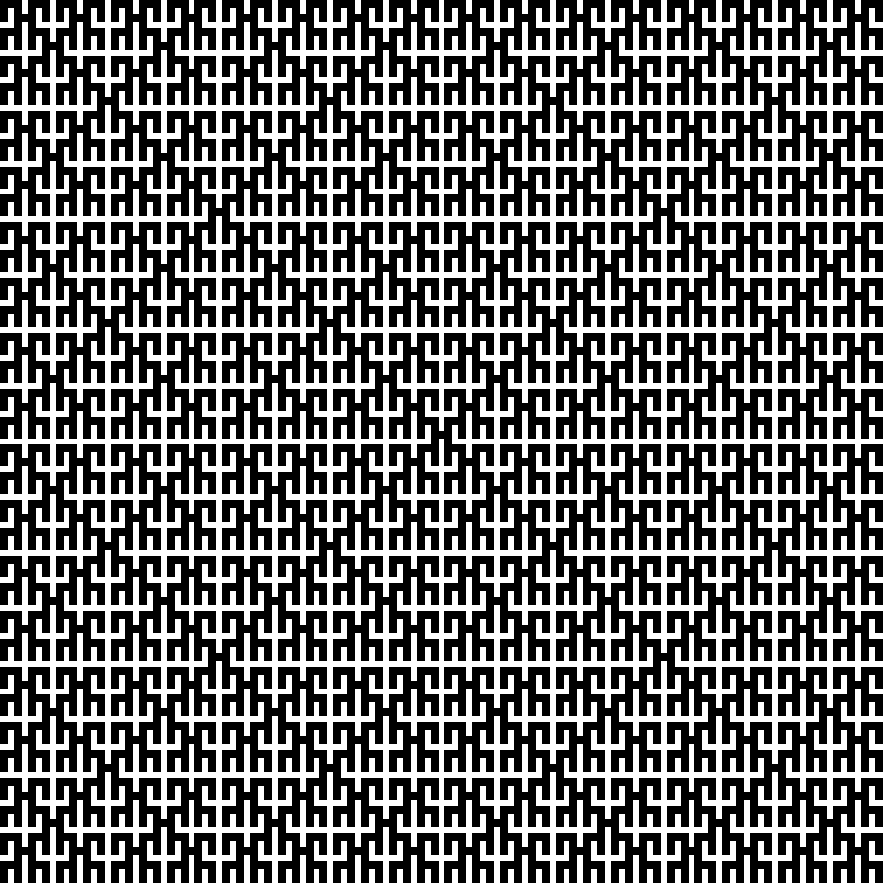}
	\captionof{figure}{Discrete approximations of the self-similar fractal trees $T_O$ (left) and $T_I$ (right). Each tree is constructed as a scaling limit of a recursively defined, self-similar spanning tree of the square lattice. In the `outer' tree $T_O$, the tree associated to the $(n+1)$th dyadic scale is formed by connecting four copies of the scale $n$ tree `around the outside' by adding edges on the centre left, centre right, and top of the square. In the `inner' tree $T_I$, the tree associated to the $(n+1)$th dyadic scale is formed by connecting four copies of the scale $n$ tree `in the middle' by adding three edges to the centre of the square. 
	}
		\label{fig:T_OI}
\end{figure*}

One very interesting possibility 
is that the \emph{intrinsic geometries} of the models share a common scaling limit across different geometries of the same dimension, even though the extrinsic scaling limits must be different. For example, it may be that large uniform lattice trees on $\Z^4$ and the Heisenberg group have a common scaling limit \emph{when considered as abstract metric trees}. The simulations presented in  Figure~\ref{fig:lattice_tree_embeddings} show that such a conjecture is at least plausible and is worthy of further investigation in future work. Still, such a conjecture would be difficult to confirm in light of the distinct extrinsic scaling limits and would not obviously explain e.g.\ the coincidence of exponents describing the \emph{extrinsic} geometry of lattice trees.

\medskip

We remark that there is an extensive literature investigating critical behaviour on \emph{hyperbolic} lattices including e.g.\ \cite{MR3819418,PhysRevE.79.011124,MR3962879,MR2745724,MR2450478,benedetti2015critical,gendiar2014mean}. These lattices are very different from the non-Euclidean lattices we consider in this paper. Indeed, hyperbolic lattices are of infinite-dimensional volume growth and are therefore expected to exhibit mean-field behaviour for both models; this has been proven rigorously for percolation on arbitrary hyperbolic lattices in \cite{MR3962879} and for lattice trees on certain hyperbolic lattices by Madras and Wu \cite{MR2745724}.  We believe our paper is the first to systematically investigate critical exponents on transitive non-Euclidean lattices \emph{below} the upper-critical dimension.

 \subsection{Self-similar fractals}
The self-similar Carnot groups arising as scaling limits of transitive graphs of polynomial growth
 can be thought of as very special examples of \emph{fractal} spaces.
 As such, it is natural to wonder to what extent the phenomena discussed above extend to more general fractals. The situation here is more complicated. We will restrict our attention in the fractal case to Bernoulli percolation, where previous works investigating the effect of fractal geometry on percolation critical probabilities and critical exponents include \cite{Bo_Ming_1988,PIRC,Lin_1997,PhysRevE.55.94,PhysRevLett.51.2347,doi:10.1080/13642818408238847,BALANKIN2019957,BALANKIN201812,MONCEAU2004310}. In these works, percolation on families of fractals with varying fractal and spectral dimensions is investigated, with the focus often on Sierpinski-type fractals. (Of course one does not sample percolation directly on the continuum fractals but rather on appropriately chosen `prefractal' graphical approximants; we discuss this further in \cref{section:FTrees}.) These results 
 demonstrate that, in contrast to our Conjecture~\ref{conj:1}, percolation critical exponents on fractals cannot depend solely on the Hasudorff dimension (which is the most popular continuum analogue of the volume-growth dimension). 

Once it is known that universality does not hold across fractals with identical Hausdorff dimensions, a next natural hypothesis is that critical exponents are instead a function of some \emph{set} of properties or dimensions which better capture the geometry of the fractal. 
A specific proposal to this effect was made by Balankin et al.\ \cite{BALANKIN201812}, who 
suggested that the critical exponents should be determined by a set of three fractal dimensions, namely the Hausdorff, spectral, and topological Hausdorff dimensions. 
The more general view that the spectral dimension is important to the determination of critical behaviour has been advocated by many authors; see the introduction of \cite{millan2021complex} for an overview.

In this paper we make a novel contribution to this problem by cross-comparing percolation critical exponents between pairs of fractals which are distinct but for which many important notions of dimension coincide. The specific examples we consider are constructed as products of various recursively-defined self-similar fractal trees. We use these examples due to the flexibility of their construction and the ease of computation of their associated fractal dimensions. The dimensions of the four different fractal products we consider and our numerical estimates of their percolation critical exponents are summarised in \cref{table:fractal_summary}.

 We begin by constructing two fractal trees $T_I$ and $T_O$, each with Hausdorff dimension 2, such that the two fractal products $H_1=T_O\times[0,1]$ and $H_2=T_I\times[0,1]$ share the same Hausdorff, spectral, topological and topological Hausdorff dimensions. Both trees $T_I$ and $T_O$ are defined as scaling limits of self-similar spanning trees of the square lattice $\mathbb{Z}^2$ as depicted in \cref{fig:T_OI}. We present strong numerical evidence that the two fractals $H_1$ and $H_2$ have distinct values of the percolation Fisher exponent $\tau$ which characterises the cluster size-distribution at criticality. This provides strong evidence against the aforementioned conjecture of Balankin et al.\ \cite{BALANKIN201812}. Moreover, since the two fractals we consider are very similar in a large number of ways beyond these dimensions, our results suggest that any universality principle applying to fractals must be much weaker than in the transitive case.

 On the other hand, a more positive picture emerges when one considers the critical exponent $\sigma$ which characterises the size of the percolation scaling window. Indeed, for $H_1$ and $H_2$
 our results were inconclusive but consistent with the hypothesis that the two values of $\sigma$ coincide. 
 In order to investigate this potential phenomenon further, we constructed and analyzed two further fractal tree products which we call $H_3$ and $H_4$.
   As with $H_1$ and $H_2$, the two fractals $H_3$ and $H_4$ share many notions of dimension despite having distinct geometries in other regards, but are `higher-dimensional' overall than $H_1$ and $H_2$. We present strong numerical evidence that $H_3$ and $H_4$ have a shared value of $\sigma$. This may be related to the phenomenon of \emph{weak-universality} as discussed in \cite{MONCEAU20041}, and weakly suggests that the exponent $\sigma$ may indeed be a function of some small set of parameters associated to the fractal. 

\medskip

\textbf{Organisation:} 
The rest of the paper is structured as follows: In \cref{section:models}, we recall the definitions of the two models we will study and the exponents we wish to compute, and describe the methodologies used in our simulations. Further details of an improvement to the invasion percolation methodology are given in Appendix \ref{section:IPA}.
In \cref{section:TLattices} we give background on the four geometries $\mathcal{H}$, $\mathcal{H}\times\mathbb{Z}$, $G_{4,3}$, and $G_{5,8}$ and present our numerical results regarding percolation and lattice trees in these geometries. In \cref{section:FTrees}, we describe the four fractals $H_1$, $H_2$, $H_3$, and $H_4$, and present the relevant numerics. 
Finally, we summarise our findings and discuss
possible directions for future work in \cref{section:OQD}.

\medskip

\textbf{Computations:} We used C++ \cite{ISO:2017:IIIa} for all simulations, which were carried out on the Cambridge Faculty of Mathematics HPC system, fawcett. We used the very fast Martinus Robin-Hood Hashing \cite{martinus} in place of standard library unordered maps.  

\section{Models and algorithms} \label{section:models}

In this section we give relevant background on percolation and lattice trees and review the methodology we use to compute the critical exponents describing these models.

		\subsection{Bernoulli bond percolation}
		\label{sec:percolation_background}
		Fix $p \in [0,1]$. Given a graph $G=(V,E)$, we attach i.i.d.\ Bernoulli random variables $(\omega_e)_{e\in E}$ of parameter $p$ to the edges of the graph and say that an edge $e$ is \emph{open} if $\omega_e=1$ and \emph{closed} if $\omega_e=0$.
		 We denote the associated product probability measure by $\mathbb{P}_p$.
		Given any vertex $v\in V$, we define the \emph{cluster}  $K_v$ of $v$ to be the set of vertices that are accessible from $o$ by paths consisting only of open edges. 
		Given an infinite graph $G$, we define the \emph{critical probability} $p_c$ to be the infimal value of $p$ for which infinite clusters exist with positive probability. Note that the value of $p_c$ depends strongly on the microscopic details of the graph and is not universal.
		
		We now introduce the exponents we consider and some relevant (non-rigorous) scaling theory, referring the reader to e.g.\ \cite[Chapter 9]{grimmett2010percolation} for further background.
		Let $o$ be a fixed vertex of $G$, which we regard as the origin. Assuming they are well-defined, the exponents $\tau$ and $\rho$ describe the distribution of the volume and (extrinsic) radius of the cluster of the origin at criticality by
		\begin{align*}
		\P_{p_c}(|K_o| \geq s) &\approx s^{2-\tau} && \text{ as $s\uparrow \infty$ and}\\
		\P_{p_c}(\operatorname{rad}(K_o) \geq r) &\approx r^{-1/\rho} && \text{ as $r\uparrow \infty$,}
		\end{align*}
		where $\operatorname{rad}(K_o)$ is the maximum distance in $G$ between $o$ and another point of $K_o$. 
		(We keep the meaning of the symbol $\approx$ intentionally vague; it should not be read as corresponding to any specific or consistent notion of asymptotic equivalence.) Below the upper-critical dimension, these exponents are expected to determine each other via the hyperscaling relation $\tau=(2d\rho-1)/(d\rho-1)$ \cite[Chapter 9]{grimmett2010percolation}.
		It is a standard assumption of scaling theory that there exists a further exponent $\sigma$ such that
		\[
		\P_{p}(|K_o| = s) \approx s^{1-\tau} g_\pm\bigl(|p-p_c|^{1/\sigma} \cdot s\bigr)
		\]
		for some rapidly decaying functions $g_-$ and $g_+$ describing the cases $p\leq p_c$ and $p \geq p_c$ respectively. In particular, this ansatz predicts that the probability $\P_{p}(|K_o| = n)$ is of the same order as its critical value when $n \ll |p-p_c|^{-1/\sigma}$ and is very small when $n \gg |p-p_c|^{-1/\sigma}$, and we think of $|p-p_c|^{-1/\sigma}$ as describing the ``typical size of a large finite cluster".
		
		The calculations we perform in this paper will utilize a slightly different approach to scaling theory, adapted from the presentation of \cite{PhysRevE.57.230}, which we now overview.
		When $G$ is transitive, we fix an origin vertex $o$ as above and write $P_{\geq s}=P_{\geq s,p_c}$ for the cluster size distribution, where $P_{\geq s,p} = \mathbb{P}_{p}(|K_o|\geq s)$. 
		For non-transitive fractals, we take the origin $o$ to a vertex selected uniformly at random (in a sense which will be made precise later), and then define $P_{\geq s,p}=\mathbf{E}[\mathbb{P}_{p_c}(\abs{K_o}=s)]$ where $\mathbf{E}[\cdot]$ denotes the expectation with respect to the random origin $o$.
		We will assume as a basis for calculations that the critical cluster size distribution is described by the ansatz
		\begin{equation}\label{eq:asymp1}
			P_{\geq s,p_c}= A_0 s^{2-\tau}(1+A_1s^{-\Omega}+\cdots)
		\end{equation}
		for some $\tau,\Omega>0$ and $A_0,A_1\in\mathbb{R}$. The exponent $\tau$ is known as the Fisher exponent, and $\Omega$ is the leading correction-to-scaling component whose impact becomes negligible for large $s$. Both these exponents are expected to be universal in the sense that they only depend on the large scale geometry of the graph, whereas the constants $A_i$ are non-universal and will also depend on the microscopic geometry of the graph.
		In order to compute $\sigma$, we will assume similarly that
		\begin{equation} \label{eq:offcrit1}
		P_{\geq s}= C_0 s^{2-\tau}(1+C_1(p-p_c)s^\sigma+\cdots)
		\end{equation}
		for $p\neq p_c$ and values of $s$ that are not too large or small.

		We use an algebraic manipulation, as in \cite{PhysRevE.57.230}, to derive a more convenient scaling relation for $\sigma$ which only relies on properties of the percolation clusters at criticality. Taking derivatives of \eqref{eq:offcrit1} with respect to $p$, we get
		\begin{equation}
		\frac{d P_{\geq s,p}}{dp} = C_0 C_1 s^{2-\tau+\sigma} +\cdots		
		\end{equation}
		If we let $g_{n,t}$ be the number of possible cluster configurations containing the origin and exactly $n$ open edges, and with $t$ closed edges adjacent to the cluster, then
		\[
		P_{\geq s,p} = \sum_{n\geq s}\sum_{t} g_{n,t} p^n (1-p)^t.
		\]
		Taking derivatives with respect to $p$ gives
		\begin{align*}
		\frac{d P_{\geq s,p}}{dp} &= \sum_{n\geq s}\sum_{t} g_{n,t} p^n(1-p)^t \big(\frac{n}{p}-\frac{t}{1-p}\big)\\&=
		\frac{\mathbb{E}[ n\mathbf{1}_{n\geq s}]}{p}-\frac{\mathbb{E}[t\mathbf{1}_{n\geq s}]}{1-p}
		\end{align*}
		so that
		\begin{equation*} 
		 \frac{\mathbb{E}_{p_c}[ n\mathbf{1}_{n\geq s}]}{p_c}-\frac{\mathbb{E}_{p_c}[t\mathbf{1}_{n\geq s}]}{1-p_c} = C_0 C_1 s^{2-\tau+\sigma} +\cdots,	
		\end{equation*}
	and hence
	\begin{equation} \label{eq:sigma3}
		\frac{\mathbb{E}_{p_c}[ n\mid{n\geq s}]}{p_c}-\frac{\mathbb{E}_{p_c}[t\mid{n\geq s}]}{1-p_c} = C_1 s^{\sigma} +\cdots.
	\end{equation}
		As in \cite{PhysRevE.57.230}, this will be used as an assumed formula to compute $\sigma$ using only information at criticality. Let us note however that the expectation on the left hand side has poor numerical properties since the associated random variable $n/p-t/(1-p)$ is heavy tailed at criticality. Indeed, \eqref{eq:sigma3} should really be interpreted as a statement about the  $p \uparrow p_c$ limit since the left hand side is not well-defined at criticality.
		
		Lastly, we define $Q_{\geq r}=\P_{p_c}(\operatorname{rad}(K_o) \geq r)$, and assume the form
		\[
		Q_{\geq r}= F_0 r^{-1/\rho} (1+F_1 r^{-\zeta}+\ldots),
		\]
		for some $\rho,\zeta>0$ and $F_0,F_1\in\R$, where $F_0$ and $F_1$ are not expected to be universal.

		\medskip

		\textbf{Methodology}. We now describe the methods we used to compute critical exponents for percolation. 
		Our first step was to estimate the value of the critical probability $p_c$. To do this, we began by employing the \emph{invasion percolation} methodology of \cite{PhysRevE.98.022120,MR3819418}, using the bulk-to-boundary ratio developed by Leath in \cite{PhysRevLett.36.921} as an estimator of the critical probability, and then using the extrapolation hypothesis developed in \cite{PhysRevE.98.022120} to further refine the resulting estimate. In fact, we implemented a simple improvement to this methodology that resulted in substantial run-time reductions and which we describe in detail in \Cref{section:IPA}. 
		
		Invasion percolation is a stochastic model for the transport of fluid through porous media \cite{Wilkinson_1983,LO,chandler_koplik_lerman_willemsen_1982}. It operates by assigning i.i.d.\ uniform random variables $U_e$ taking values in $[0,1]$ to the edges of some graph $G$ with root vertex $o$. 
		We then define the sequences $(e_n)_{n\geq 1}$, $(V_n)_{n\geq 0}$, $(E_n)_{n\geq 0}$, and $(F_n)_{n\geq 0}$ recursively 
		as follows:
		\begin{enumerate}
			\item Start with $V_0=\{o\}$, $E_0=\varnothing$, and $F_0 = \{\{x,o\}:x\sim o\}$.
			\item At each step $n\geq 1$, let $e_n$ be the element of $F_{n-1}$ minimizing $U_e$, let $E_n=E_{n-1}\cup\{e\}$, let $V_n$ be the set of vertices adjacent to at least one edge of $E_n$, and let $F_n$ be the set of edges that have at least one endpoint in $V_n$ but do not belong to $E_n$.
		\end{enumerate}
		We call $V_n$ the \emph{invasion cluster} up to time $n$, and $F_{n}$ the \emph{frontier} at time $n$. 
		The \emph{bulk-to-boundary ratios} are given by the random sequence
		\[a_n = \frac{|E_n|}{|E_n|+|F_n|} = \frac{n}{|E_n|+|F_n|}.\]
		It is proven in \cite[Chapter 11]{LP:book} that $\limsup_{n\to\infty} U_{e_n} = p_c$ almost surely, and it is believed that
		\[
		a_n \to p_c \qquad \text{ as $n \to \infty$};
		\]
		this has been verified rigorously \cite{cmp/1104114182} for $\mathbb{Z}^2$, but it is expected  to hold for a much wider range of graphs.
		Assuming that this limiting relation holds, one may estimate the critical probability $p_c$ by running invasion percolation for a long time and computing the resulting bulk-to-boundary ratio. 

		In \cite{MR3819418} this method was improved via the following extrapolation argument. For invasion percolation on the binary tree, the bulk to boundary ratio can be shown to satisfy the asymptotics
		\[\E a_n \approx \frac{p_c}{1+An^{-1}}\]
		for some constant $A$. The authors of  \cite{MR3819418} conjecture and verify numerically that for Euclidean lattices one has the analogous formula
		\begin{equation}\label{eq:pc}
		\mathbb{E}{a_n} \approx \frac{p_c}{1+An^{-\delta}}\end{equation}
		with a high degree of accuracy for some positive constants $A$ and $\delta$. 
		We will assume that such a formula also holds in our settings and, following \cite{MR3819418}, use the ansatz
		\[
	\mathbb{E}{a_n} = \frac{p_c}{1+An^{-\delta}(1+Bn^{-\delta'} + C n^{-\delta''} + \cdots)}
		\]
		as a basis from which to calculate $p_c$; this allows us to gather data for a relatively small number of time-steps and then use curve fitting to give an estimate of $p_c$.

		The use of a heap or sorted list for storing/extracting values on the frontier lets us compute $S_n$ in time $O(n\log n)$ and memory $O(n)$. The simple improvement we outline in Appendix \ref{section:IPA} reduces the size of the sorted list used for the frontier by a power and thus significantly reduces the running time.

		The major advantages of using invasion percolation for computing $p_c$ are as follows:
		\begin{enumerate}
			\item It does not require us to store large blocks of the relatively high dimensional lattices on which we simulate percolation - instead we need only store a number of edges or vertices which is linear in the number of steps of the algorithm thus far.
			\item It does not require us to assume \emph{a priori} values of any critical exponents, unlike the methods of \cite{PhysRevResearch.2.013067}.
			
			\item It does not require detailed understanding of the geometry of the graph under consideration, unlike the wrapping method used in \cite{PhysRevE.64.016706} or the multi-scale analysis used in \cite{BALANKIN201812}.
		\end{enumerate}
	Invasion percolation also allows us to narrow in on a relatively precise value of $p_c$ with far smaller memory and time requirements than by starting from scratch utilizing a logarithmic search with the Leath algorithm as in \cite{PhysRevE.57.230}.

		Having estimated $p_c$ via invasion percolation as described above, we then used the \emph{Leath algorithm} \cite{leath1976cluster} to generate a population of samples for the cluster at the origin. For each sample, we recorded both the cardinality of the cluster and the number of boundary edges.
		For the lattices, we least-mean-squares fitted the parameters in the following equations, approximately valid at $p\approx p_c$, to give estimates of $\tau$ and $\sigma$:
		\begin{enumerate}[labelindent=0.5cm,labelwidth=\widthof{Q2a.},itemindent=1em,leftmargin=!]
			\item[\textsf{Q1a.}] $\log_2 P_{\geq s} = (2-\tau)\log_2(s) + B s^{-\Omega} + C$
			\item[\textsf{Q2a.}] $\log_2\left(\frac{\mathbb{E}[ n\mid{n\geq s}]}{p}-\frac{\mathbb{E}[t\mid{n\geq s}]}{1-p}\right) = \sigma\log_2(s) +D$.
		\end{enumerate}
		For the fractals, we plotted
			\begin{enumerate}[labelindent=0.5cm,labelwidth=\widthof{Q2b.},itemindent=1em,leftmargin=!]
			\item[\textsf{Q1b.}] $\log_2 P_{\geq s}$ against  $\log_2(s)$
			\item[\textsf{Q2b.}] $\log_2\left(\frac{\mathbb{E}[ n\mid{n\geq s}]}{p}-\frac{\mathbb{E}[t\mid{n\geq s}]}{1-p}\right)$ against  $\log_2(s)$.
		\end{enumerate}
	for $p \approx p_c$ and calculated the gradient of the approximately linear sections at large $s$ to give estimates of $\tau$ and $\sigma$.
		
		In the case of the non-Euclidean lattices, the values of $\tau$ and $\sigma$ we obtained were very close to the values in the literature for the corresponding Euclidean lattices. Having obtained these estimates, we then utilized the methodology of \cite{PhysRevE.57.230,PhysRevResearch.2.013067} to improve our value for the critical probability and give further credence to our conclusion. To this end, we sampled the cluster at the origin using the Leath algorithm at a range of values of $p$ near our initial estimate of $p_c$ and plotted the following graphs:
	\begin{enumerate}[labelindent=0.5cm,labelwidth=\widthof{G2.},itemindent=1em,leftmargin=!]
		\item[\textsf{G1.}] $s^{\tau_E-2} P_{\geq s}$ against $s^{-\Omega_E}$
		\item[\textsf{G2.}] $s^{\tau_E-2} P_{\geq s}$ against $s^{\sigma_E}$,

	\end{enumerate}
	where $\tau_E,\sigma_E,\Omega_E$ are the corresponding estimates of the Euclidean exponents calculated in previous literature.
	For the first graph, looking at relation $\eqref{eq:asymp1}$,  we expect that the curve does not deviate from its linear trajectory for small $s^{-\Omega}$ when $p$ is close to $p_c$, and for the second graph, looking at relation $\eqref{eq:offcrit1}$, we expect a plateau for large $s^{\sigma}$ when $p$ is close to $p_c$. We observed that this was indeed the case, lending further credibility to the accuracy of our estimates.
	
	We calculated estimates of the extrinsic exponent for each transitive lattice by recording the maximum extrinsic distance of any vertex visited in runs of the Leath algorithm. We plotted the following curve and calculated the gradient of its approximately linear final segment to give $-1/\rho$:
	\begin{enumerate}[labelindent=0.5cm,labelwidth=\widthof{Q2b.},itemindent=1em,leftmargin=!]
		\item[\textsf{Q3.}] $\log_2 Q_{\geq r}$ against  $\log_2(r)$.
	\end{enumerate}

\noindent 	For the percolation on products of fractal trees, where we did not have reference values of $\tau$ and $\sigma$ to compare with, we instead refined our values of the critical exponents and $p_c$ 
 by plotting Q1b over very large ranges of $s$ for a selection of probabilities $p$ and finding the value of $p$ which gave the smallest deviation from linearity for medium and large $s$ as in \cite{PhysRevE.57.230,WSPL1}.

		\subsection{Lattice Trees}
		\label{sec:latticetreesbackground}

		Given a transitive connected graph $G$ and a fixed vertex $o$ of $G$, a \emph{lattice tree} is a finite connected subgraph of $G$ that contains $o$ and is a tree, i.e.\ does not contain any cycles. We let $T_n$ be the set of $n$-vertex lattice trees; The uniform lattice tree of size $n$ is then just the random variable given by selecting one of these tree uniformly at random.
		We will study how the following quantities depend on the tree size $n$:
		\begin{itemize}
		\item \textbf{The mean branch size, $B(n)$}: If we take an edge from the lattice tree and delete it, the branch size is the cardinality of the smaller of the two resultant subtrees. The mean branch size is the expectation of this quantity over the lattice tree and over an edge picked uniformly at random from the lattice tree.
		\item \textbf{The intrinsic longest path, $I(n)$}: This is the expected length of the longest path in the lattice tree, where the length of the path is given by the \emph{intrinsic metric}, i.e. the graph metric of the tree.
		\item \textbf{The extrinsic displacement of the longest path, $E(n)$}: This is the expected extrinsic distance between the two endpoints of some maximal-length intrinsic path in the tree. (The method used to pick a particular such path when it is non-unique is described below; the details of this should not be important.)
		\end{itemize}
		Here we are using the extrinsic displacement of the longest path as an easier-to-compute substitute for the true extrinsic diameter of the tree, which we expect to be of the same order.
	Assuming they are well-defined, the exponents $\rho$ and $\nu$ describe the asymptotics of $B(n)$, $I(n)$, and $E(n)$ via
	\[
	B(n) \approx I(n) \approx n^\rho \qquad \text{ and } \qquad E(n) \approx n^\nu.
	\]

%
%
	We remark as a point of general interest that the exact equality $\nu=0.5$ is believed to hold for \emph{three-dimensional} lattice trees. This equality has been proven rigorously for \emph{branched polymers} \cite{MR2549378,MR2031859}, which are believed to be in the same universality class as lattice trees.

	\medskip
	
	\textbf{Methodology.} In order to sample approximately uniform lattice trees, we employed a combination of two Markov chain Monte Carlo (MCMC) algorithms: the \emph{cut-and-paste} algorithm developed in \cite{Rensburg_1992} and the \emph{cycle-breaking} algorithm described in \cite{fredes2021models}.
	 We computed the critical exponents $\nu$ and $\rho$ by the same method described in \cite{PhysRevE.58.3971,Rensburg_1992} where they were calculated for Euclidean lattices of dimensions $2$ through $7$.
	
	The MCMC algorithm involves evolving some arbitrary initial tree $t_0$ by applying a sequence of randomly chosen operations to form a process $(t_i)_{i\geq 1}$ on the space of lattice trees.	
	The possible operations and their probabilities are chosen such that, when restricted to the set of $n$-vertex lattice trees $T_n$, the resultant process is Markovian, irreducible and aperiodic, and has the uniform distribution as its invariant distribution. Standard Markov chain theory then implies that the process will converge to the uniform measure on $T_n$ as $i \to \infty$. After an initial mixing period, the process is sampled at regular intervals, and measurements of interest are calculated and recorded. In the absence of bounds on the mixing/relaxation time, we fixed the number of steps of the algorithm between samples in such a way that the autocorrelation of the measured quantities across the samples was found to be negligible.

 The cut-and-paste (\textsf{CP}) algorithm involves snipping the tree at an edge chosen uniformly at random, applying a uniformly chosen random symmetry to the smaller of the resultant sub trees, choosing a gluing site on the larger of the resultant subtrees, and attempting to glue the  smaller subtree onto the larger subtree at the gluing site; if the resulting subgraph is no longer a tree then we reject the move and return the tree we started with. 
  Let us now give a detailed description of this algorithm in the case that the base graph is $\mathbb{Z}^d$. We write $S^d$ for the group of symmetries of $\Z^d$ fixing the origin and let $\Gamma_G(x)$ be the neighbourhood of $x$ in the graph $G$. Given a lattice tree $t_i$, the cut-and-paste algorithm produces a new lattice tree $t_{i+1}$ via the following randomized procedure:
	\begin{enumerate}[labelindent=0pt,labelwidth=\widthof{CP7.},itemindent=0em,leftmargin=!]
		\item[\textsf{CP1.}] Select a vertex $u\in t_i$ and select a vertex $v\in \Gamma_G(u)$  uniformly at random. If $uv$ is not an edge of the tree, return $t_{i+1}=t_i$ and halt.
		\item[\textsf{CP2.}] Delete edge $uv$ from the tree to form smaller subtree $s_{i+1}$ and larger subtree $S_{i+1}$. Relabel $u$, $v$ if necessary so that $u$ is contained in the smaller subtree.
		\item[\textsf{CP3.}] Select a vertex $w$ uniformly at random from $S_{i+1}$, and choose a vertex $z$ uniformly from $\Gamma_G(w)$.
		\item[\textsf{CP4.}] If $z$ is contained in $S_{i+1}$, return $t_{i+1}=t_i$. Otherwise form an edge between $z$ and $w$.
		\item[\textsf{CP5.}] Apply a randomly chosen symmetry from $S^d$ to the smaller subtree $s_{i+1}$.
		\item[\textsf{CP6.}] Translate the transformed smaller subtree so that vertex $u$ coincides with vertex $z$, and identify $u$ and $z$.
		\item[\textsf{CP7.}] If the resultant union of the two trees forms a valid tree, then accept this tree as $t_{i+1}$ and halt. Otherwise return $t_{i+1}=t_i$ and halt.
	\end{enumerate}

As far as we are aware, the cut-and-paste algorithm has only previously been used in the case of Euclidean lattices. While the algorithm remains valid for general Cayley graphs, where translations are always well-defined, the other Cayley graphs we consider will typically have few, if any, non-trivial symmetries fixing their respective origins. This lack of non-trivial symmetries does not stop the algorithm from working or being well-defined, but is likely to seriously increase the mixing time of the relevant Markov chain. See \cite{Rensburg_1992} for a proof that the \textsf{CP} algorithm does indeed produce uniform lattice trees.

To increase the mixing rate of the Markov chain at minimal computational cost, we combined the above with another algorithm, the ``BreakCycle" (\textsf{BC}) algorithm, as developed in the recent paper  \cite{fredes2021models}. Given a lattice tree $t_i$, this algorithm produces a lattice tree $t_{i+1}$ via the following randomized procedure:
\begin{enumerate}[labelindent=0pt,labelwidth=\widthof{BCb1.},itemindent=0em,leftmargin=!]
	\item[\textsf{BC1.}] Select a vertex $u\in t_i$ and select a neighbouring vertex $v\in \Gamma_G(u)$ uniformly at random. If $uv$ is an edge of the tree, return $t_{i+1}=t_i$ and halt.
	\item[\textsf{BC2.}]  If $v$ is in $t_i$ go to \textsf{BCa,} else 
	go to \textsf{BCb}.
	\item[\textsf{BCa1.}] Add $uv$ to $t_i$ to form $t_i^\prime$, and identify the unique cycle $C\subset E$ in $t_i^\prime$ using a breadth-first search.
	\item[\textsf{BCa2.}] Delete an edge uniformly at random from $C\setminus uv$ to give $t_{i+1}$ and halt.
	\item[\textsf{BCb1.}] Select a vertex $u^\prime$ in $t_i$ uniformly at random, and a vertex $v^\prime\in \Gamma_G(u^\prime)$ also uniformly at random. If $u^\prime v^\prime$ is not an edge of $t_i$ then restart. If $t_i\cup uv \setminus u^\prime v^\prime$ forms a valid tree, then set this to $t_{i+1}$ and halt, otherwise set $t_{i+1}=t_i$ and halt.
\end{enumerate}
We observe that in the case  $v\notin t_i$, the algorithm effectively reduces to the CP algorithm where the edge chosen is a leaf edge, and the smaller tree is just a single vertex.

We expect the distribution of the length of cycles broken/formed by the BC algorithm to follow a power law, and the breaking of large cycles can trigger large scale structural changes in the graph which results in relatively fast mixing. We refer the reader to \cite{fredes2021models} for in-depth analysis and proof that the algorithm does indeed produce uniform lattice trees.

We note that we progress to the second stage of the \textsf{CP} algorithm only if the initial vertices $u$ and $v$ form an edge in $t_i$. On the other hand, we progress to the second stage of the \textsf{BC} algorithm only if $uv$ is not an edge of $t_i$. Thus \textsf{CP} and \textsf{BC} progress under disjoint conditions, and we use this fact to combine them into a new MCMC algorithm, which we will term the \emph{Cut-and-Paste Break-Cycle algorithm} (\textsf{CPBC}), which preserves the aperiodicity and irreducibility of both individual algorithms:
\begin{enumerate}[labelindent=0em,labelwidth=\widthof{CPBC1.},itemindent=0em,leftmargin=!]
	\item[\textsf{CPBC1.}] Select a vertex $u\in t_i$ uniformly at random, and then select a neighbouring vertex $v\in \Gamma_G(u)$ also uniformly at random.
	\item[\textsf{CPBC2.}] If $uv$ is an edge, go to \textsf{CP2}, otherwise, go to \textsf{BC2}.
\end{enumerate}
A few lines of algebra establishes that this algorithm also preserves the uniform  invariant distribution of its two constituents.
	
One we have sampled the lattice tree, we must calculate the exponents.	
We used breadth-first search and dynamic programming techniques to calculate the mean branch size.
To find an intrinsic longest path, we use the method introduced by Dijkstra around 1960: We choose a vertex $v$ in the tree (at random, the choice being immaterial), and then find a vertex $u$ in the tree with maximum intrinsic distance from $v$ using breadth-first search. We then find a vertex $u^\prime$ in the tree with maximum intrinsic distance from $u$, and record the intrinsic distance between $u$ and $u^\prime$. 
The fact that this produces a pair of vertices at maximal intrinsic distance from each other is proven formally in \cite{BULTERMAN200293}.
Once this is done, we compute the extrinsic distance between $u$ and $u^\prime$ either exactly or using an approximating quasi-norm as discussed in the next section, with the details being context-dependent. 
In each case, we averaged the outputs of these computations over a large number of runs to estimate $I(n)$, $E(n)$, and $B(n)$, plotted log-log plots of these quantities against $n$, 
and calculated estimates of $\nu$ and $\rho$ by measuring the gradients of the final sections of the resulting curves.

\section{Transitive Lattices} \label{section:TLattices}

In this section we define the various Cayley graphs we consider and report the outcomes of our simulations on these Cayley graphs. 
Given a finitely generated group $\Gamma$ and a finite set $S$ which generates $\Gamma$, 
the (right) \emph{Cayley graph} $\textrm{Cay}(\Gamma,S)$ is defined to be the undirected graph  with vertex set $\Gamma$ and edge set $\{\gamma, \gamma s\}: \gamma \in \Gamma, s \in S \cup S^{-1}\}$. 
Cayley graphs are always transitive since each element $\gamma$ of $\Gamma$ defines an automorphism of $\textrm{Cay}(\Gamma,S)$ by left multiplication. The graph metric on $\textrm{Cay}(\Gamma,S)$ is also known as the \emph{word metric} and can be expressed as 
\begin{multline*}d_{S}\left(\gamma_{1}, \gamma_{2}\right)=\min \left\{n \geq 0: \exists s_{1}, \ldots, s_{n} \in S \cup S^{-1} \right. \\ \left.\textrm{ such that }\gamma_{2}=\gamma_{1} s_{1} \cdots s_{n}\right\},\end{multline*} and observe that this coincides with the graph metric. For each of the groups we consider, the word metric is comparable to a \emph{quasi-norm} that is much easier to compute. We will use these quasi-norms in place of the word metric when computing distances on $G_{4,3}$ and $G_{5,8}$.

\medskip

Recall that two metric spaces $(X,d_X)$ and $(Y,d_Y)$ are said to be \emph{quasi-isometric} if there exist positive constants $\alpha$ and $\beta$ and a function $\phi:X\to Y$ such that $\alpha^{-1} d_X(x,y)-\beta \leq d_Y(\phi(x),\phi(y)) \leq \alpha d_X(x,y) + \beta$ for every $x,y\in X$ and for every $y\in Y$ there exists $x \in X$ with $d_Y(y,\phi(x))\leq \beta$. It is easily seen that different Cayley graphs of the same finitely generated group are quasi-isometric to each other and that e.g.\ $\Z^d$ is quasi-isometric to $\R^d$ for each $d\geq 1$.

\medskip

When $\Gamma$ is a group, the \emph{lower central series} of $\Gamma$ is defined recursively by $\Gamma_1 = \Gamma$ and $\Gamma_{i+1}=[\Gamma_i,\Gamma]=\langle \{[a,b]:a\in \Gamma_i,b\in \Gamma\}\rangle$. The group $\Gamma$ is said to be \emph{nilpotent} if there exists $s \geq 1$, known as the \emph{step} of $\Gamma$, so that $\Gamma_s$ is abelian and hence that $\Gamma_i =\{\mathrm{id}\}$ for every $i>s$. The Bass-Guivarc'h formula \cite{bass72poly-growth,guivarch73poly-growth} states that if $\Gamma$ is a torsion-free nilpotent group then $\Gamma$ has volume growth dimension $\sum_{i=1}^s i r_i$ where $r_i$ is the rank of the abelian group $\Gamma_i/\Gamma_{i+1}$. The quantity $\sum_{i=1}^s i r_i$ is also known as the \emph{homogeneous dimension} of the group. It is a consequence of Pansu's theorem \cite{MR741395} that both the step $s$ and the sequence $(r_1,\ldots,r_s)$ are quasi-isometry invariants of nilpotent groups.

\begin{table*}
\centering
\makebox[\textwidth][c]{
\begin{tabular}{|l||l| l | l | l | } 
		\hline
		Group & \makecell{Set} & \makecell{Multiplication rule} & \makecell{Generators} & Quasi-norm\\
		\hline \hline
		$\cH$ \makecell{\phantom{.}\\\phantom{.}} & $\Z^3$ & \makecell{$(a_1,b_1,c_1)\cdot(a_2,b_2,c_2)=(a_1+a_2,b_1+b_2,c_1+c_2+a_1b_2)$} & $\{a,b\}$ & $|a|+|b|+|c|^{1/2}$\\
		\hline
		$\cH \times \Z$ & $\Z^4$ & \makecell{$(a_1,b_1,c_1,d_1)\cdot(a_2,b_2,c_2,d_2)=$\\\hspace{2.57cm}$(a_1+a_2,b_1+b_2,c_1+c_2+a_1b_2,d_1+d_2)$} & $\{a,b,d\}$ & $|a|+|b|+|c|^{1/2}+|d|$\\
		\hline
		$G_{4,3}$ & $4\Z \times (2\Z)^3$ & \makecell{$(a_1,b_1,c_1,d_1)\cdot(a_2,b_2,c_2,d_2)=$\\\hspace{0.2cm}$
	(b_1d_2+\tfrac{1}{2} d_2^2 c_1+a_2+a_1,c_1 d_2+b_1+b_2,c_1+c_2,d_1+d_2)$}&$\{2b,2c,2d\}$&$|a|^{1/3}+|b|^{1/2}+|c|+|d|$\\
	\hline
	$G_{5,8}$ & $\Z^5$ & \makecell{$(a_1,b_1,c_1,d_1,e_1)\cdot (a_2,b_2,c_2,d_2,e_2)=$\\$
	(a_1+a_2 + b_1d_2,b_1+b_2,-d_1 e_2+c_1+c_2,d_1+d_2,e_1+e_2)$}& $\{b,d,e\}$ &$|a|^{1/2}+|b|+|c|^{1/2}+|d|+|e|$\\
	\hline
	\end{tabular}
	}
	\caption{Definitions and basic properties of the Cayley graphs we consider}
	\label{table:group_defs}
\end{table*}

\medskip

We chose four non-Euclidean groups to study, namely $\mathcal{H}$, $\mathcal{H}\times\mathbb{R}$, $G_{4,3}$ and $G_{5,8}$. In some cases we also carried out simulations on $\Z^4$ and $\Z^5$ so that we could directly compare our results to the Euclidean case.
The upper critical dimension of percolation is $6$, so we limited our study to the most interesting dimensions of four and five where mean-field behaviour does not hold but there is more than one quasi-isometry class of geometries to consider.
The upper critical dimension for lattice trees is $8$, meaning that more interesting possibilities are available. 
We chose to study the two seven-dimensional groups $G_{4,3}$ and $G_{5,8}$ since they were highly distinct from the other examples we considered, being neither abelian, generalized Heisenberg, nor products thereof. 
These groups are defined as lattices in the nilpotent Lie groups corresponding to the nilpotent Lie algebras notated in \cite{MR2303198} and \cite[Table 1]{isenrich2020coneequivalent} as $\mathscr{L}_{4,3}$ and $
\mathscr{L}_{5,8}$.
The multiplication rules and generating sets of these groups were computed using Maple.
A complete taxonomy of possible low-dimensional geometries can be found in \cite[Tables 1-4 and Figure 5]{isenrich2020coneequivalent}.

\medskip

We briefly introduce each of the groups we consider, with most of the relevant information summarised in \cref{table:group_defs}.

\medskip

\textbf{The Heisenburg Group $\mathcal{H}$.}
The discrete Heisenburg group can be defined as the set of integer-valued upper-triangular $3\times3$ matrices under matrix multiplication.
We identify each matrix $M\in \mathcal{H}$ with an element of $\mathbb{Z}^3$ via the bijection $\phi:\mathbb{Z}^3\rightarrow\mathcal{H}$ given by
\[
\phi\big((a,b,c)\big) = \begin{pmatrix}
	1 & a & c\\
	0 & 1 & b\\
	0 & 0 & 1
\end{pmatrix},
\] and use these coordinates to represent  elements of the group. These are known as the \emph{Mal'cev} coordinates. 
Multiplication of two elements is therefore given by:
\[(a_1,b_1,c_1)\cdot (a_2,b_2,c_2) =(a_1+a_2,b_1+b_2,c_1+c_2+ a_1b_2). \]
The Heisenburg group is generated by the elements 
$a = (1,0,0)$ and $b = (0,1,0)$ as witnessed by the identity
\[
(x,y,z) = b^y [a,b]^z a^x,
\]
where $[a,b]$ is the commutator $aba^{-1}b^{-1}$.
We will work with the right-Cayley graph $\Gamma_{\mathcal{H}}=\textrm{Cay}(\mathcal{H},\{a,b\})$. (Note that this is not the Cayley graph depicted in \cref{fig:Heisenberg}, which has generating set $\{a,b,c\}$.)
The graph metric on this Cayley graph is equivalent \cite[3.1.6]{nilpotentLG} to the quasi-norm 
\[
\left\|(a,b,c)\right\| = \abs{a}+\abs{b}+\abs{c}^{1/2}.
\]
 We will also make use of a formula for computing graph distances in this Cayley graph that is described in \cite{wordh}; this formula is too long to reproduce here but is easily implemented on a computer.  The Heisenberg group has step $2$ and $(r_1,r_2)=(2,1)$.

 \medskip

All of the above mentioned facts have obvious consequences for the product space $\mathcal{H}\times \Z$, for which we will consider the Cayley graph generated by $a=(1,0,0,0)$, $b=(0,1,0,0)$, and $d=(0,0,0,1)$. This group has step $2$ and $(r_1,r_2)=(3,1)$.

\medskip

\textbf{The seven-dimensional geometry  $G_{4,3}$}.
The group  $G_{4,3}$ is defined as a lattice in the nilpotent Lie group corresponding to the Lie algebra notated in \cite{MR2303198} as $\mathscr{L}_{4,3}$. Concretely, the group is defined as the set
$4\mathbb{Z}\times2\mathbb{Z}\times2\mathbb{Z}\times2\mathbb{Z}$ 
equipped with the multiplication operation
\begin{multline*}(a_1,b_1,c_1,d_1)\times(a_2,b_2,c_2,d_2) =\\
	(b_1d_2+\frac{1}{2} d_2^2 c_1+a_2+a_1,c_1 d_2+b_1+b_2,c_1+c_2,d_1+d_2),
\end{multline*} which has identity element $(0,0,0,0)$. The group is generated by the elements
$2b=(0,2,0,0)$, $2c=(0,0,2,0)$, and $2d=(0,0,0,2)$
  as witnessed by the formula
\[(4x,2y,2z,2w) = [2b,2d]^x (2d)^w (2b)^y (2c)^z.\]
We work with the Cayley graph
$\textrm{Cay}(G_{4,3},\{2b,2c,2d\})$, whose word metric is comparable to the quasi-norm
\[
\left\|(a,b,c,d)\right\| = \abs{a}^{1/3}+\abs{b}^{1/2} + \abs{c}+\abs{d}\]
 The group $G_{4,3}$ has step $3$ and $(r_1,r_2,r_3)=(2,1,1)$.

\medskip

\textbf{The seven-dimensional geometry $G_{5,8}$}.
The group  $G_{5,8}$ is defined as a lattice in the nilpotent Lie group corresponding to the Lie algebra notated in \cite{MR2303198} as $
\mathscr{L}_{5,8}$. 
Concretely, the group is defined as the set
$\Z^5$  
equipped with the multiplication operation
\begin{multline*}(a_1,b_1,c_1,d_1,e_1)\times(a_2,b_2,c_2,d_2,e_2) =\\
	(a_1+a_2 + b_1d_2,b_1+b_2,-d_1 e_2+c_1+c_2,d_1+d_2,e_1+e_2),
\end{multline*} which has identity element $(0,0,0,0,0)$. The group is generated by the elements $b = (0,1,0,0,0)$, $d = (0,0,0,1,0)$, and $e = (0,0,0,0,1)$  as witnessed by the formula
\[(x,y,z,w,v) = [e,d]^z e^v d^w [b,d]^x b^y.\]
We work with the Cayley graph
$\Gamma_{G_{5,8}}=\textrm{Cay}(G_{5,8},\{b,d,e\})$, whose word metric is comparable to the quasi-norm

\[
\left\|(a,b,c,d,e)\right\| = \abs{a}^{1/2}+\abs{b} + \abs{c}^{1/2}+\abs{d}+\abs{e}
\]
 The group $G_{5,8}$ has step $2$ and $(r_1,r_2)=(3,2)$.

\begin{figure*}
	\centering
	\begin{minipage}[b]{0.475\textwidth}
	\centering
	\includegraphics[width=1\textwidth]{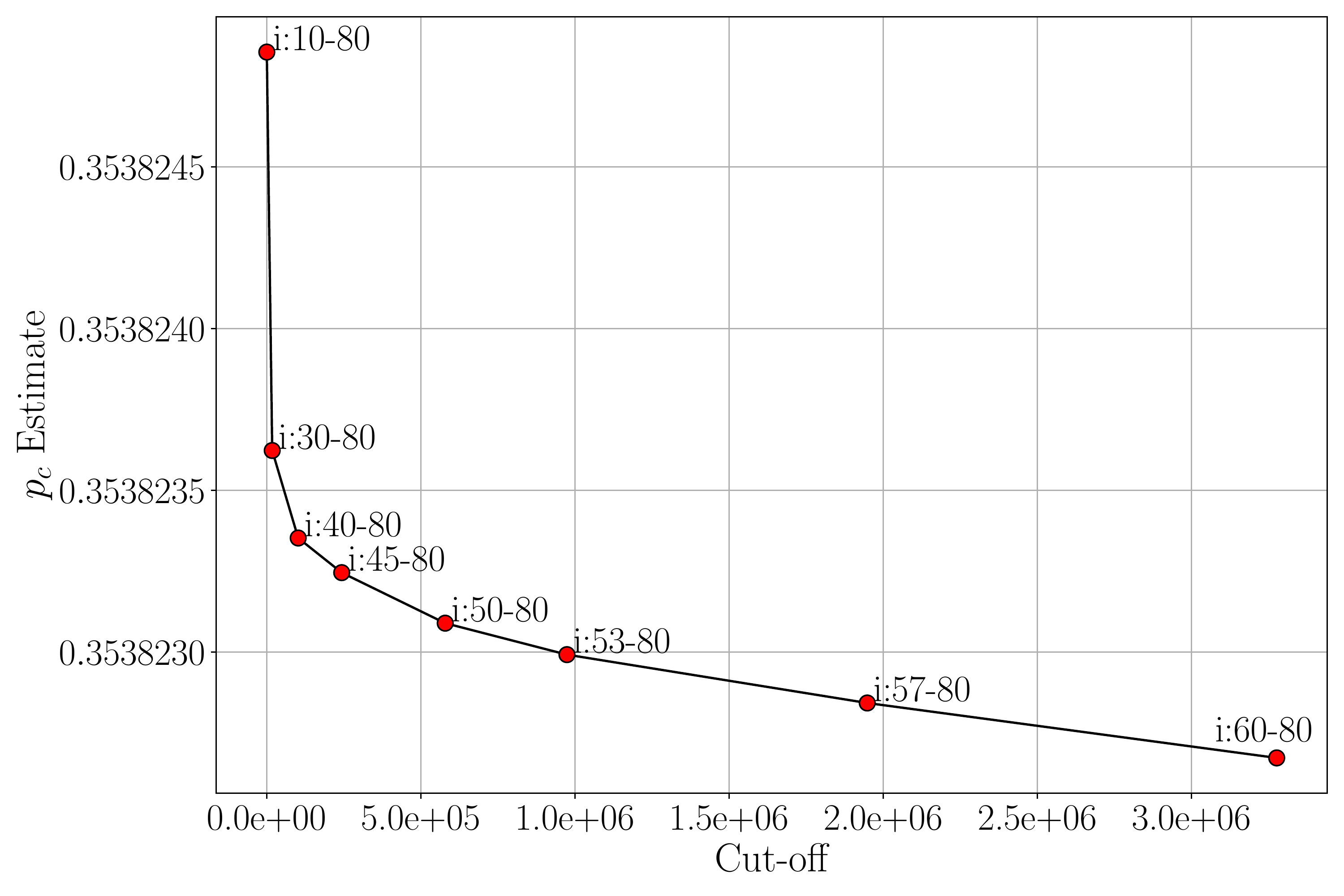}
	\subcaption{The Heisenberg group}
	\label{fig:4dpcextrapwcutoff}
	\end{minipage}
	\hfill
	\begin{minipage}[b]{0.475\textwidth}
	\centering
	\includegraphics[width=1\textwidth]{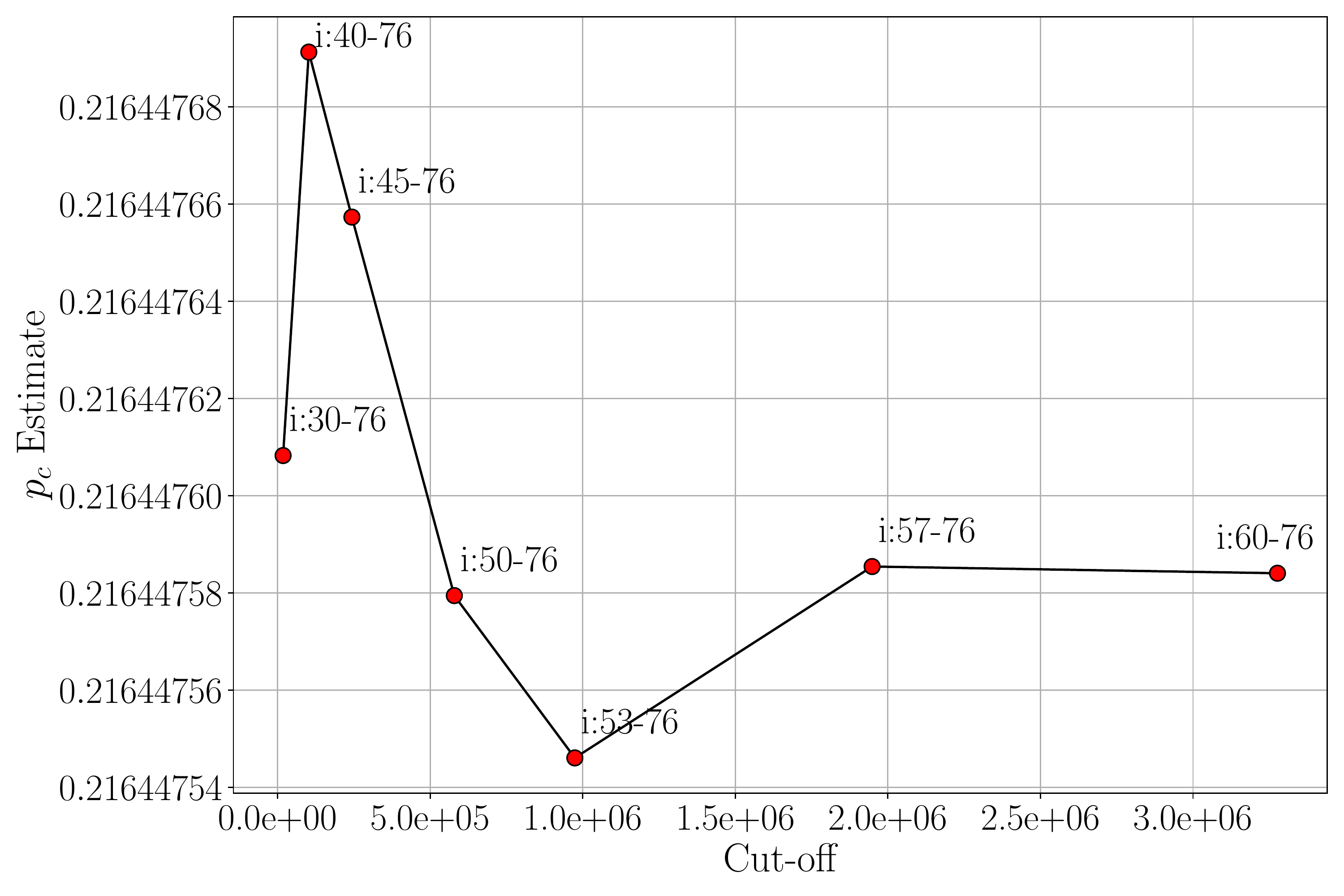}
	\subcaption{The product space $\cH\times\Z$.}
	\label{fig:5dpcextrapwcutoff}
	\end{minipage}
	\caption{Estimated percolation thresholds for $\mathcal{H}$ and $\mathcal{H}\times \Z$ with varying numbers of excluded initial points. A label of the form $i:a-b$ indicates that the fit was calculated with $a_{\lfloor{100\times2^{i/4}}\rfloor},i =a\ldots b$.}
\end{figure*}

\begin{figure*}
	\centering
	\begin{minipage}[b]{0.475\textwidth}
	\centering
	\label{fig:4dpcleath1}
	\includegraphics[width=1\textwidth]{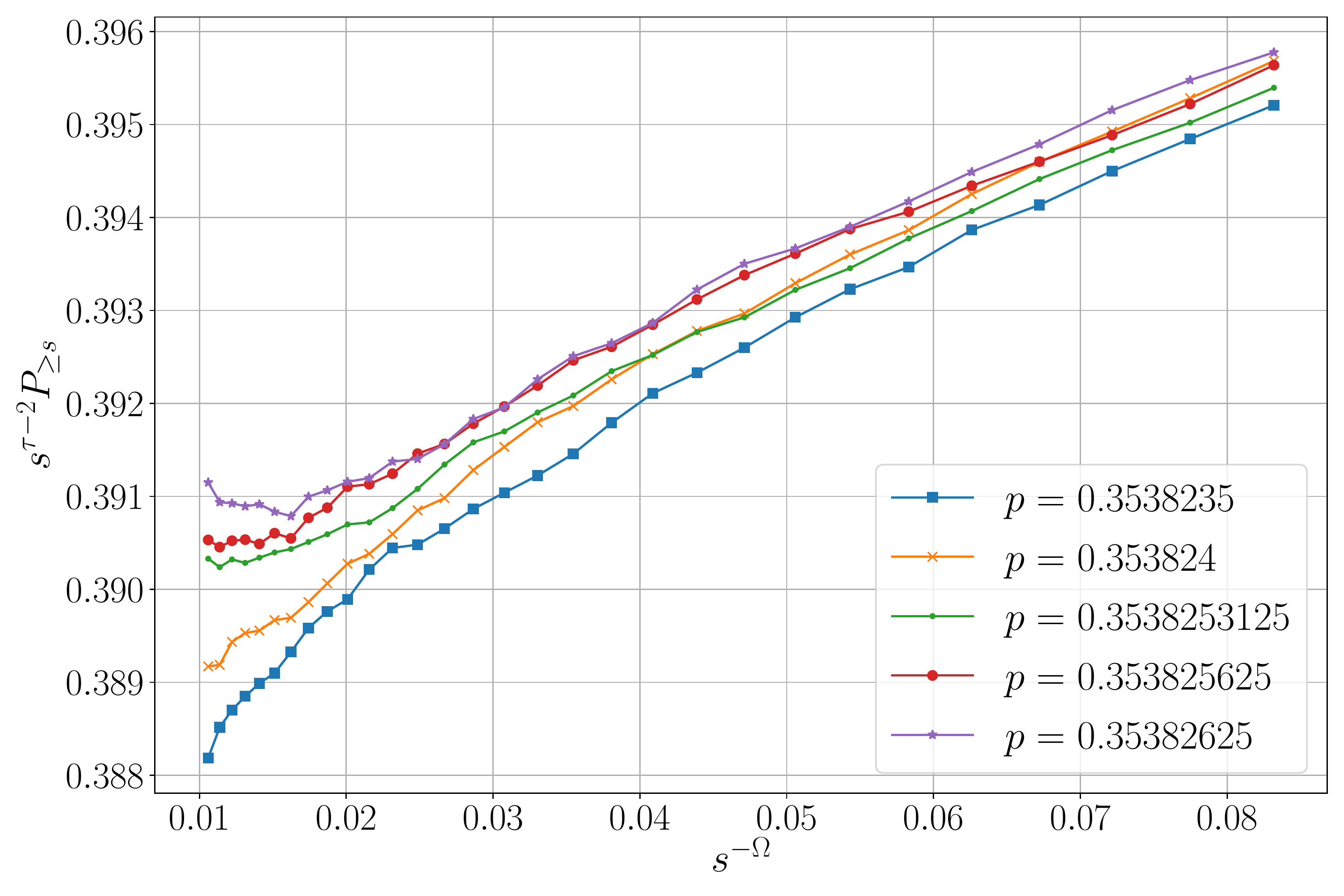}
	\subcaption{Plot of $s^{-\Omega}$ against $s^{\tau-1}P_{\geq s}$ on $\mathcal{H}$. Smaller deviations from linearity for small $s^{-\Omega}$ indicates that $p$ is closer to $p_c$.}
	\end{minipage}
	\hfill
		\begin{minipage}[b]{0.475\textwidth}
			\centering
	\label{fig:4dpcleath2}
	\includegraphics[width=1\textwidth]{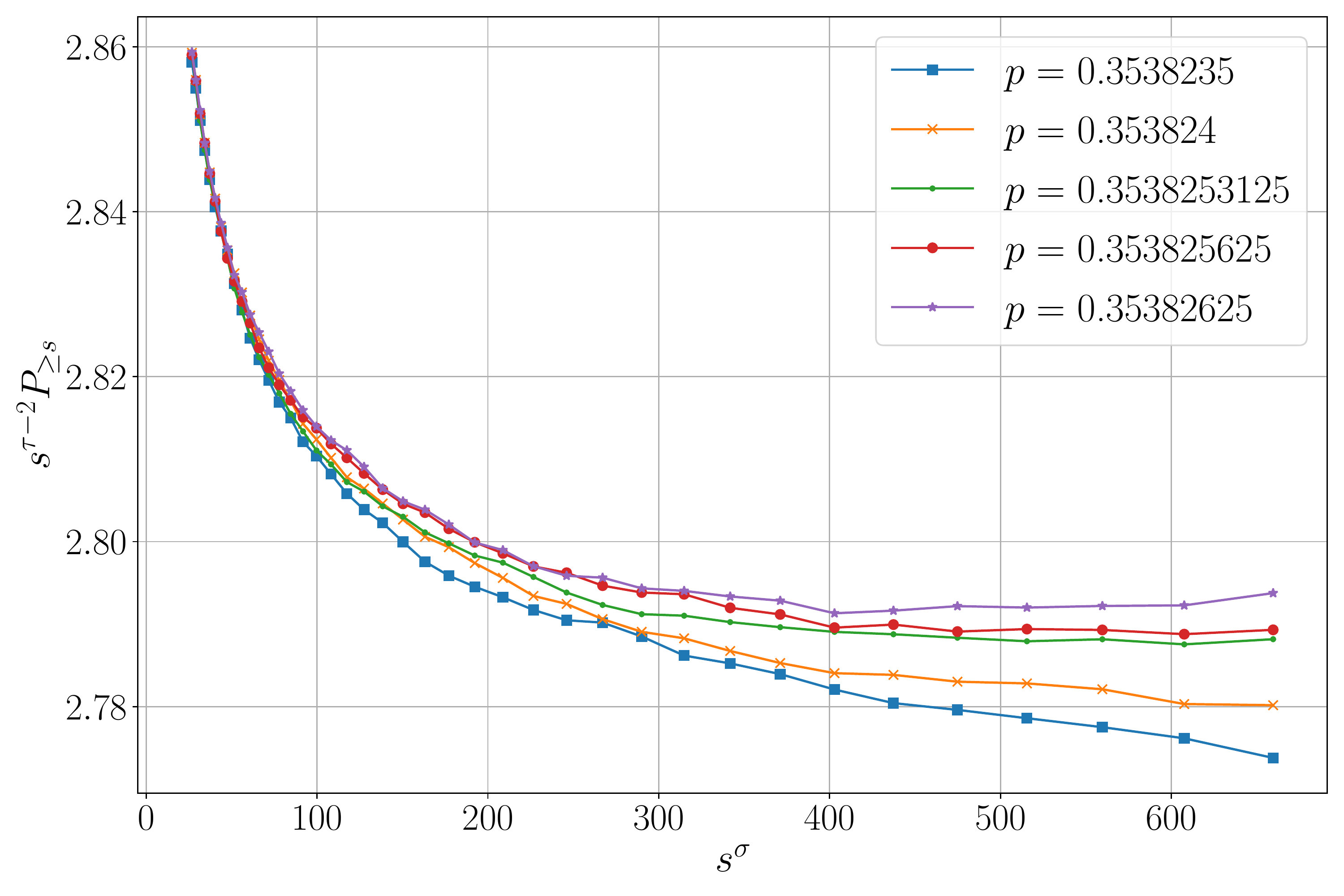}
	\subcaption{Plot of $s^{\sigma}$ against $s^{\tau-1}P_{\geq s}$ on $\mathcal{H}$. A plateau at large $s^\sigma$ indicates that $p$ is close to $p_c$.}
	\end{minipage}

	\begin{minipage}[b]{0.475\textwidth}
	\centering
	\label{fig:4dpcleath1}
	\vspace{0.5em}
	\includegraphics[width=1\textwidth]{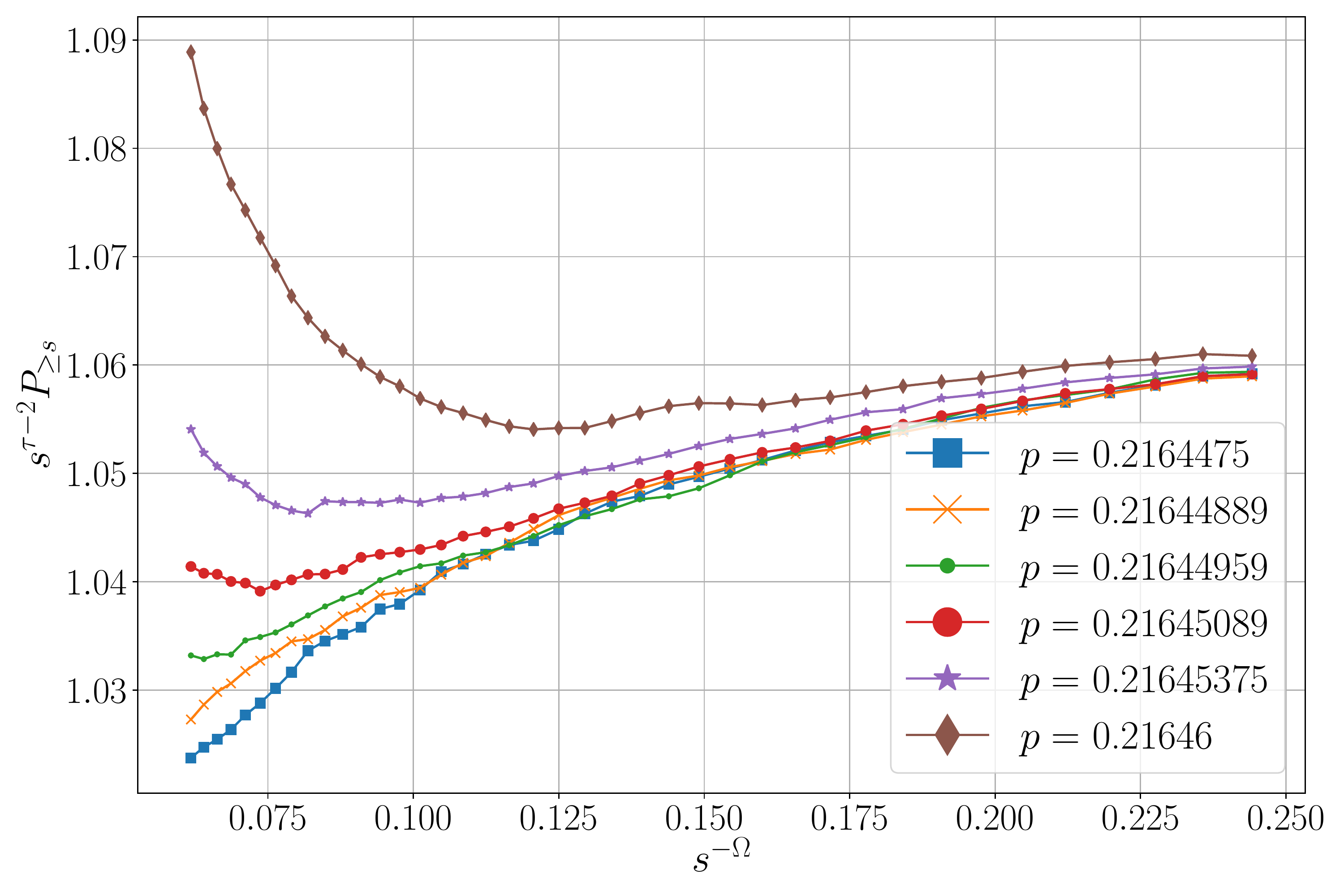}
	\subcaption{Plot of $s^{-\Omega}$ against $s^{\tau-1}P_{\geq s}$ on $\mathcal{H}\times \Z$. Smaller deviations from linearity for small $s^{-\Omega}$ indicates that $p$ is closer to $p_c$.}
	\end{minipage}
	\hfill
		\begin{minipage}[b]{0.475\textwidth}
			\centering
			\vspace{0.5em}
	\label{fig:4dpcleath2}
	\includegraphics[width=1\textwidth]{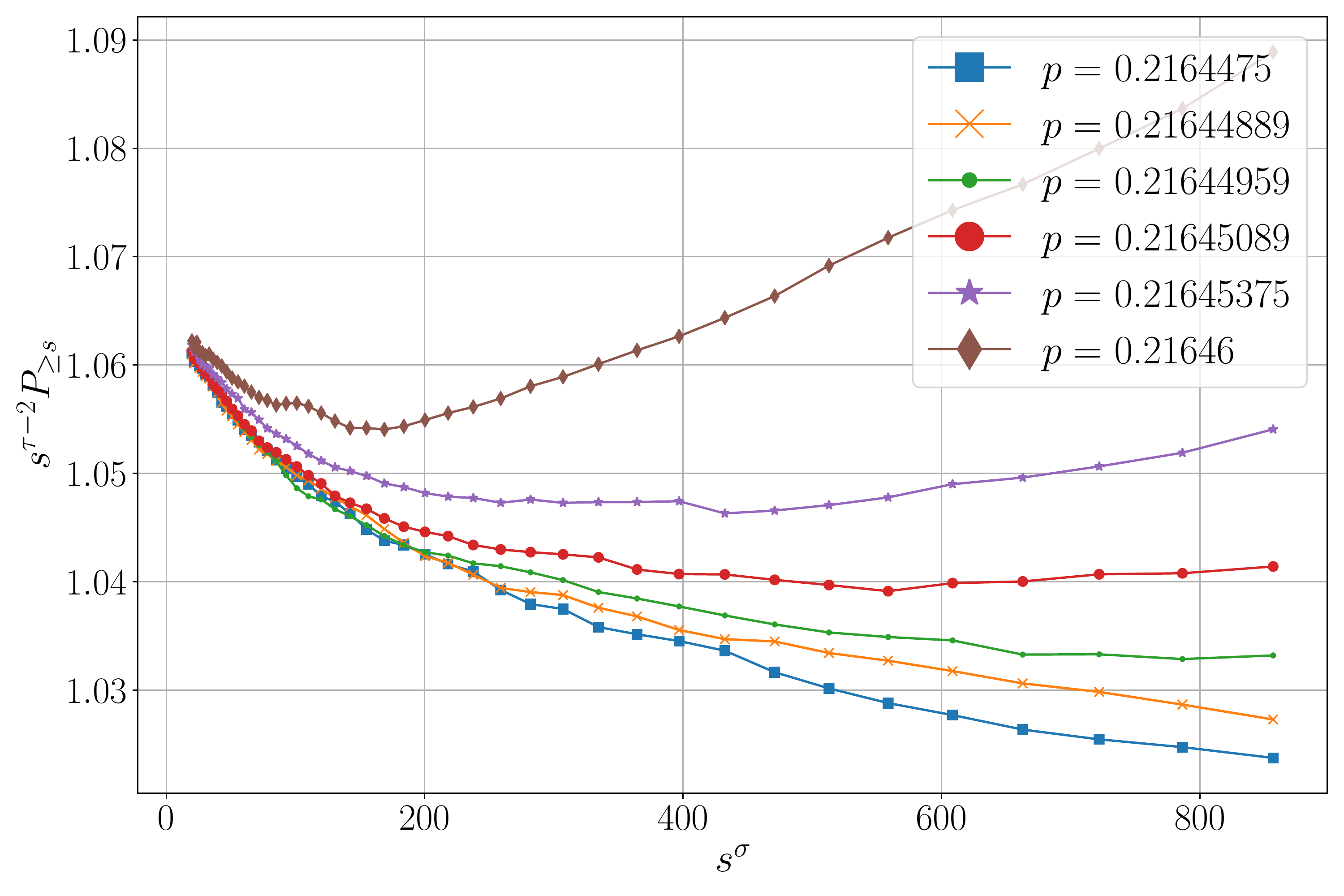}
	\subcaption{Plot of $s^{\sigma}$ against $s^{\tau-1}P_{\geq s}$ on $\mathcal{H}\times \Z$. A plateau at large $s^\sigma$ indicates that $p$ is close to $p_c$.}
	\end{minipage}
	\caption{Runs of the Leath algorithm on $\mathcal{H}$ and $\mathcal{H}\times\Z$ for different values of the percolation probability $p$. }
	\label{fig:Leath}
\end{figure*}

\subsection{Results for Percolation}
\label{sec:percolation_results}

\begin{figure*}
\vspace{-0.5em}
	\centering
	\begin{minipage}[b]{0.475\textwidth}
		\centering
		\includegraphics[width=1\textwidth]{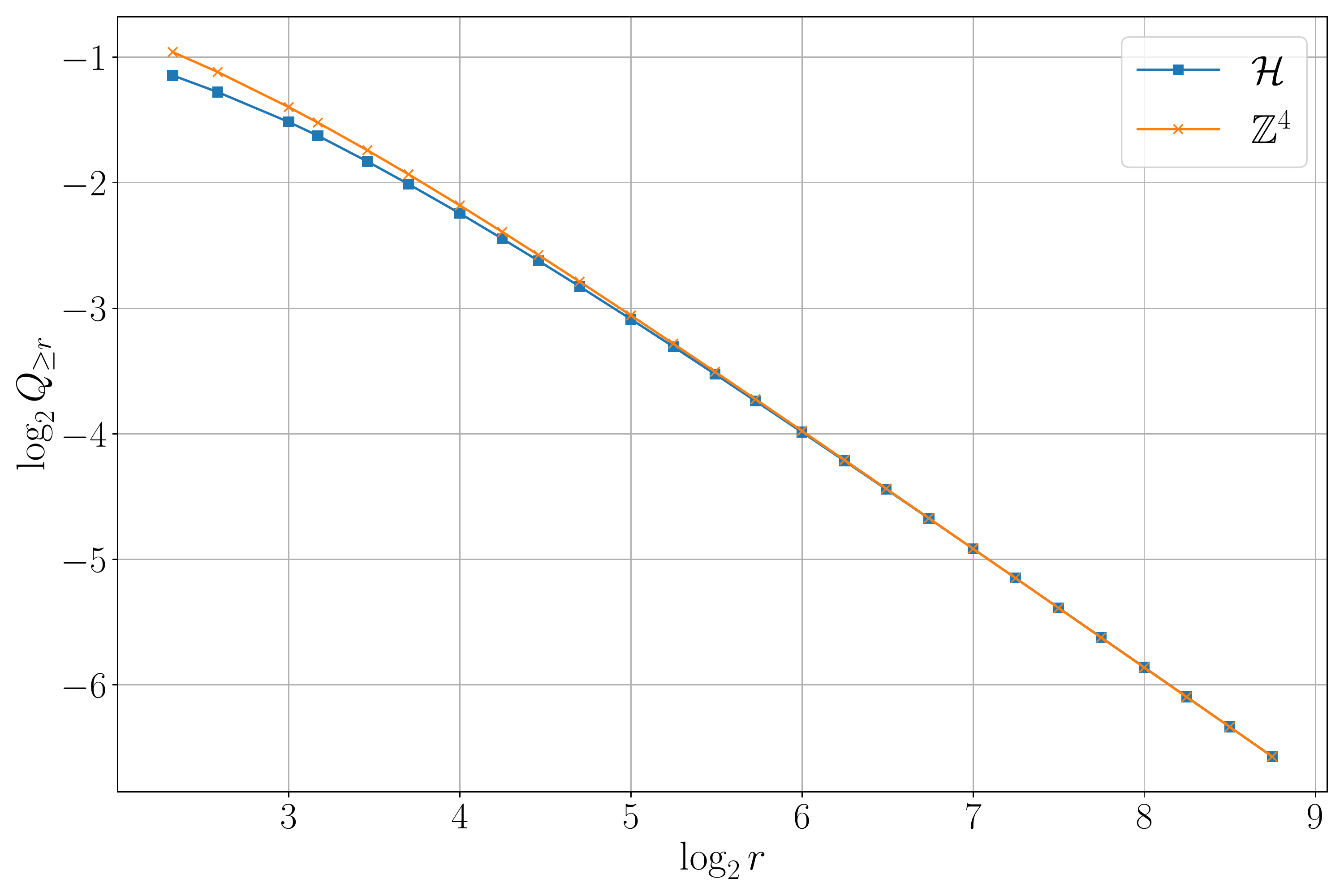}
		\subcaption{Four-dimensional lattices.}
		\label{fig:4dextrinsic}
	\end{minipage}
	\hfill
	\begin{minipage}[b]{0.475\textwidth}
		\centering
		\includegraphics[width=1\textwidth]{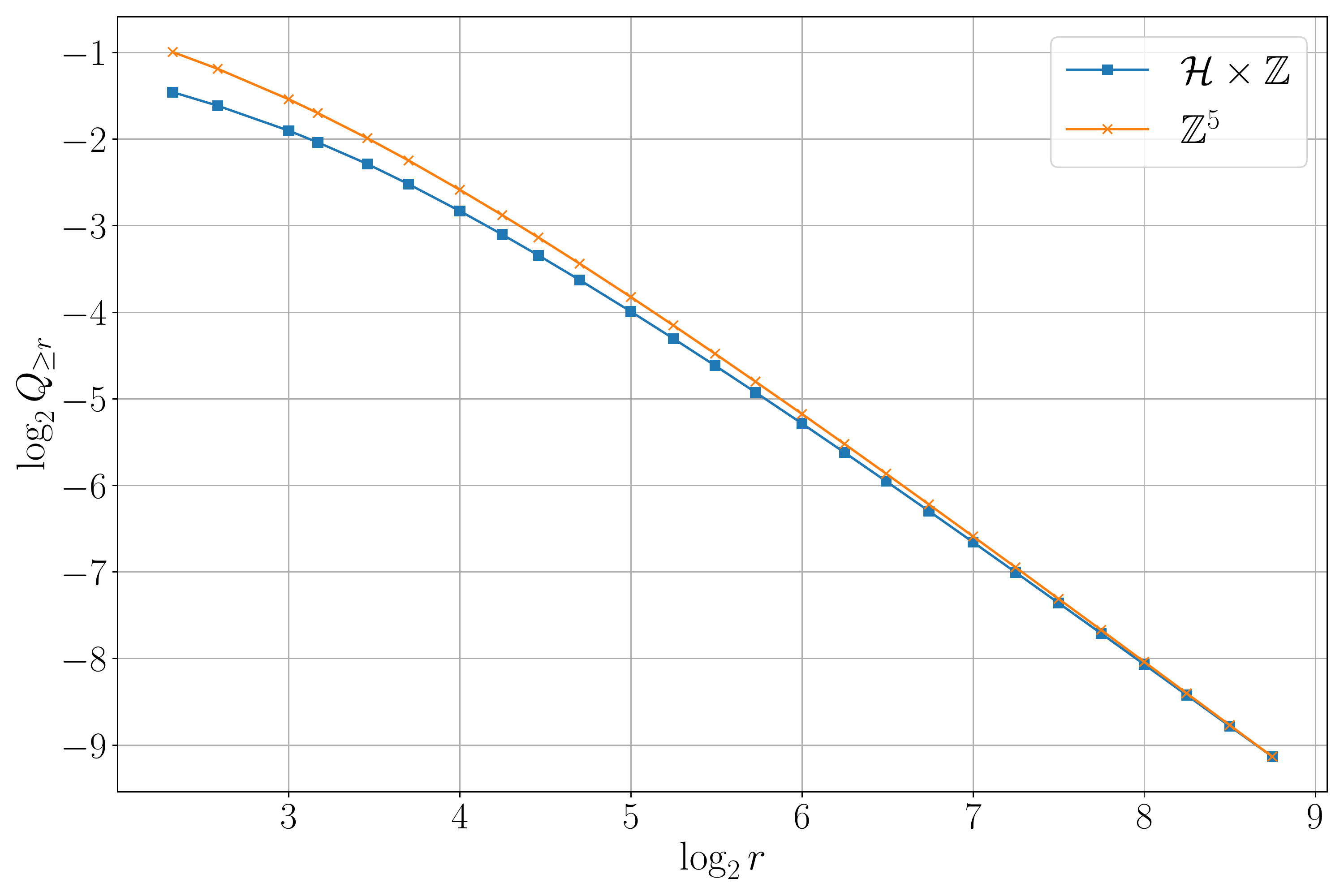}
		\subcaption{Five-dimensional lattices.}
		\label{fig:5dextrinsic}
	\end{minipage}
	\caption{Log-log plots of the radius tail $Q_{\geq r}$ against $r$ for the transitive lattices $\mathcal{H}$, $\mathbb{Z}^4$, $\mathcal{H}\times\mathcal{Z}$, and $\mathbb{Z}^5$, produced using runs of the Leath algorithm at their respective critical probabilities. The curves are vertically translated to allow for easy comparison of their final gradients, which are seen to be in close agreement in both cases.}
	\label{fig:extrinsicloglog}
	\vspace{-0.5em}
\end{figure*}

We now describe the simulations we carried out for percolation on $\cH$ and $\cH \times \Z$ and the results that we obtained.

\medskip

\textbf{Estimating $\mathbf{p_c}$}.
In each case, we began with a small number of initial runs of invasion percolation, as described in Appendix \ref{section:IPA}, in order to approximate the constants $F$ and $z$ in \eqref{eq:dbfit} to achieve a speed up for further runs.

For the Heisenberg group $\mathcal{H}$ we then generated approximately $9\times10^5$ samples, each with a total number of $\lfloor{100\times2^{67/4}}\rfloor=11,021,797$ steps, and a further $100,000$ samples each with a total number of $\lfloor{100\times2^{80/4}}\rfloor=104,857,600$ steps.
We recorded and averaged the sampled bulk-to-boundary ratios $a_n$ at $n=\lfloor{100\times2^{i/4}}\rfloor$ for $5\leq i\leq80$, for a total of 76 points. We then used weighted least mean squares to fit the parameters $p_c,A,\delta$ in \ref{eq:pc} to the data. We noticed that removing the points at small $n$ (at the beginning of the runs) shifted our estimate of $p_c$, lessening the effect of finite-size effects. We plotted the effect of removing small values of $n$ in \cref{fig:4dpcextrapwcutoff} and extrapolated from the resulting data to obtain the estimate $p_c\approx 0.3538225\pm0.0000010$.

We then repeated this procedure for $\mathcal{H}\times\mathbb{Z}$. This time we generated approximately $6\times10^5$ samples each with a total number of $\lfloor{100\times2^{67/4}}\rfloor=11,021,797$ steps, and a further approximately $200,000$ samples each with a total number of $\lfloor{100\times2^{76/4}}\rfloor=52,428,800$ steps. We recorded and averaged the sampled bulk-to-boundary ratios $a_n$ at $n=\lfloor{100\times2^{i/4}}\rfloor$ for $5\leq i\leq76$,  for a total of 72 points.
Again, plotting the effect of removing small values of $n$ in \cref{fig:5dpcextrapwcutoff} and extrapolating yielded the estimate $p_c\approx 0.2164476\pm0.0000001$.

\medskip

\textbf{Estimating intrinsic exponents}. Having obtained these estimates for the critical probability, we sampled percolation at $p=0.3538225$ for $\mathcal{H}$ and $p=0.2164476$ for $\mathcal{H}\times\mathbb{Z}$ using the Leath algorithm. In each case we collected approximately $10^8$ samples each with $2^{20}$=1,048,576 time steps. We calculated $P_{\geq s}$ and $\E[n/p-t/(1-p)|n\geq s]$ empirically from these samples and fit the data to the ansatz equations presented in \textsf{Q1a} and \textsf{Q1b} we obtained the estimates $\tau=2.315$ and $\sigma=0.4758$ for $\cH$ and $\tau=2.420$ and $\sigma=0.4988$ for $\cH \times \Z$. All these results were in close agreement with previously derived values for $\Z^4$ and $\Z^5$ (see Table \ref{table:summary}), giving weight to the claim of \cref{conj:1}.

\medskip

\textbf{Refinement and confirmation}. 
Next, we refined our values for the critical probability, and simultaneously add extra weight to the claim that the critical exponents are shared by the Euclidean and non-Euclidean lattices employing methods as outlined in \cite{PhysRevE.57.230,PhysRevResearch.2.013067}

We ran the Leath algorithm at multiple values of $p$, with between $10^8$ and $10^9$ samples per value of $p$, and with each run having $2^{20}=1,048,576$ steps. As presented in \cref{fig:Leath}, we then plotted graphs of $s^{\tau-2} P_{\geq s}$ against $s^{-\Omega}$ and against $s^\sigma$ as described in \textsf{G1} and \textsf{G2}, but where we used the values of $\tau$, $\sigma$, and $\Omega$ for $\Z^4$ and $\Z^5$ as computed in \cite{PhysRevE.64.026115}, \cite{PhysRevD.92.025012}, and \cite{PhysRevResearch.2.013067} respectively. 
If \cref{conj:1} is true, then as explained in \cite{PhysRevE.57.230},   the plots against $s^{-\Omega}$ should look approximately linear when $p=p_c$ while the plots against $s^\sigma$ should plateau for large $s$ when $p=p_c$.
As such, the figures indicate that the critical probability for $\mathcal{H}$ lies between $p=0.353824$ and $p=0.3538253125$ while the critical probability for $\mathcal{H} \times \Z$ lies between $p=0.21644 889$ and $p=0.21644 959$. In each case, the fact that we do indeed see approximately linear behaviour in the plots against $s^{-\Omega}$ and a large-$s$ plateau in the plots against $s^\sigma$ strongly suggests that \cref{conj:1} is true and e.g.\ the values of $\tau$, $\sigma$, and $\Omega$ are the same for $\cH$ and $\Z^4$.

 \begin{tablehere} 
 	\centering
 	\vspace{1em}
 	\begin{tabular}{|l||l|l|}
 		\hline
 		& $\mathcal{H}$ & $\mathcal{H}\times\mathbb{Z}$ \\ \hline \hline
 		$p_c$ &   $0.3538247\pm0.0000007$&   $0.21644925\pm0.00000036$        \\ \hline
 	\end{tabular}
 	\caption{Critical probability estimates.}
 	\label{table:pcsume}
 \end{tablehere}

\medskip

\textbf{Estimating extrinsic exponents.}
Finally, we ran the Leath algorithm on the two non-Euclidean graphs $\mathcal{H}$, $\mathcal{H}\times\Z$ and the two Euclidean graphs $\Z^4$ and $\Z^5$, but this time, instead of running for a fixed number of steps, we halted the algorithm when it first visited a vertex with extrinsic distance $2^9=512$ away from the origin.
For the non-Euclidean lattices, we ran the algorithm at the previously calculated critical percolation estimates displayed in Table \ref{table:pcsume}, and used $p_c = 0.1601312$ for $\Z^4$, extracted from  \cite{PhysRevResearch.2.013067,PhysRevE.98.022120}, and $p_c=0.11817145$ for $\Z^5$, extracted from \cite{PhysRevE.98.022120}. For each of these graphs we then plotted $\log_2 Q_{\geq s}$ against $\log_2 s$ for $s=2^{i/4} $, $10\leq i\leq 36$.  We calculated the gradients of the final sections of the curves to give the estimates $\rho= 1.051$ for $\mathcal{H}$, $\rho = 1.053$ for $\mathbb{Z}^4$, $\rho = 0.703$ for $\mathcal{H}\times\mathbb{Z}$, and $\rho = 0.684$ for $\mathbb{Z}^5$.
The large finite-size effects, especially in the five-dimensional case, meant that the computational resources available to us were insufficient to compute $\rho$ to a high level of precision. Using the scaling relation $\tau = 1 + d/(d-1/\rho)$, we computed secondary estimates
 $\tau=2.312$ for $\Z^4$ and $\mathcal{H}$, and  $\tau=2.413$ for $
\Z^5$ and $2.397$ for $\mathcal{H}\times\mathbb{Z}$.

\subsection{Results for Lattice Trees}
\label{sec:lattice_tree_results}

We now describe the simulations we carried out of lattice trees on $\cH$, $\Z^4$, $\cH \times \Z$, $\Z^5$, $G_{4,3}$, and $G_{5,8}$ and the results that we obtained. (We ran our own simulations on $\Z^4$ and $\Z^5$ for better comparability with our non-Euclidean simulations since the simulations of \cite{PhysRevE.58.3971} used much smaller tree sizes.)

For each of the graphs that we considered, we initialized the \textsf{CPBC} MCMC algorithm with 
tree sizes $s=\lfloor 10000\times20^{i/10}\rfloor$ from $i=-5$ to $i=10$ for the four and five dimensional lattices, and up to $i=13$ for the seven dimensional lattices. We collected between $100,000$ samples and $500,000$ samples for each tree size. For a tree of size $s$, we evolved the algorithm for an initial $4s$ steps, and then collected a sample every $2s$ steps thereafter.
The initial trees of size $s$ were taken to be paths with $s/2$ vertices lying along a suitable coordinate axis with additional edges coming off each vertex in another fixed coordinate direction. An estimate of the extrinsic exponent $\nu$ was calculated by finding the gradients of the final section of the relevant $\log$-$\log$ curve. An estimate of the intrinsic exponent $\rho$ was found by first averaging the two $\log$-$\log$ curves for branch-size and intrinsic radius, before taking the gradient.

\section{Self-Similar Fractals} \label{section:FTrees}
In this section we give a brief introduction to the self-similar fractals we consider, and describe our results concerning critical percolation on them.

\medskip

The fractals we consider will be defined as the scaling limits of sequences of `prefractal' graphs generated by an initial seed graph and  a recursive rule describing how the generation $n+1$ prefractal is constructed from copies of the generation $n$ prefractal. In addition to the continuum fractal scaling limit, we can also take the  \emph{Benjamini-Schramm limit} of this growing sequence of prefractal graphs, which describes how the graph looks in the vicinity of a uniform random vertex. In each of the cases we consider, the Benjamini-Schramm limit exists and is an infinite, locally finite random rooted graph, so that we can define the critical probability $p_c$ and critical exponents $\tau$ and $\sigma$ with respect to this infinite limit graph.

\medskip

\textbf{Recursive rules}. 
Each of the fractal \emph{trees} we consider will be constructed using a hierarchical coordinate system in the following way, which makes their Benjamini-Schramm limit easy to describe.
Let $N\in\mathbb{N}$ and define $\mathbb{X} = \mathbb{Z}_N^\infty$, where $\Z_N=\Z/N\Z$. We call $\mathbb{X}$ the \emph{coordinate space} of the fractal and write points in $\mathbb{X}$ as $x=(\ldots,x_1,x_0)$.  The number $N$ \linebreak

\begin{Figure}
	\centering
	\vspace{-0.5em}
	\begin{minipage}[b]{0.975\linewidth}\centering\vspace{0.5em}
	\includegraphics[width=1\textwidth]{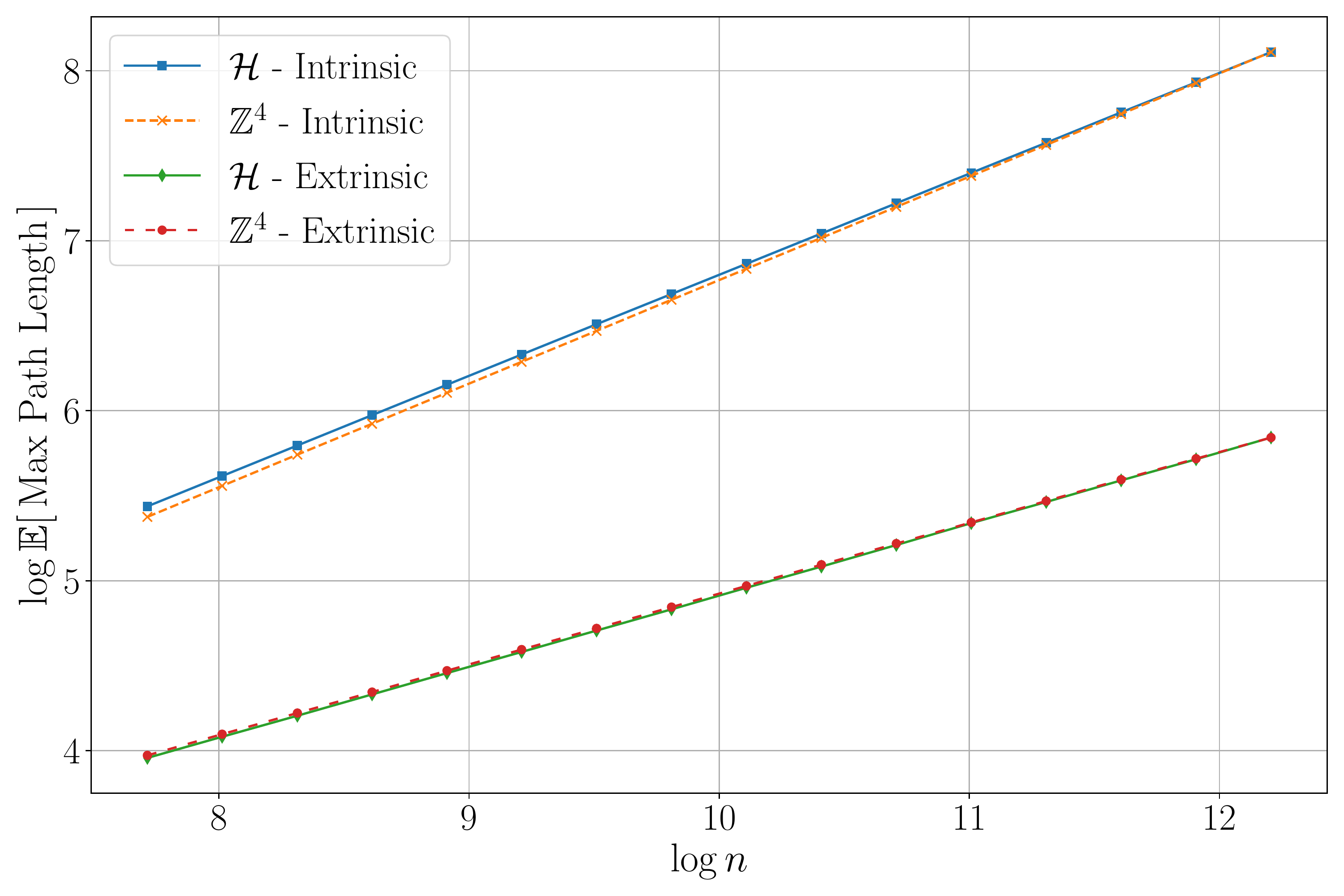}\vspace{0.5em}
	{(a) $\mathcal{H}$ and $\Z^4$.}
	\end{minipage}
		\begin{minipage}[b]{0.975\linewidth}\centering\vspace{1em}
	\label{fig:4dLAE}
	\includegraphics[width=1\textwidth]{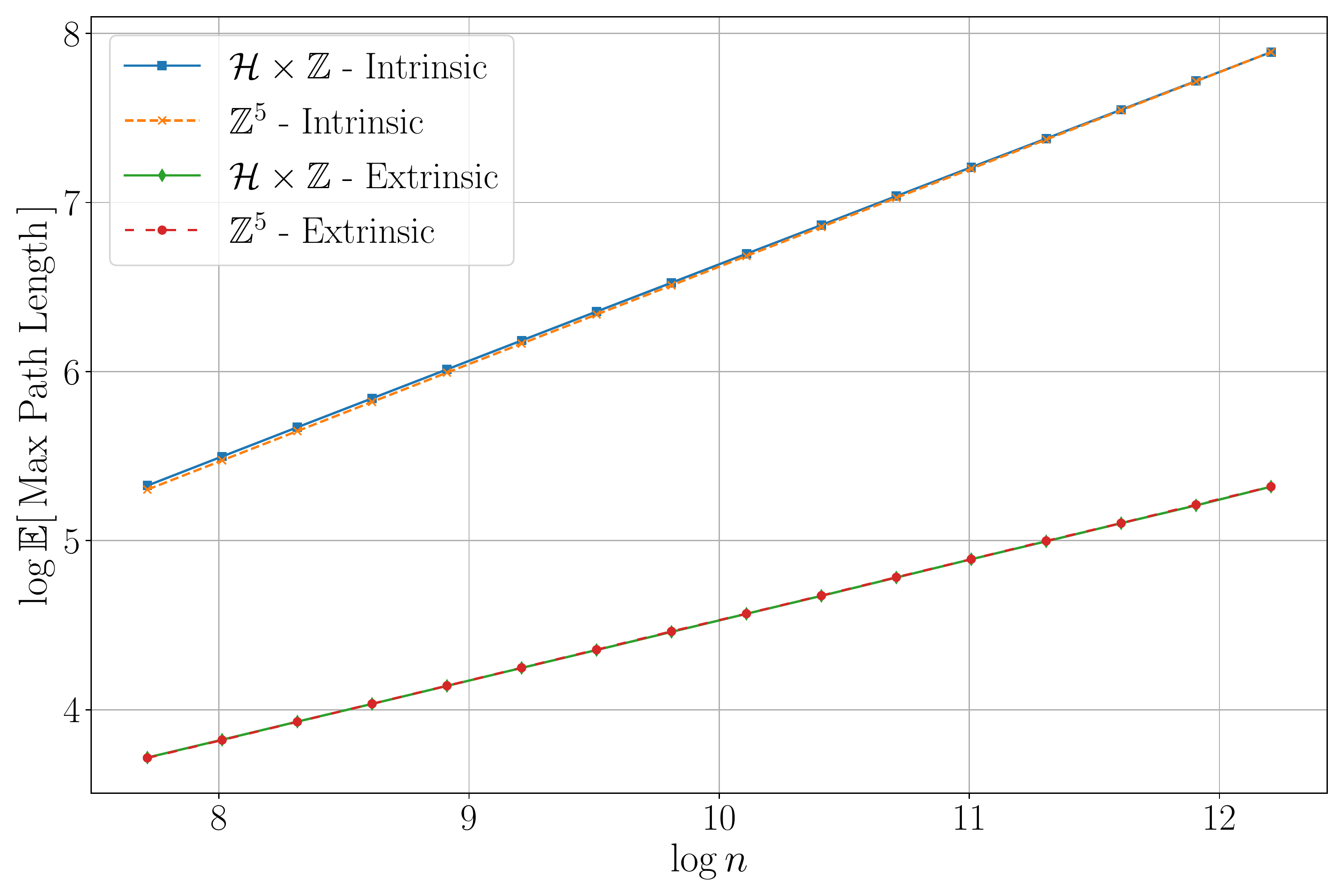}\vspace{0.5em}
	{(b) $\mathcal{H}\times \Z$ and $\Z^5$.}
	\end{minipage}
	\begin{minipage}[b]{0.975\linewidth}\centering\vspace{1em}
	\label{fig:4dLAE}
	\includegraphics[width=1\textwidth]{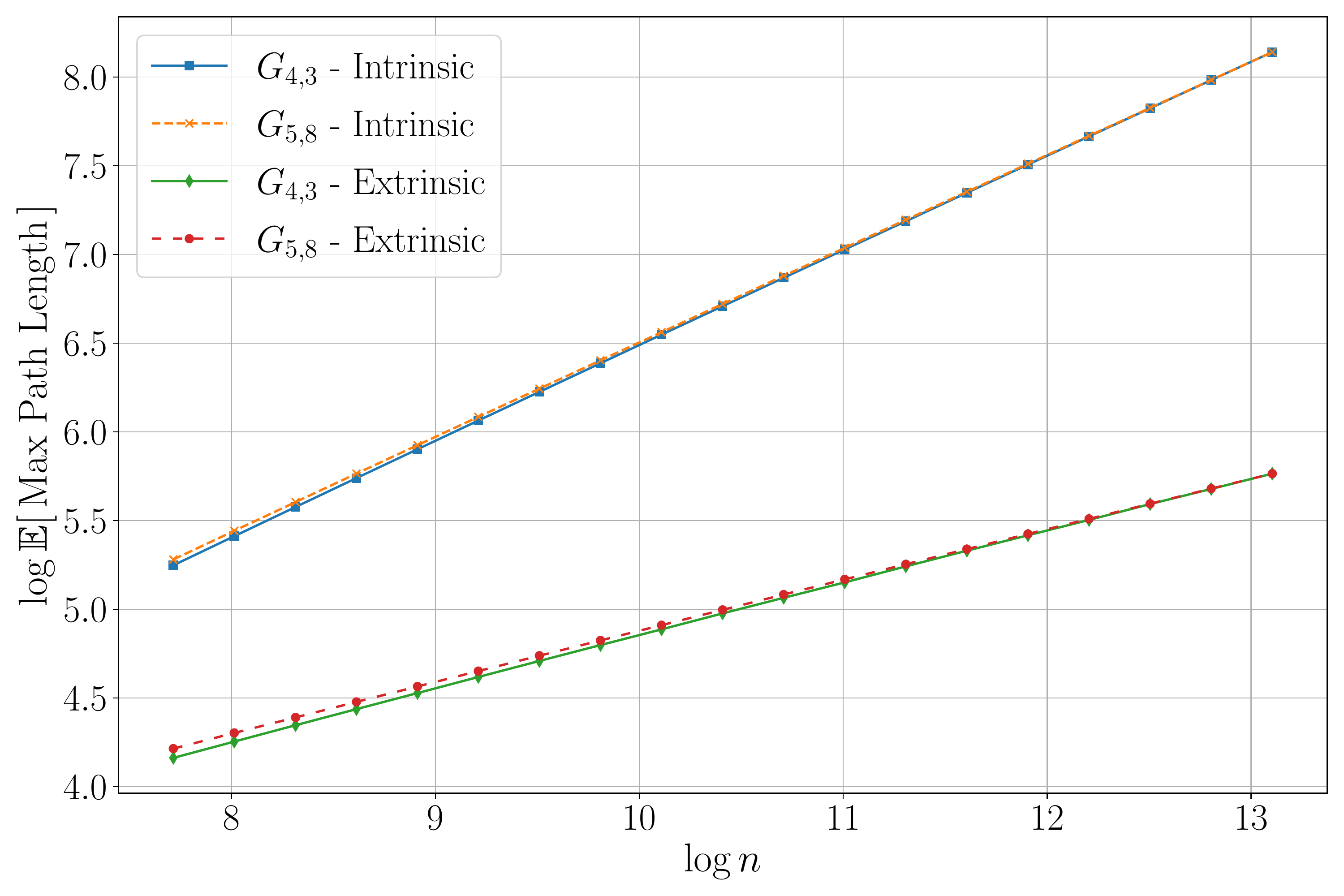}\vspace{0.5em}
	{(c) $G_{4,3}$ and $G_{5,8}$.}
	\end{minipage}
	\vspace{0.5em}
	\captionof{figure}{Log-log plots of mean intrinsic and extrinsic distances between the end-points of maximum (intinisic) length paths in the lattice tree as functions of tree size. In each case, the vertical positioning of the curves have been adjusted for ease of comparison of the gradients of the final segments.}
\end{Figure}

\begin{figure*}
	\centering
	\includegraphics[width=1\textwidth]{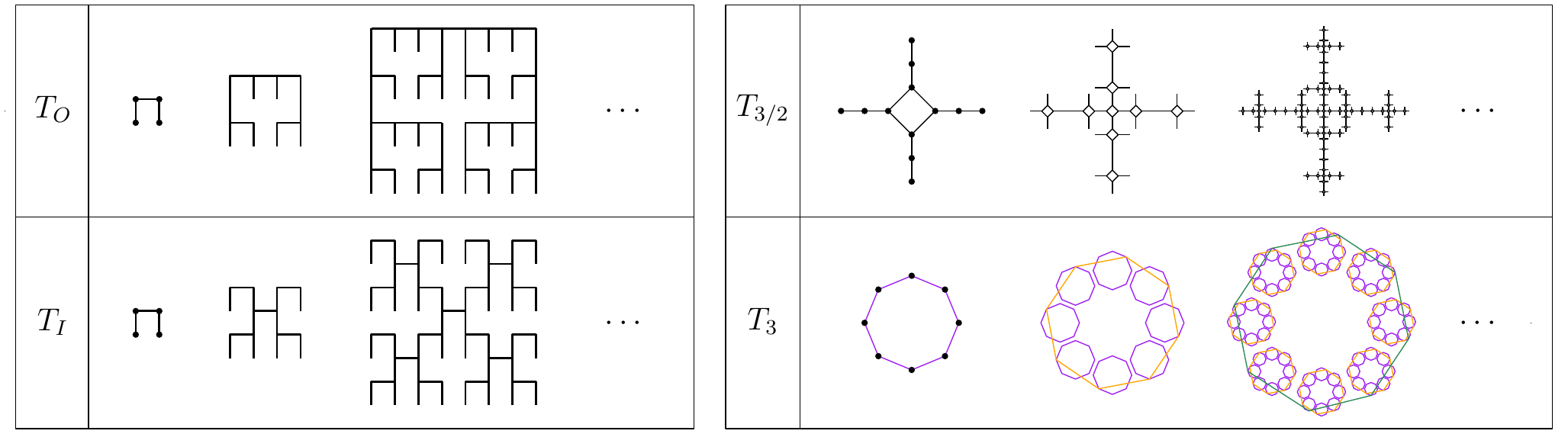}
	\captionof{figure}{Illustration of the recursive construction of graphical approximants to the fractal trees we consider. The graphs used to approximate $T_{3/2}$ and $T_3$ are not trees but \emph{quasi-trees}, with bounded-length cycles that disappear in the continuum limit. Note also that we have drawn the edges of the graphs approximating $T_3$ with different lengths in order to represent them cleanly in the plane; as a result, these drawings do not accurately represent the intrinsic geometry of the graphs in question. The colours are included to aid visualisation since the drawing is not planar. The tree $T_4$ is similar but is constructed from $16$-gons rather than octagons.}
	\label{fig:HIO}
\end{figure*}

 \noindent will represent the number of `marked points' that are used to specify specify how to construct the prefractal in one generation from the prefractal at the previous generation.
A \emph{seed graph} is defined to be a connected, undirected graph with vertex set $\Z_N$.
Given an undirected graph $G_g$ on $\mathbb{Z}_N\times\mathbb{Z}_N$, we define the \emph{contraction} $C[G_g]$ of  $G_g$ to be the graph with vertex set $\mathbb{Z}_N$, and with two vertices $i,j\in\mathbb{Z}_N$ connected if an only if there exist $a,b\in\mathbb{Z}_N$ such that $(i,a)\sim (j,b)$ in $G_g$. We say that an undirected graph $G_g$ on $\Z_N$ is a \emph{generator graph} if its contraction $C[G_g]$ is connected.

 Given a seed graph $G_s$ and a generator $G_g$, we define the fractal graph $\bbG=\bbG(G_g,G_s)$ to be the graph with vertex set $\mathbb{X}$ and where two distinct points $x=(\ldots,x_1,x_0)$ and $y=(\ldots,y_1,y_0)$ in $\bbX$ are connected by an edge if one of the following two conditions hold:
 \begin{itemize}
 \item $x_i=y_i$ for every $i\geq 1$ and $x_0 \sim y_0$ in $G_s$, or
 \item $m = \inf \{i \geq 1 : x_j = y_j $ for every $j \geq i\}$ is finite and strictly larger than one, $x_i=x_{m-1}$ and $y_i=y_{m-1}$ for every $0\leq i \leq m-1$, and  $(x_m,x_{m-1})$ is adjacent to $(y_m,y_{m-1})$ in $G_g$.
 \end{itemize} 
 Note that $x,y\in \bbX$ belong to the same connected component of $\bbG$ if and only if $x_i=y_i$ for all sufficiently large $i$, so that $\bbG$ has uncountably many connected components. For each $n$ the finite subgraphs of $\bbG$ induced by the sets $\Lambda_n(y)=\{x:x_i = y_i $ for every $i \geq n\}$ have isomorphism class that does not depend on the choice of $y\in \bbX$, and we define $G_n$ to be a graph with this isomorphism class. This ensures that $G_1$ is equal to the seed graph $G_s$, while for each $n\geq 1$ we can form $G_{n+1}$ by attaching edges between $N$ copies of $G_n$ according to the combinatorics of the generator $G_g$. The Benjamini-Schramm limit of the graph sequence $(G_n)_{n\geq 1}$ is equal to the rooted graph $(G_\infty,o)$ defined by taking $o\in \bbX$ to have i.i.d.\ uniform coordinates in $\Z/N\Z$ and taking $G_\infty$ to be the connected component of $o$ in the uncountably infinite graph $\bbG$.  

\medskip

 Algorithmically, this representation of the infinite-volume prefractal $(G_\infty,o)$ has the advantage that the initial sequence of coordinates $(o_1,\ldots,o_k)$ typically determines the isomorphism class of a large neighbourhood around $o$, and we can sample more terms of this sequence on an as-needed basis as we explore the percolation cluster of $o$ or run invasion percolation from $o$. 

\begin{table*}
	\centering
	\centering
\begin{tabular}{|l||l|l|l|}
\hline
	& $N$ & Seed edges $E[G_s]$ & Generator edges $E[G_g]$ \\
	\hline 
\makecell{\phantom{.}\\\phantom{.}}$T_3$ & $8$ & $\{\{i,i+1\} :0\leq i\leq 7\}$ & $\{\{(i,i+1),(j,j+1)\}:  0\leq i,j\leq 7, i-j=\pm1\}$ \\
\hline
\makecell{\phantom{.}\\\phantom{.}}$T_4$ & $16$ & $\{\{i,i+1\} :0\leq i\leq 15\}$ & $\{\{(i,i+1),(j,j+1)\}:  0\leq i,j\leq 15, i-j=\pm1\}$ \\
\hline
\makecell{\phantom{.}\\\phantom{.}} $T_{3/2}$ & $8$ & $\{\{2i,2i+2\} :0\leq i\leq 3\}$ & $\{\{(2i,2i+5),(2j,2j+5)\}: 0\leq i,j \leq 3, i-j=\pm1\}$ \\
\hline
\end{tabular}
\caption{Formal encodings of the fractal trees $T_3$, $T_4$, and $T_{3/2}$ used in our explicit recursive scheme for constructing fractals. All addition is computed modulo $N$.}
\label{table:seedsandgenerators}
\end{table*}

\medskip

The fractal trees $T_O$ and $T_I$ presented in \cref{fig:T_OI} are both easily represented via this recursive scheme with $N=4$: In both cases we take the seed graph to have edge set $\{\{0,1\},\{1,2\},\{2,3\}\}$. For the `outer' tree $T_O$ we take the generator graph to have edge set $\{\{(0,1),(1,0)\},\{(1,2),(2,1)\},\{(2,3),(3,2)\}\}$, while for the `inner' tree $T_I$ we take the generator graph to have edge set $\{\{(0,2),(1,3)\},\{(1,3),(2,0)\},\{(2,0),(3,1)\}\}$. We encourage the reader to work through this simple example to see how our fractal encoding scheme works in practice. The reader may also find it enlightening to consider how the infinite line graph $\Z$ can be expressed as a Benjamini-Schramm limit of graphs defined through a similar recursive scheme.

\medskip

Besides $T_O$ and $T_I$ we will also consider three further fractal trees which we call $T_3$, $T_4$, and $T_{3/2}$.  In fact, it will be convenient to consider graphical approximants of these trees that are not themselves tree but are quasi-trees in the sense that they include cycles of bounded length which disappear in the continuum limit. 
 The formal definitions of these fractal trees in terms of our recursive scheme are stated in \cref{table:seedsandgenerators} with graphical representations of the first three generations given in \cref{fig:HIO}.

\subsection{Fractal dimensions}

Let us now briefly review the definitions and background on the dimensions we consider, referring the reader to \cite{MR3236784} for further background.

We begin with the Hausdorff dimension and topological dimension, which are both classical.
Given a non-empty metric space $X$, the $d$-dimensional Hasudorff outer measure of a set $S\subset X$ is defined as 
\begin{multline*}
	\mathcal{H}^d (S) = \liminf_{r\rightarrow 0}\big\{\sum_Ir_i^d:\text{ there is a cover of } S\\ \text{ by balls of radii }0<r_i<r\big\}.
\end{multline*}
The \emph{Hausdorff dimension} of $X$ is then 
\[
\dim_H X=\inf\{d\geq 0:\mathcal{H}^d(X) <\infty\}.
\]
In non-pathological examples one typically has that a continuum fractal has Hausdorf dimension $d$ if and only if the Benjamini-Schramm limit of its prefractal approximants has volume-growth dimension $d$ in the sense that $|B(o,r)|\approx r^d$ as $r\to\infty$.
Moreover, in non-pathological examples one also has that the Hausdorff dimension is additive in the sense that $\dim_H X \times Y = \dim_H X + \dim_H Y$; see \cite[Chapter 7]{MR3236784} for precise theorems to this effect. All the examples we consider will have the very strong property of being \emph{Ahlfors regular}, which ensures that the Hausdorff dimension is indeed additive for these examples.

\medskip

Suppose that we construct a fractal via a recursive rule as discussed at the beginning of this section, and let $(G_n)_{n\geq 1}$ be the associated sequence of prefractal graphical approximants so that $G_n$ has $N^n$ vertices for some $N \geq 2$. If the ratio of diameters of $G_{n+1}$ and $G_n$ tends to $N^\alpha$ as $n\to\infty$, then the associated continuum fractal will typically have Hausdorff dimension $\log N / \log N^\alpha = 1/\alpha$. Again, all the examples we consider are sufficiently well-behaved that these heuristics can easily be turned into rigorous proofs with a little work. See \cite[Section 9.2]{MR3236784} for detailed justifications of various related formulae. It follows from these considerations that both fractal trees $T_O$ and $T_I$ have Hausdorff dimension $2$: At each succesive scale of approximation the number of vertices is multiplied by four while the diameter roughly doubles. Similarly, in $T_3$ and $T_4$ the diameter roughly doubles in each generation while the volume increases by a factor of $8$ or $16$ as appropriate, so that these trees have Hausdorff dimensions $\log 8/\log 2=3$ and $\log 16/\log 2 =4$ respectively. Finally, in $T_{3/2}$ the diameter roughly quadruples at each scale while the volume increases by a factor of $8$, so that $T_{3/2}$ has Hausdorff dimension $\log 8/\log 4 = 3/2$.

\medskip

The \emph{toplogical dimension} (a.k.a.\ lower inductive dimension) $\dim_t X$ of a separable metric space $X$ is defined inductively by $\dim_t \varnothing = -1$ and
\begin{multline*}
	\dim_t X=\inf\{d:X \text{ has a basis } U \text{ such that  }\\\dim_t\partial U\leq d-1\text{ for every }U\in U\}.
\end{multline*}
Note that \emph{real trees} such as $\R$, $[0,1]$, and the fractal trees we consider always have topological dimension $1$. The topological dimension is not additive in general but always satisfies the inequality $\dim_t X \times Y \leq \dim_t X + \dim_t Y$ \cite[Theorem 1.5.16]{MR0482697}. Moreover, if $Y$ is a subspace of $X$ then $\dim_t Y \leq \dim_t X$, a fact referred to as the \emph{subspace theorem} \cite[Theorem 1.1.2]{MR0482697}. If $X=T_1 \times T_2 \times \cdots \times T_k$ is a product of real trees then it follows that $\dim_t X \leq k$, and since $X$ contains a copy of the space $[0,\varepsilon]^k$ for some $\varepsilon>0$ it follows from the subspace theorem that $\dim_t X = k$. This equality determines the topological dimension for all the examples we consider.

\medskip

The \emph{toplogical Hausdorff dimension} is a much more recent notion of dimension that was introduced by Balka, Buczolich and Elekes \cite{BALKA2015881}. The  topological Hausdorff  dimension $\dim_{tH} X$ of a non-empty metric space $X$ is defined to be
\begin{multline*}
	\dim_{tH}X = \inf\{d:X\text{ has a basis }\mathcal{U}\text{ such that }\\\dim_H \partial U\leq d-1\text{ for every }U\in \mathcal{H} \},
\end{multline*}
where $\dim_H \varnothing$ is defined to be $-1$. It is proven in \cite[Theorem 4.21]{BALKA2015881} that
\begin{equation} \label{eq:tHA}
\dim_{tH}(X\times[0,1]) = 1+ \dim_H(X),
\end{equation}
for every non-empty and separable metric space $X$. This allows us to always reduce the computation of the topological Hausdorff dimension to that of the Hausdorff dimension by working only with products with $[0,1]$; this corresponds to taking products with $\Z$ for the relevant Benjamini-Schramm limits.

\medskip

It remains to introduce the \emph{spectral dimension}, which is most easily defined for the infinite Benjamini-Schramm limit $(G_\infty,o)$ associated to the fractal. Indeed, an infinite connected graph $G$ is said to have spectral dimension $d_s=\dim_s G$ if the simple random walk return probabilities $p_n(v,v)$ satisfy
\[
p_{2n}(v,v) = n^{-d_2/2 + o(1)}
\] 
as $n\to \infty$ for each vertex $v$ of $G$. (In principle one can define the spectral dimension of a continuum fractal directly by first defining Brownian motion on that fractal, but this is a very delicate matter in general.)
It is easily seen from the definition that the spectral dimension is additive with respect to products in the sense that if $G$ and $H$ are two infinite, connected graphs then $\dim_s G \times H = \dim_s G + \dim_s H$.
Most fractal \emph{trees} have spectral and Hausdorff dimensions related by the formula 
\[
\dim_s = \frac{2\dim_H}{\dim_H+1},
\] 
and it is not difficult to justify that this equality does indeed hold for all the fractal trees we consider. (Indeed, this equality should hold whenever the \emph{effective resistance} between a vertex and the boundary of the ball of radius $r$ grows like $r^{1-o(1)}$, and for trees this holds whenever sunsequential limits do not have vertices of infinite degree; this can be deduced from the same methods used in \cite[Section 8]{MR4055195}.) Thus, $T_O$, $T_I$, $T_{3/2}$, $T_3$, and $T_4$ have spectral dimensions $4/3$, $4/3$, $6/5$, $3/2$, and $8/5$ respectively.

\subsection{Equidimensional fractal products} 
\label{sec:TA}

 We now define the two pairs of equidimensional fractal products on which we will study percolation.
The first pair is given by \[H_1=T_O\times[0,1] \quad \text{ and } \quad H_2=T_I\times[0,1].\]
It follows from the above discussion that these two fractals both have Hausdorff dimension $3$, topological dimension $2$, spectral dimension $7/3$ and topological Hausdorff dimension $3$. 
The second pair is given by \[H_3=T_{3/2}\times T_{3/2}\times T_4\times [0,1] \text{ and } H_4=T_{3}\times T_{3}\times[0,1]\times[0,1].\]
These two fractals both have Hausdorff dimension
\[
\frac{3}{2}+\frac{3}{2}+4+1 = 3+3+1+1= 8,
\] topological dimension $4$, spectral dimension 
\[
\frac{6}{5}+\frac{6}{5}+\frac{8}{5}+1=\frac{3}{2}+\frac{3}{2}+1+1=5,
\]
 and topological Hausdorff dimension $8$. This pair of examples is interesting to study in part because the two fractals $H_3$ and $H_4$ seem to have `the same dimensions for different reasons', with different components of their defining products making up different proportions of their shared Hausdorff and spectral dimensions. This would seem to make them a prime candidate for a failure of universality, although in the end the large finite-size effects made it difficult for us to compare the exponents in the two cases.

\medskip

Again, we do not work directly with continuum fractals, but instead consider the Benjamini-Schramm limits defined via the recursive schemes specifying the trees $T_O$, $T_I$, $T_3$, $T_4$, and $T_{3/2}$ above. Thus, for example, when we simulate percolation on $H_3$ we are really simulating percolation on the product of two independent copies of the Benjamini-Schramm limit associated to $T_{3/2}$, a further independent copy of the Benjamini-Schramm limit associated to $T_4$, and one copy of $\Z$.
 Since we \linebreak 

\begin{Figure}
	\centering
	\includegraphics[width=0.975\textwidth]{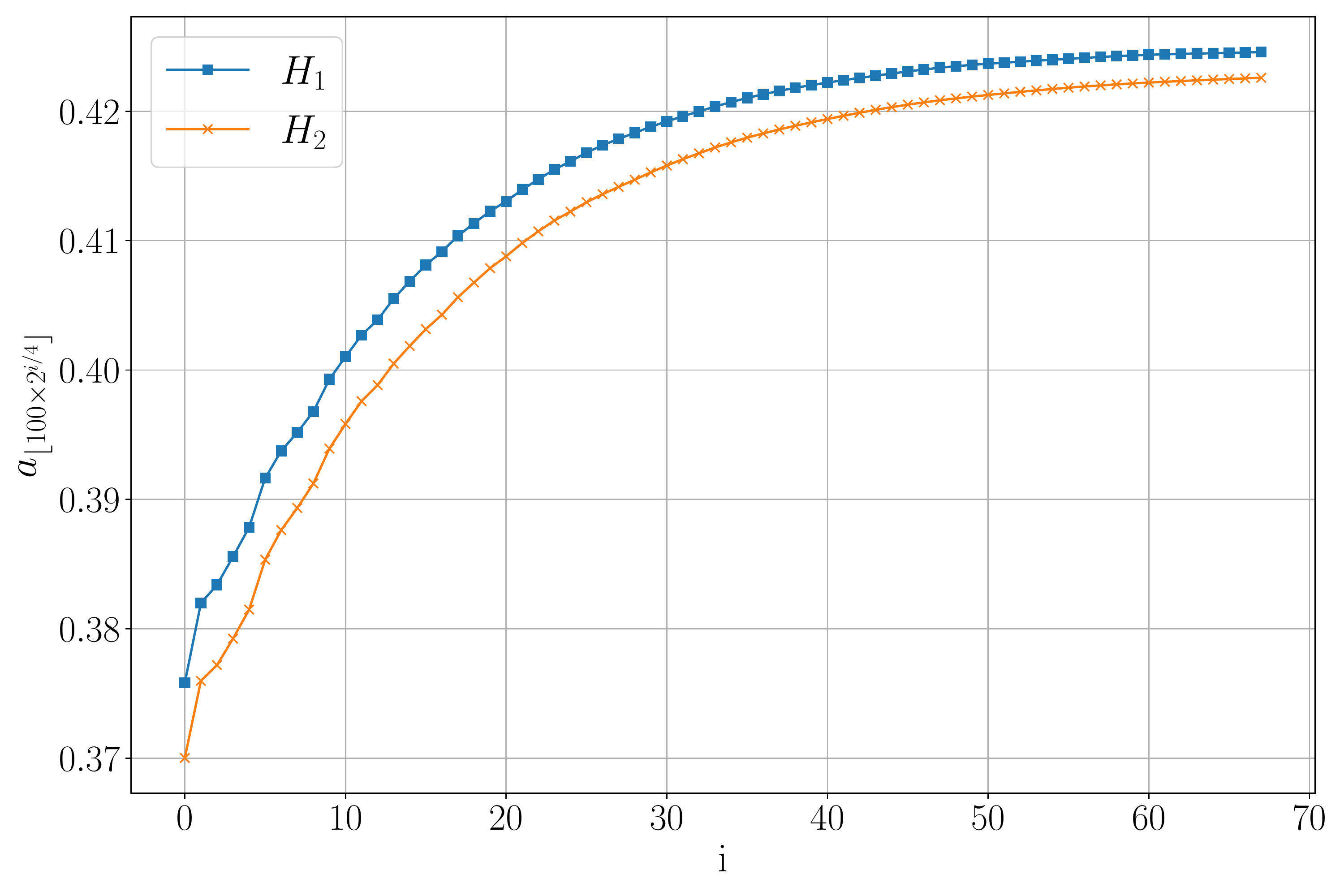}
	\captionof{figure}{The average bulk-to-boundary ratios obtained by invasion percolation for H1 and H2.}
	\label{fig:H1H2inv}
\end{Figure}

 \noindent   will always use these same graphical approximations, for clarity of exposition we will abuse the terminology by speaking simply of 'percolation on $H_1$' and so on.

\subsection{Results} \label{section:results}

We now discuss the results of our simulations of percolation on the self-similar fractals beginning with the equidimensional pair $H_1$ and $H_2$.
As with the transitive lattices, we began by running invasion percolation with approximately $10^6$ samples, each time recording the bulk-to-boundary ratios $a_n$ at $n=\lfloor 100\times 2^{i/4}\rfloor,\ 0\leq i\leq 67 $. The outcome of these simulations is recorded in \cref{fig:H1H2inv}. 


In order to estimate $p_c$ from this data we carried out a similar analysis to the transitive case, varying the amount we cut-off at the beginning before curve-fitting. This gave the initial estimates $p_c=0.4249$ for $H_1$ and $p_c=0.4232$ for $H_2$.


We emphasise that the oscillations in figure $\ref{fig:H1H2inv}$ are in fact a feature rather than noise. This was a consistent appearance throughout our simulations for fractals, both for invasion percolation and the Leath algorithm. For the former the oscillations decayed, while for the latter they grew. For the former it meant more initial data had to be discarded, and for the latter it made estimating linearity more difficult. This became 
 more of a problem for some of the higher dimensional fractals, in particular in relation to runs of the Leath algorithm. 


Having obtained an initial estimate for $p_c$, we then sampled the percolation cluster for $H_1$ and $H_2$ at differing values of $p$, and produced log-log plots of the average volume tail $P_{\geq s}=\mathbf{E}\left[\mathbb{P}_p(|K_o|\geq s)\right]$ against $s$ at different values of $p$ as in $\textsf{Q1b}$, taking $s=\lfloor{2^{i/4}}\rfloor$ with maximal $i$ ranging up to $27$ and with between $10^6$ and $10^8$ samples for each value of $p$. The outcomes of this investigation are recorded in \cref{fig:H1H2loglog} below.
Drawing tangents along the curves of \cref{fig:H1H2loglog} reveals that $p_c=0.42545\pm0.0005$ for $H_1$ and $p_c=0.423225\pm0.00025$ for $H_2$. It is interesting to note that invasion percolation gave a far more precise reading for $H_2$ than $H_1$.  


\begin{Figure}
	
	\includegraphics[width=0.975\textwidth]{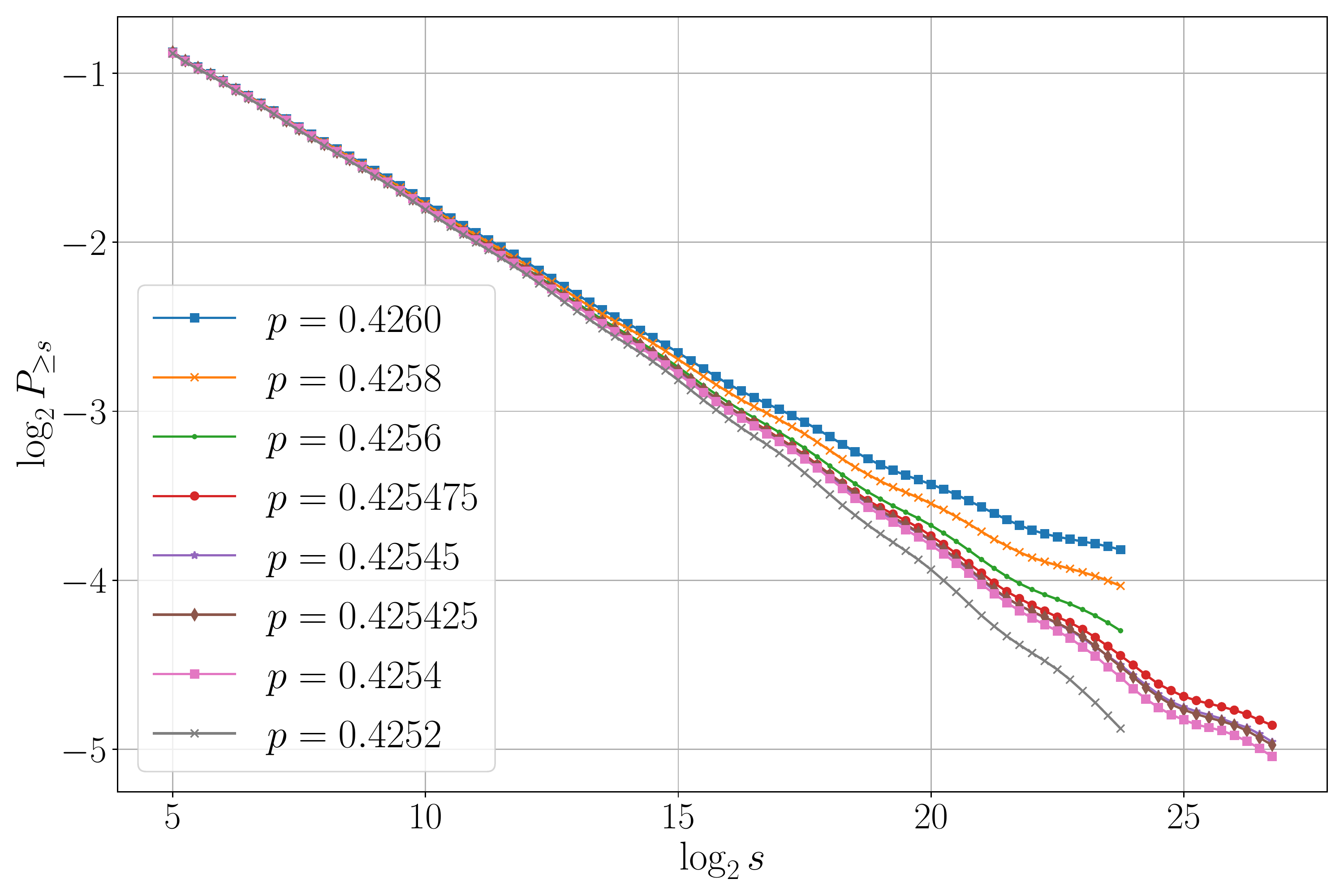}\\

	\includegraphics[width=0.975\textwidth]{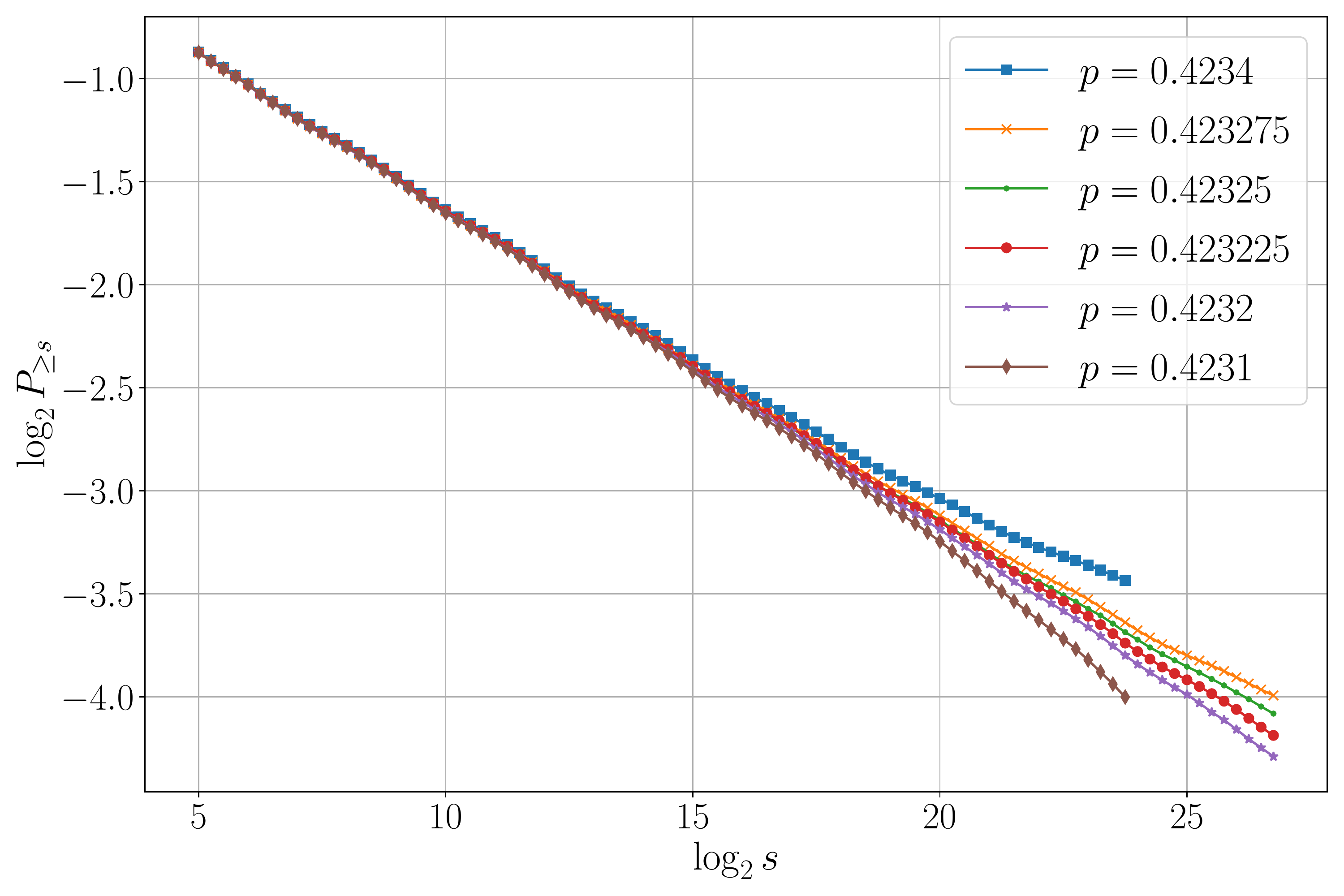}
	\captionof{figure}{Log-log plots of the volume tail distribution for $H_1$ (top) and $H_2$ (bottom) at different values of the percolation probability $p$. Smaller deviations from linearity indicate that $p$ is closer to the critical probability $p_c$.}
	\label{fig:H1H2loglog}
	\vspace{0.5cm}
		\includegraphics[width=1\textwidth]{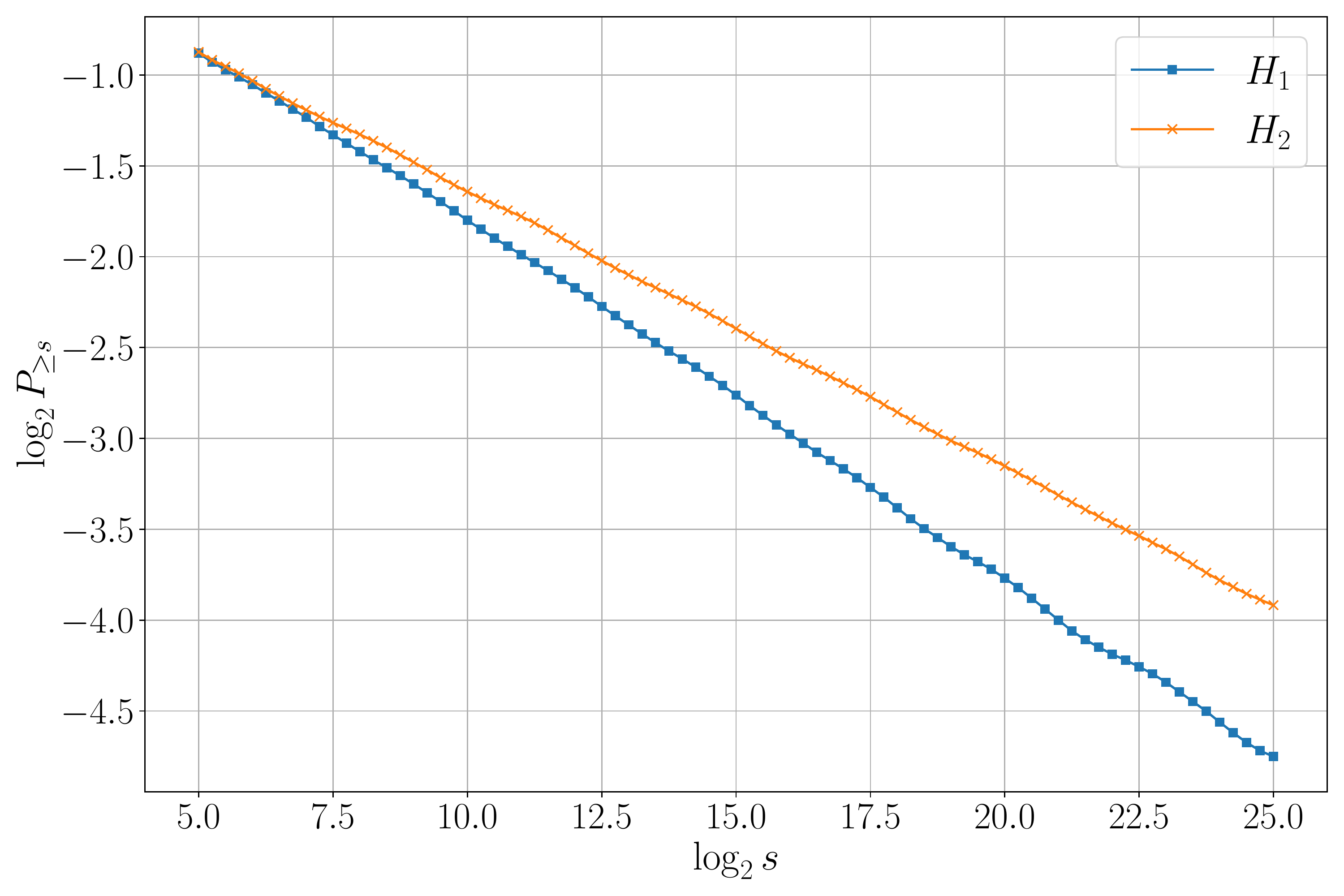}
	\captionof{figure}{Log-log plots of the volume tail distribution for $H_1$ and $H_2$ at the estimated critical probabilities of $p=0.42545$ for $H_1$ and $p=0.423225$ for $H_2$. The two lines clearly have distinct slopes, leading strong evidence to the claim that $H_1$ and $H_2$ have distinct values of the critical exponent $\tau$.}
	\label{fig:H1H2taucomp}
\end{Figure}

\begin{Figure}
	\centering
	\includegraphics[width=0.975\textwidth]{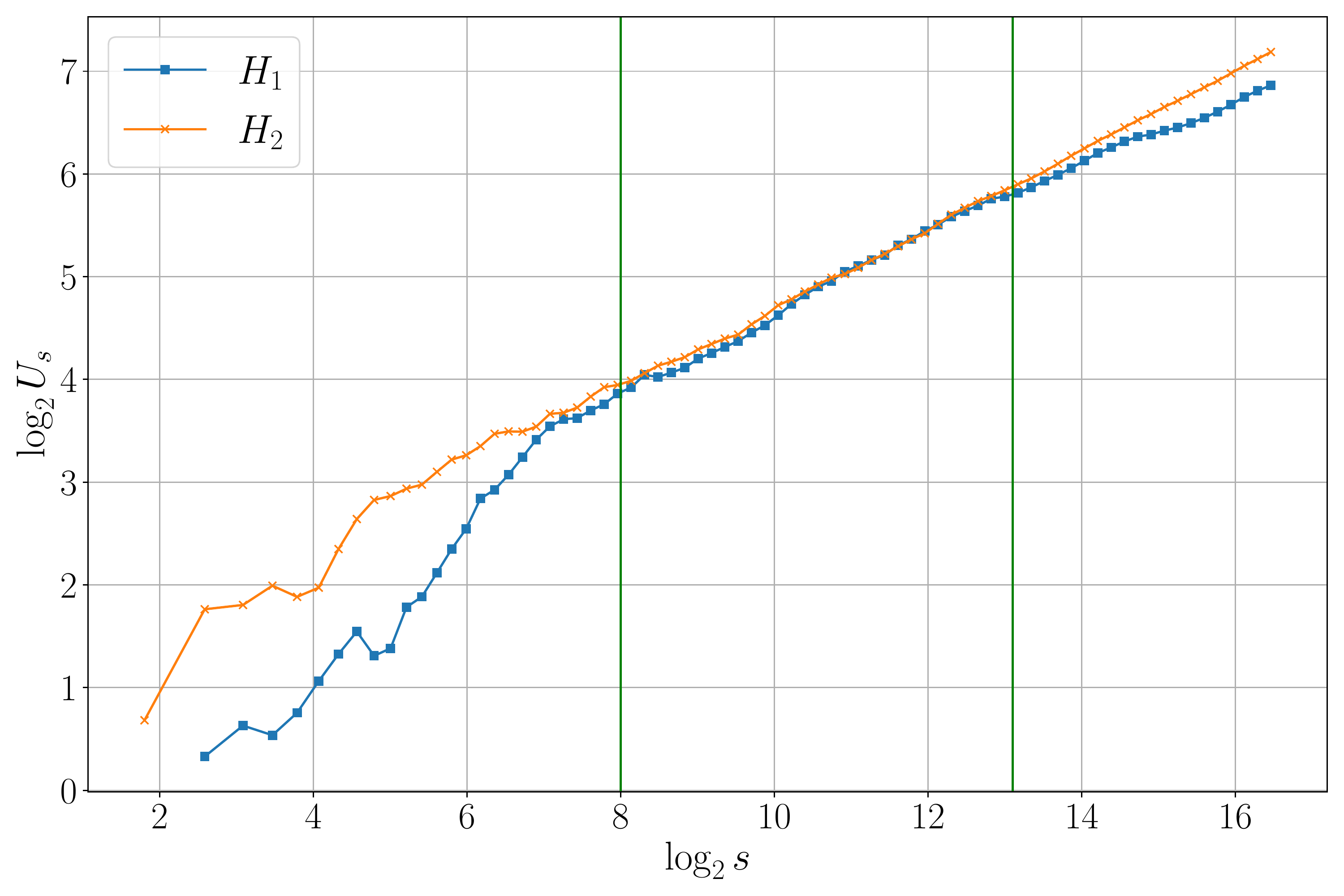}
	\captionof{figure}{Plot of $\log_2 U_s$ against $\log_2 s$ with $U_s$ calculated by averaging $U_{s,p}$ at  $p=0.4242,\,0.4246,0.4248$ for $H_1$ and  $p=0.4231,\, 0.4234$ for $H_2$.}
		\label{fig:H1H2sigma}
\end{Figure}

\noindent

\medskip

Finding the gradient of the these critical log-log plots gave $\tau=2.195\pm 0.005$ for $H_1$ and a more precise reading of $2.151\pm0.001$ for $H_2$, where uncertainties were estimated by varying the portions of the midsections of the curves over which the gradients were calculated and calculating over multiple curves corresponding to probabilities within the aforementioned ranges for $p_c$.  The more prominent presence of the oscillations for $H_1$ made the reading for $\tau$ less precise. A direct visual comparison of the two critical log-log volume tail plots is provided in \cref{fig:H1H2taucomp}.

\medskip

We now turn to estimating the exponent $\sigma$, for which our results are less clear.
Unfortunately, we found the method based on the ansatz \eqref{eq:sigma3} to work very poorly for these graphs, with the resulting expectation requiring a prohibitively large number of samples to stabilize.
As such, we resorted to a more ad-hoc analysis to estimate $\sigma$.
 First, we rearranged the ansatz formula  \eqref{eq:offcrit1} to obtain that
\begin{align} \label{eq:ups}
u_{s,p}:=\frac{\log P_{\geq s,p}-\log P_{\geq s,p_c}}{p-p_c}= C_1 s^{\sigma}+\ldots.
\end{align}
Then, for each fractal, we took the average $U_s$ of $u_{s,p}$ over a selection of near-critical $p$ and used curve-fitting over $s$ to output a value of $\sigma$. The outcome of this investigation is recorded in \cref{fig:H1H2sigma}.
As can be seen from this figure, the results of this investigation are inconclusive at best, with large non-linearities in the curve for $H_1$ preventing us from getting a reliable estimate of $\sigma$ in this case.
There seems to be a section of alignment, but not enough to confirm or disconfirm that the $\sigma$ critical exponents are the same. The deviation at the end may indicate a different $\sigma$ exponent, or it could be due to the the imprecision of our estimate of $p_c$, or it could indicate that $p-p_c$ is large enough that the approximations in the derivation of \eqref{eq:ups} are not valid.

\medskip

\enlargethispage{1\baselineskip}

\textbf{$H_3$ and $H_4$.} We now turn to our results for the equidimensional fractal products $H_3$ and $H_4$.
Running invasion percolation and extrapolating as above (see \cref{fig:H3H4inv}) gave the estimates $p_c=0.11705$ for $H_3$ and $p_c=0.11326$ for $H_4$. Having obtained this estimate we then sampled the percolation \linebreak

\begin{Figure}
	
	\label{fig:H3H4inv}
	\includegraphics[width=0.975\textwidth]{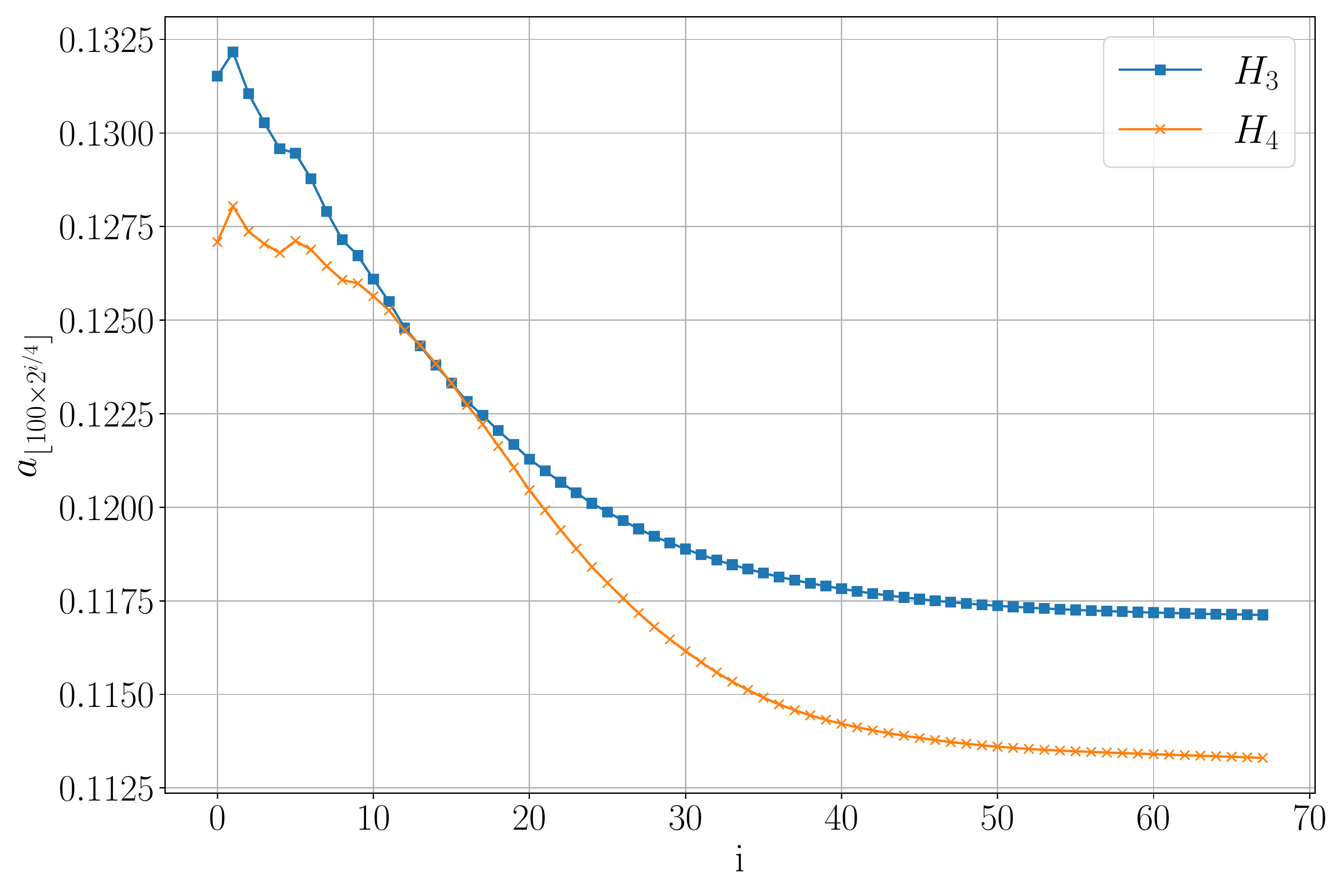}
	\captionof{figure}{The average bulk-to-boundary ratios obtained by invasion percolation for $H_3$ and $H_4$. Note the very pronounced finite-size effects for $H_4$.}
\vspace{0.75cm}
	\includegraphics[width=0.975\textwidth]{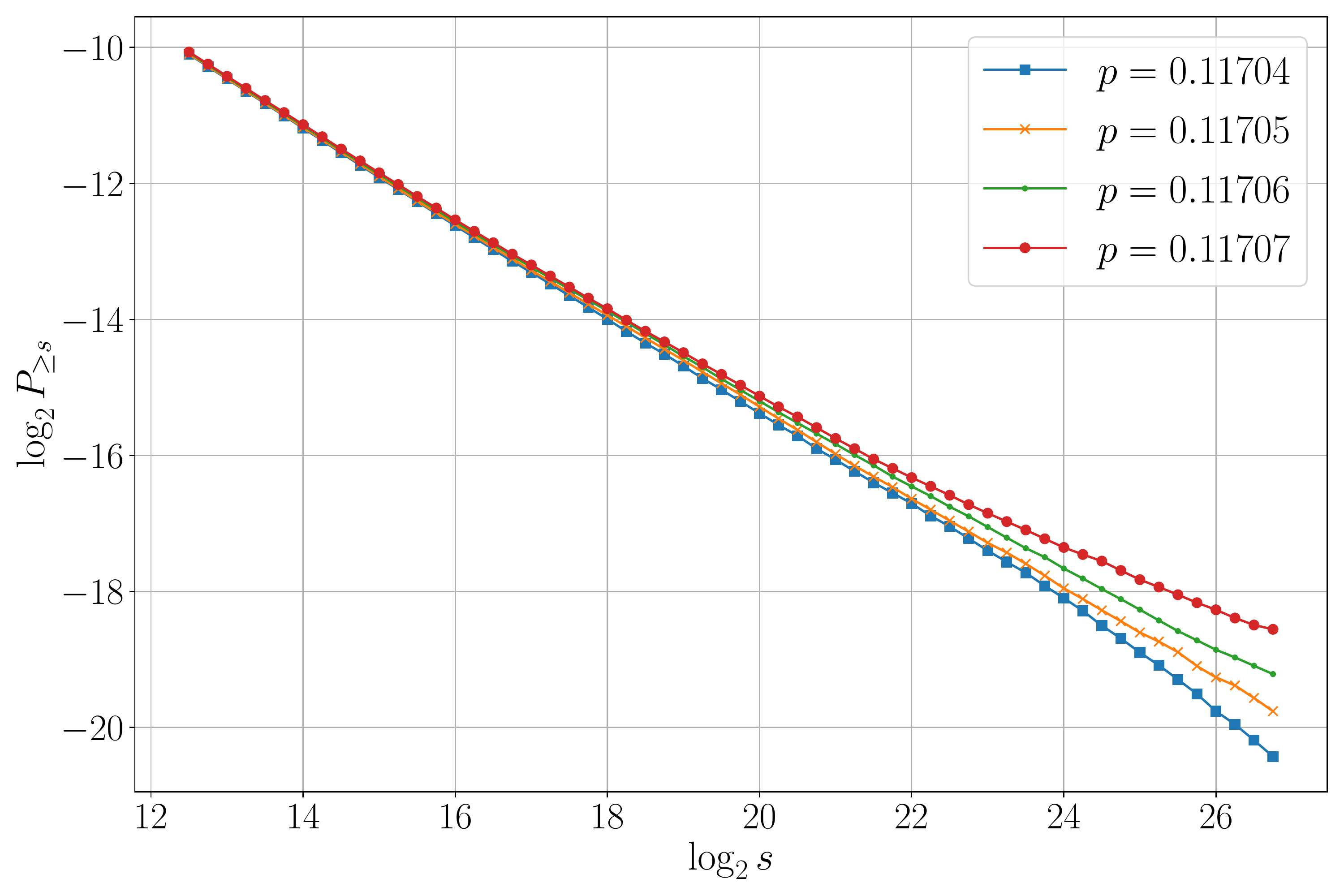}\\
	\includegraphics[width=0.975\textwidth]{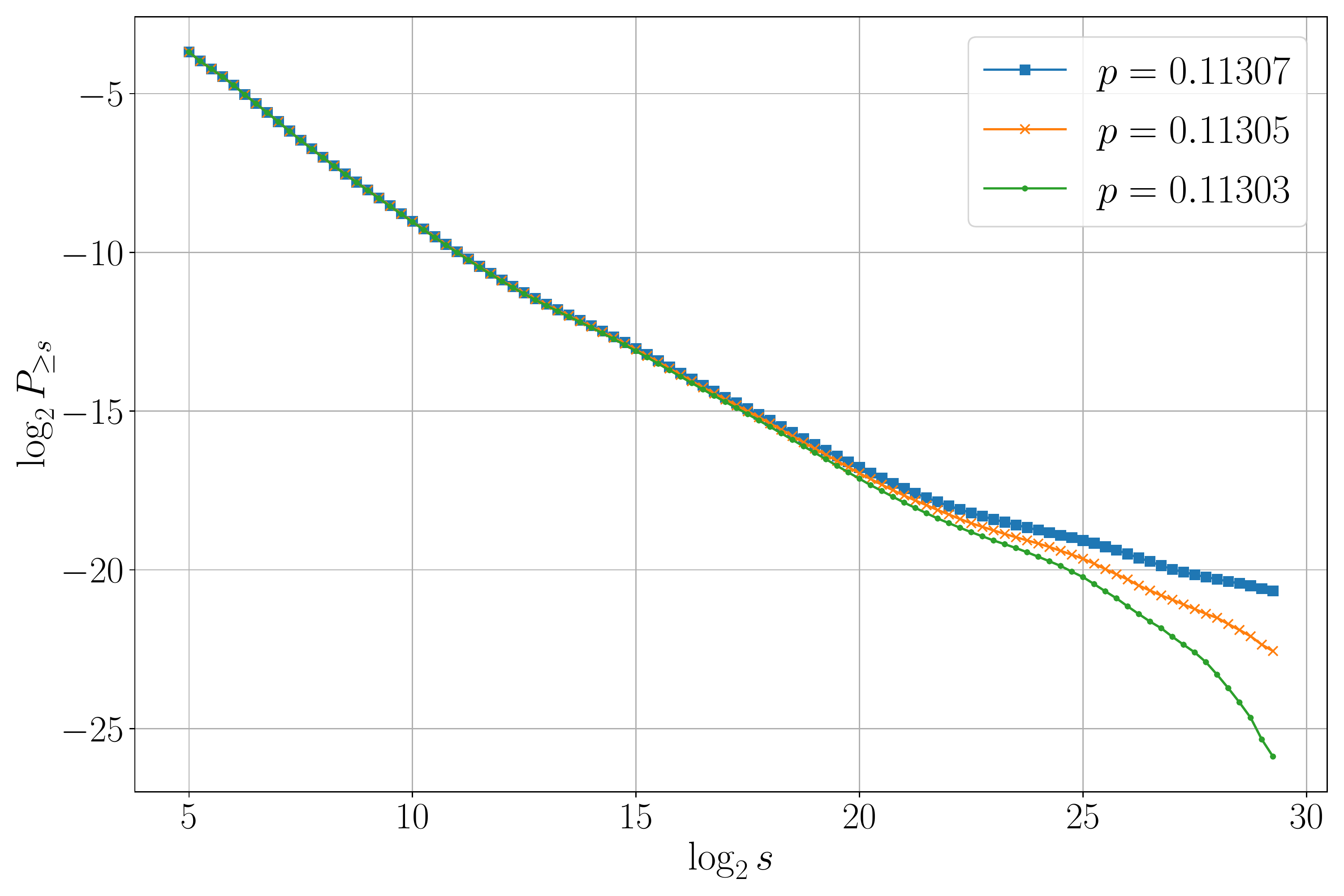}
	\captionof{figure}{Log-log plots of the volume tail distribution for $H_3$ (top) and $H_4$ (bottom) with different values of the percolation probability $p$. Smaller deviations from linearity indicate that $p$ is closer to $p_c$. The large deviations in linearity present in all the curves plotted for $H_4$
makes it difficult to compare the two curves or to reliably estimate the relevant value of $\tau$ in this case.
	}
	\label{fig:H3H4loglog}
\end{Figure}

\begin{Figure}
	\includegraphics[width=0.975\textwidth]{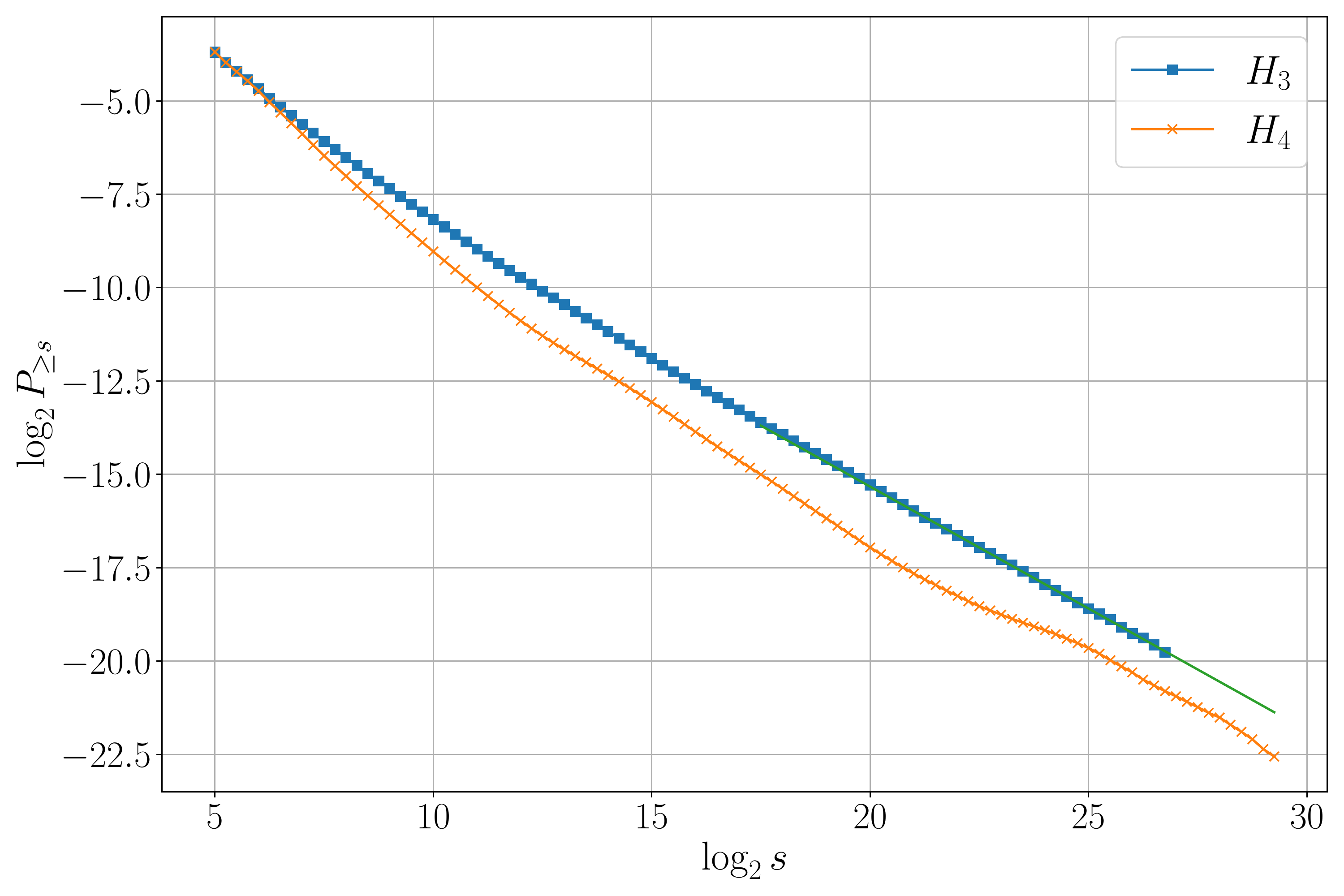}
	\captionof{figure}{Log-log plots of the volume tail distribution for $H_3$ and $H_4$ with $p=0.11705$ for $H_3$ and $p=0.11305$ for $H_4$. The green line is an extrapolation of the linear section of the graph for $H_3$ in order to ease comparisons with the curve for $H_4$. }
	\label{fig:H3H4taucomp}
\vspace{0.3cm}
	\includegraphics[width=0.975\textwidth]{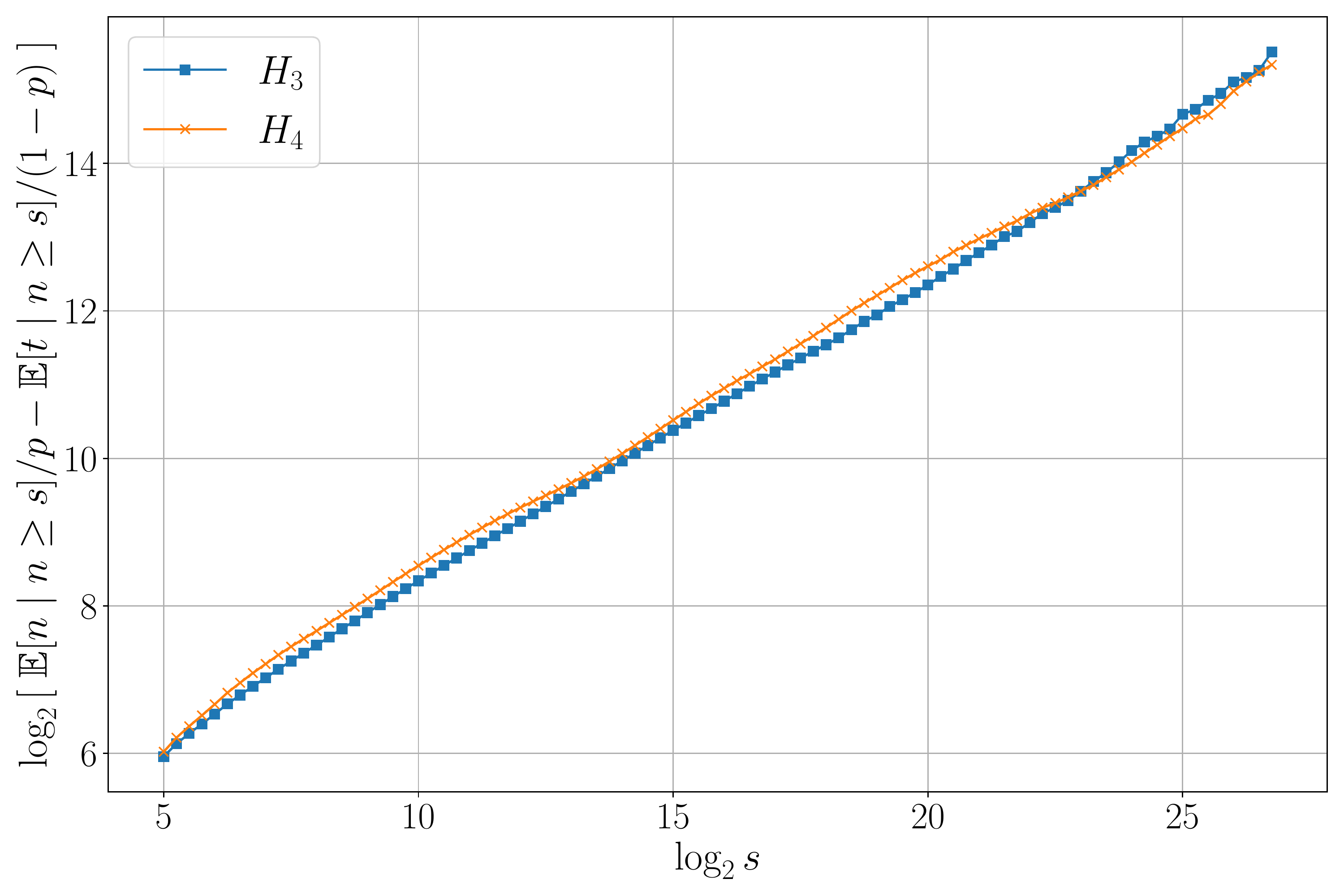}
	\captionof{figure}{Graphs of \textsf{Q2b} with $p=0.11705$ for $H_3$ and $p=0.11305$ for $H_4$. }
		\label{fig:H3H4sigmacomp}
\end{Figure}

\noindent  cluster at a varierty of nearby values of $p$ and produced log-log  plots 
of both the volume tail distribution and the quantity $\E[n/p-t/(1-p) | n \geq s]$ as in $\textsf{Q2b}$, where we used between $5\times 10^8$ and $3\times 10^9$ samples to estimate each of the relevant quantities. The outcomes of these investigations are recorded in \cref{fig:H3H4loglog,fig:H3H4taucomp,fig:H3H4sigmacomp}.

\medskip

Plotting tangents to the final segments of each of the curves in Figure \ref{fig:H3H4loglog} and finding the closest linear fit with the midsection of the curve gave $p_c=0.11705\pm0.00001$ for $H_3$ and $p_c=0.11305\pm0.00002$ for $H_4$. Invasion percolation therefore gave an extremely accurate estimate for $H_3$ but a much less accurate estimate for $H_4$. This was due to the prominent oscillatory behaviour in the bulk-to-boundary ratios, most likely due to the presence of the $T_4$ tree in $H_4$. 

\medskip

By considering tangents, we estimated $\tau=2.66\pm 0.01$ for $H_3$ and $\sigma=0.41\pm 0.01$ for both $H_3$ and $H_4$, suggesting that these two fractals share the same value of the exponent $\sigma$.
Unfortunately, the large deviations from linearity in \cref{fig:H3H4taucomp}

\noindent prevented us from obtaining an estimate on $\tau$ for $H_4$ to any reasonable level of accuracy, and it is unclear whether one should expect $H_3$ and $H_4$ to share a common value of this exponent.

\section{Discussion and Open Questions} \label{section:OQD}
\textbf{Summary.}
In this paper we presented strong numerical evidence in support of our conjecture that the critical exponents governing critical percolation and lattice trees on transitive lattices of polynomial volume growth depend only on the dimension and not on any other  features of the large-scale geometry.
For self-similar fractals, we showed that the situation is more complicated: we presented examples of two fractals having the same Hausdorff, spectral, topological, and topological Hausdorff dimensions, but which have distinct numerical values of the percolation Fisher exponent $\tau$. On the other hand, we do not rule out that the exponent $\sigma$ is determined by these dimensions. This may be related to the phenomenon of \emph{weak-universality} as discussed in \cite{MONCEAU20041} and deserves closer investigation in future work.

\medskip
\noindent  \textbf{Open Questions.}
We now present a collection of open problems and directions for future research:

\begin{enumerate}
	
	\item Provide theoretical reasoning either in support of or against Conjecture \ref{conj:1}. Is there a reason these exponents might be extremely close without being exactly the same?
	\item Does the conjecture hold for other models with an upper-critical dimension $d_c>4$ such as lattice animals, minimal spanning trees, and invasion percolation?
	\item Are the exponents describing  logarithmic corrections at the upper-critical dimension $d_c$ independent of the choice of $d_c$-dimensional transitive graph? This question is also interesting for models with $d_c=4$ such as the Ising model,
	  self-avoiding walk, and the uniform spanning tree.
	\item Further investigate the extent to which the exponent $\sigma$ is constrained by the dimensions we consider. Do $H_1$ and $H_2$ have the same value of $\sigma$? Is there a reason why $\sigma$ would be less sensitive to the geometry than $\tau$? 
	Is this related to the phenomenon of weak universality?
	
\end{enumerate}

Of course there are many other directions that one might pursue in relation to our work.
In addition to the endless variety of fractals, there are also many other transitive graphs of polynomial growth for which the problems studied in this paper are interesting \cite{MR2303198,isenrich2020coneequivalent}. There are also many other exponents associated to the models that one could seek to estimate. In particular the exponents characterising the intrinsic radii of critical percolation clusters, and the exponent characterising the sub-exponential correction to growth of the number of lattice trees of size $n$ \cite{PhysRevE.67.036116,MR2549378,MR2031859}.

Finally, let us remark that our focus in this paper has been to investigate a large number of different examples rather than devoting too much computing time to a very in-depth analysis of any particular example. It may be worthwhile in the future to subject our conclusions to further scrutiny by selecting one or two of the quantities we investigated to be the subject of a more intensive study.

\subsection*{Acknowledgements}
The work of TH was supported by ERC starting grant 804166 (SPRS). 
NH  was  supported  by  the  doctoral  training  centre,  Cambridge  Mathematics  of Information (CMI).
We thank Romain Tessera for very helpful correspondence on the quasi-isometric classification  of nilpotent groups.

\footnotesize{
	\bibliography{bibliography}
\bibliographystyle{abbrv}
}

\normalsize{}
\appendix

\section{Improvements to the invasion percolation algorithm} \label{section:IPA}
In this appendix we outline an improvement to the invasion percolation algorithm which ended up giving us a speed-up of up to $30\%$.
We first recall the definition of invasion percolation. We continue to use the same definitions introduced in \cref{sec:percolation_background}, including the edge labels $(U_e)_{e\in E}$, the edge $e_n$ invaded through at time $n$, and the frontier $F_n$ at time $n$.

To find $e_n$ efficiently at each step of the simulation we maintain a heap $H$ of all the values in the frontier $F_n$. Maintaining a heap of size $m$ has computational complexity $O(\log m)$ at each step. Unfortunately, at least for $d$-dimensional lattices, we can expect the cardinality of the frontier to grow linearly in $n$ \cite{cmp/1104114182}, and so at step $n$ we incur computational cost $O(\log n)$. However, many of the edges in the frontier, which have weights significantly larger than $p_c$, will never be extracted, and therefore sit uselessly in the heap incurring an additional computational cost without benefit. The method we present and analyse below allows us to decide which edges we need to actually add to the heap and which we do not and, similarly, which elements of the heap we need to preserve, and which can be thrown away. This allows us to substantially diminish the section of the frontier we store in the heap with very little extra work. The resultant heap size is still polynomial in $n$ but with power $<1$, and the logarithm translates this into a constant factor speed-up.

The random variables $(U_e)$  couple Bernoulli percolation across all $p\in[0,1]$. We let $B_p=\{e:U_e\leq p\}$ be the edges open under percolation probability $p$, and we refer to any cluster of edges in $B_p$ as a $p$-cluster. The invasion percolation process explores in such a way that for any $p\in(0,1)$ we explore the entirety of the $p$-cluster we are in before exiting to explore a new $p$-cluster, which we again explore the entirety of before exiting, and so on. In particular, for each $p>p_c$ the invasion process eventually reaches an infinite $p$-cluster which it then remains inside in perpetuity.
For $p>p_c$, we let $T_p$ denote the time at which the exploration process first enters an infinite $p$-cluster, and we let $p(n)$ denote the inverse of $T_p$, i.e. $p(n) = \inf\{p>p_c:T_p<n\}=\inf\{p>p_c:\forall i\geq n, e_i>p\}$. 
 We observe that once we have entered one of these supercritical infinite $p$-clusters, edges $e$ in the frontier with $U_e>p$ will never be extracted from the frontier. We thus define the \emph{active frontier} $A_n = F_n \cap \{e:U_e\leq p(n)\}$ and call the complement  $F_n \setminus A_n$ the 
 \emph{inactive frontier}.

For each $p<p_c$, we eventually leave any $p$-cluster we are exploring, so that if we define $\tau_{p,i}$ to be the $i$th time invasion percolation exits a $p$-cluster then $\tau_{p,i}<\infty$ for every $p<p_c$ and $i\geq 1$. At time $\tau_{p,i}$ there do not exist any edges in the frontier with value less than $p$.
Moreover, conditional on the weights of the edges we have added to the cluster, the weight of each edge $e$ in the frontier is a uniform random number between $1$ and the maximal weight of an edge that was added to the cluster since $e$ was added to the frontier. 
As noted above, we have as in \cite{cmp/1104114182} that the frontier grows linearly in time. The size of the active frontier at time $\tau_{p,i}$ can therefore be bounded 
\[
\abs{A_{\tau_{p,i}}} = O\left( \tau_{p,i} \cdot |p(\tau_{p,i})-p|\right).
\]
Now, if we assume the ansatz
\[
\tau_{p,i}-\tau_{p,i-1}\sim \abs{p-p_c}^{-\alpha},\ T_{p}\sim \abs{p-p_c}^{-\beta},
\] for some $\alpha,\beta>0$, 
we get for $p$ near $p_c$ that
\[
\abs{A_{\tau_{p,i}}} = O\left( \tau_{p,i} \cdot\left( p_c-p +{\tau_{p,i}}^{-1/\beta}\right) \right).
\]
 Taking $p_i<p_c$ to be chosen so that $\abs{p_i-p_c}\sim i^{-\gamma}$ for some large $\gamma$, we then approximate  $\tau_{p,i}\sim i \abs{p-p_c}^{-\alpha}$, so $\abs{p-p_c}\sim \tau_{p,i}^{-\gamma/(1+\alpha\gamma)}$, and thus
\[\abs{A_{\tau_{p_i,i}}}=O\left( \tau_{p,i}^{1-1/\alpha}+\tau_{p,i}^{1-1/\beta}\right).
\]
This suggests that the active frontier is smaller than the frontier by a power of $n$ as $n\to \infty$.

\medskip

\textbf{Implementation.}
The above analysis suggests that if we knew $T_p$ exactly, we could store just the active frontier, and so utilize a heap of size approximately $Cn^{\max\{1-\alpha^{-1},1-\beta^{-1}\}}$, rather than $Cn$. Thus, taking logarithms, we could achieve a multiplicative speed up of the running time proportional to ${\max\{1-\alpha^{-1},1-\beta^{-1}\}}^{-1}$. 
Although we do not know $p_c$ or $T_p$ exactly, this suggests the following improvement to the invasion percolation algorithm.
At each time step $n$, we will attempt to estimate $p(n)$ via the ansatz
\[p(n) =  a_n + Fn^{-z}, \]
where $a_n$ is the bulk-to-boundary ratio and $F$ and $z$ are constants. We estimate the constants $F$ and $z$
by running  invasion  percolation $m$ times, each with run length $k$, and label by $e_{j,i}$ the $i$th frontier values of run $j$,  and by $a_{j,i}$, the $i$th bulk-to-boundary ratio of run $j$,  we can find $F$ and $z$ such that
\begin{equation} \label{eq:dbfit}
	Fn^{-z}\geq b_n := \sup_{j\leq m} ({e_{j,n} - a_{j,n}}),
\end{equation}
where we only consider $n\leq N$ for some $N<<k$. (In practice we set $m=100,\, k=10^6$ and visually estimated a suitable $N$.) We then slightly modulate the obtained values ($F$ upwards, $z$ downwards) to account for the finite number of runs and the finite run length.

During further runs of invasion percolation, we then calculated $Fn^{-z}$ at each time step, and did not add edges to the frontier which had values $U_e\geq a_n+Fn^{-z}$. Additionally, every $10^6$ time steps, we did a sweep of the frontier, and deleted any remaining frontier edges which had values $\geq Fn^{-z}$.
A nice feature of this method is that it is not very important to get accurate estimates of $F$ and $z$. Indeed, the algorithm will always give an exact sample of invasion percolation but may be forced to halt if a poor choice of $F$ and $z$ leads the active frontier to become empty; if we choose $F$ and $z$ so that this is never observed to happen, we speed up our sampling of invasion percolation without distorting the distribution of the process in any way. The fact that the algorithm halts when $F$ and $z$ are chosen incorrectly can also be used to 
automate the calculation of $F$ and $z$.

\end{multicols}

\end{document}